\documentclass[letterpaper,twocolumn,10pt]{article}
\usepackage{zhanggroup}

\usepackage{booktabs}
\usepackage{multirow}
\usepackage{amsthm}
\usepackage{amsmath}
\usepackage{amssymb}
\usepackage{xspace}
\usepackage{xcolor}
\usepackage{colortbl}
\usepackage{graphicx}
\usepackage[absolute]{textpos}
\usepackage{caption,subcaption}
\usepackage{picture,graphics}

\hypersetup{
  colorlinks,
  linkcolor={blue!60!green},
  citecolor={green!60!blue},
  urlcolor={orange!60!red}
}

\newcommand{\refappendix}[1]{\hyperref[#1]{Appendix~\ref*{#1}}}

\theoremstyle{definition}
\newtheorem{definition}{Definition}[section]
\theoremstyle{remark}
\newtheorem*{remark}{Remark}

\definecolor{mygray}{gray}{.9}
\newcommand{\mypara}[1]{\smallskip\noindent\textbf{#1:}}

\newcommand{\drec}{\ensuremath{\mathcal{D}_{\textit{rec}}}\xspace}

\newcommand{\dtar}{\ensuremath{\mathcal{D}_{\textit{tar}}}\xspace}

\newcommand{\algo}{\ensuremath{\mathcal{A}}\xspace}
\newcommand{\knowledge}{\ensuremath{\mathcal{K}}\xspace}

\begin{document}

\date{}

\title{\Large \bf SoK: Data Reconstruction Attacks Against Machine Learning \\Models: Definition, Metrics, and Benchmark} 

\author{
{\rm Rui Wen\textsuperscript{1}\thanks{The first two authors made equal contributions.}}\ \ \
{\rm Yiyong Liu\textsuperscript{2}\textsuperscript{\textcolor{blue!60!green}{$\ast$}}}\ \ \
{\rm Michael Backes\textsuperscript{2}}\ \ \
{\rm Yang Zhang\textsuperscript{2}\thanks{Corresponding author}}
\\
\textsuperscript{1}\textit{Institute of Science Tokyo}\ \ \ 
\textsuperscript{2}\textit{CISPA Helmholtz Center for Information Security}
}

\maketitle

\begin{abstract}
Data reconstruction attacks, which aim to recover the training dataset of a target model with limited access, have gained increasing attention in recent years.
However, there is currently no consensus on a formal definition of data reconstruction attacks or appropriate evaluation metrics for measuring their quality.
This lack of rigorous definitions and universal metrics has hindered further advancement in this field. 
In this paper, we address this issue in the vision domain by proposing a unified attack taxonomy and formal definitions of data reconstruction attacks. 
We first propose a set of quantitative evaluation metrics that consider important criteria such as quantifiability, consistency, precision, and diversity. 
Additionally, we leverage large language models (LLMs) as a substitute for human judgment, enabling visual evaluation with an emphasis on high-quality reconstructions. 
Using our proposed taxonomy and metrics, we present a unified framework for systematically evaluating the strengths and limitations of existing attacks and establishing a benchmark for future research. 
Empirical results, primarily from a memorization perspective, not only validate the effectiveness of our metrics but also offer valuable insights for designing new attacks.
\end{abstract}

\section{Introduction}
\label{section:intro}

The prosperous development of machine learning techniques has been witnessed in the past decade, which fertilizes its application in real-world scenarios.
However, training a machine learning model for privacy-crucial tasks, such as person identification~\cite{BTS13,YSLXSH22}, disease prediction~\cite{KA18}, and financial risk prediction~\cite{CG162}, demands a large volume of data which is not only valuable but also sensitive.
As a result, model owners tend to release the model only, taking it for granted that the training data will not be leaked.

However, a series of works demonstrate that with limited access to a target model, the adversary is capable of inferring partial/complete information about the model's training samples.
As a representative, membership inference attacks (MIA)~\cite{SSSS17,SZHBFB19,LZ21,WLBZ24,LLWHYZFZ24,HWWBSZ21,LLHYBZ22,WWBBHSZ24,HLXCZ22,LHYZ24, ZYWBZ24} denote a line of works that aim to infer whether a specific data sample is in the target model's training dataset.
The disclosure of the membership status is a severe privacy breach as this information could indicate certain sensitive properties of the target sample.

A more challenging privacy attack is data reconstruction, which aims to recover the entire training dataset of a target model. 
This attack is considered the ultimate privacy breach, as it exposes all information about every sample. 
While membership inference (MIA) shares some similarities with data reconstruction, several key differences make data reconstruction a stronger attack. 
MIA is a sample-level attack, determining the membership status of individual samples, while data reconstruction is a dataset-level attack aimed at extracting the entire training dataset. 
From another angle, MIA is a decision problem, whereas data reconstruction is a search problem. 
In theory, MIA could aid in data reconstruction if the adversary has a large candidate dataset containing all the target samples and can perfectly predict membership. 
However, these assumptions are too strong, and to our knowledge, no one has attempted to use MIA for data reconstruction.

Owing to the crucial role of data reconstruction in the privacy domain, numerous attacks have been proposed in recent years. 
For example, Fredrikson et al.~\cite{FJR15} propose the first data reconstruction attack, namely model inversion, which requires white-box access to the target model.
Later, Yang et al.~\cite{YZCL19} relax this assumption by leveraging a training-based approach.
Zhang et al.~\cite{ZJPWLS20} adopt a Generative Adversarial Network (GAN)~\cite{GPMXWOCB14} to enhance the reconstruction quality.
However, some researchers~\cite{SSSS17,MSCS19} have claimed that model inversion can only recover a representative sample for each class of the target model, thus not an ideal data reconstruction attack.
On the other hand, for certain training paradigms like federated learning~\cite{ZLH19} and online learning~\cite{SBBFZ20}, some researchers have shown that they are able to reconstruct individual training samples.
Moreover, all these works have used different types of evaluation metrics (see \autoref{section:existingmetrics} for more details).

Despite all the efforts, one of the major problems in the field of data reconstruction is that there does not exist a rigorous and unified definition for the attack.
Moreover, the community has no consensus on what are the proper metrics for attack evaluation.
This predicament roots in the diversity of the attack scenarios, e.g., various threat models and different training paradigms.
At the same time, the lack of universal metrics exacerbates this problem since no existing metric is able to reflect all aspects of the reconstructions.
We argue that without a clear-stated definition and metrics, it is hard to conduct further investigation in this direction.

In this paper, we take the first step in tackling this problem by 1) providing a definition of data reconstruction, 2) proposing a set of evaluation metrics, and 3) performing a large scale of experiments to establish a benchmark for further study.
Specifically, our contributions can be summarized as follows:
\begin{itemize}
    \item Definition: We recapitulate the reconstruction goal and attack information in existing work and formally provide an attack taxonomy.
    Based on this, we quantitatively and rigorously define the data reconstruction attack.
    \item Metrics: We formalize the desiderata for evaluation metrics and, based on this framework, propose two sets of metrics that address both macro and micro aspects of reconstruction, while emphasizing the importance of diversity in reconstruction quality. 
    Additionally, we mitigate the limitations of using visualization as an evaluation tool by incorporating large language models (LLMs), with this metric specifically designed to assess high-quality reconstructions.
    \item Evaluation: We present a unified framework for data reconstruction attacks. 
    Especially from the perspective of memorization, we conduct a thorough evaluation of ten reconstruction attacks under various attack scenarios. 
    Our experiments demonstrate the effectiveness of the proposed metrics and provide a comprehensive analysis of existing attacks.
\end{itemize}

\mypara{Implications}
In this work, we propose the first rigorous definition of data reconstruction and develop a set of metrics.
Our evaluation can serve as a benchmark for the current state-of-the-art approaches.
We believe our results pave the way for further investigation into data reconstruction.
We will share our code to facilitate research in the field in the future.

\section{Background}
\label{section:pre}

\subsection{Machine Learning Models}

Machine learning algorithms aim to construct models that can accurately predict given inputs.
These models are typically represented by a parameterized function $f_{\theta}: \mathcal{X}\rightarrow\mathcal{Y}$, where $\mathcal{X}$ denotes the input space and $\mathcal{Y}$ denotes the output space containing all possible predictions.
To determine parameters $\theta$ that lead to optimal performance, a common approach is to minimize the following objective function using backpropagation:
\[
\min_{\theta}\mathcal{L}(f_{\theta}(x),y)
\]
where $\mathcal{L}$ denotes the classification loss, $(x,y)\in\mathcal{X}\times\mathcal{Y}$ are samples used to train the target model, constituting the training dataset.

Recent research shows that given access to a trained target model $f_{\theta}$, the adversary might exploit it through techniques such as membership inference~\cite{SSSS17,SZHBFB19,LWHSZBCFZ22,LZ21,CCNSTT22,HWWBSZ21,LZBZ22,LLHYBZ22} and data reconstruction attacks~\cite{FJR15,DeepDream,YMALMHJK20,ZJPWLS20,YZCL19,SBBFZ20,ZLH19,HVYSI22}.
These attacks focus on inferring micro-level information about individual training samples, potentially leading to a direct privacy breach.

\subsection{Data Reconstruction Attacks}
\label{section:existrec}

The data reconstruction attack aims to recover the target dataset with limited access to the target model, with the aid of additional knowledge possessed by the adversary.

Existing attacks broadly fall into three categories: optimization-based, training-based, and analysis-based. 
In this paper, we thoroughly investigate ten representative reconstruction attacks, analyzing their performance and limitations. 
These attacks are briefly introduced below:

\subsubsection{Optimization-based Attack}

Most existing reconstruction attacks can be classified into this particular attack type. 
Such attacks aim to reconstruct the training dataset by iteratively optimizing the input until the desired class achieves a high likelihood score.

Recent attacks have incorporated generative models to enhance the quality of reconstruction, employing different architectural choices~\cite{CKJQ21,WFLKZM21} and loss functions~\cite{SHCAK22}.  We select seven representative attacks in our paper.

\mypara{MI-Face~\cite{FJR15}}
Fredrikson et al.\ take the first step in reconstructing training samples from a trained model.
Given a class $y$, the method first initializes a sample, and then updates $x$ to maximize the likelihood/probability of belonging to that class, i.e.,
\[
\min_{x}\mathcal{L}(f_{\theta}(x),y)
\]
where $\mathcal{L}$ is the classification loss.
Similar to training a machine learning model, MI-Face also uses backpropagating.  
But the difference is MI-Face focuses on optimizing $x$ rather than the parameters $\theta$ in normal training.
The backpropagation process demands white-box access to the target model, and the reconstructed sample always converges to the most confident $x$ near the initial point.  
Because $x$ is high-dimensional, the reconstruction quality highly depends on the initialization.

\mypara{DeepDream~\cite{DeepDream}}
DeepDream was originally proposed to interpret machine learning models.
However, this approach can also be utilized to improve MI-Face to acquire better results.
The key idea is introducing regularization terms that force reconstructed samples to share similar statistics to natural images by penalizing each sample's total variance and $\ell_2$ norm:
\[
\min_{x}\mathcal{L}(f_{\theta}(x),y)+\alpha_{\text{tv}}\mathcal{R}_{\text{tv}}(x)+\alpha_{\ell_2}\mathcal{R}_{\ell_2}(x)
\]
where $\mathcal{R}_{\text{tv}}$ and $\mathcal{R}_{\ell_2}$ denote total variance and $\ell_2$ norm, respectively.

\mypara{DeepInversion~\cite{YMALMHJK20}}
DeepInversion further improves DeepDream by adding another loss.  
Its intuition is that the batch normalization layers encoded statistical information about the training samples.
Thus, minimizing the distance between reconstructed statistics and those stored in the target model helps.

\mypara{Revealer~\cite{ZJPWLS20}}
Contrary to exploiting information encoded in the target model, Revealer leverages an auxiliary dataset to train a Generative Adversarial Network (GAN) $G$ that generates samples $x$.  
Now instead of optimizing $x$, Revealer optimizes the input random seed $z$ to $G$ (and the reconstructed sample would be $x=G(z)$):
\[
\min_{z}\mathcal{L}(f_{\theta}(G(z)),y)
\]
The intuition is to utilize GAN to force the output $G(z)$ to always look `real' on any optimized $z$, at the same time, guarantee high confidence in the prediction.

\mypara{KEDMI~\cite{CKJQ21}}
KEDMI enhances the GAN-based approach in two primary ways. 
First, it optimizes the process of extracting knowledge from the auxiliary dataset by modifying the GAN's training objective. 
Specifically, it leverages labels assigned by the target model to the auxiliary dataset. 
The discriminator is trained to not only distinguish between real and fake samples but also to differentiate among the labels, enabling more nuanced learning.
Second, instead of focusing on reconstructing single data points, KEDMI targets the reconstruction of the entire data distribution. 
To achieve this, it explicitly parameterizes the training data distribution and approximates the reconstructed distribution using its distributional parameters, such as the mean ($\mu$) and standard deviation ($\sigma$).

\mypara{PLGMI~\cite{YCZZYZ23}}
PLGMI decouples the search space by training a conditional GAN (cGAN). 
It utilizes pseudo-labels generated by a top-n selection strategy to steer the training process, combined with a max-margin loss, which collectively improves the effectiveness of the attack.

\mypara{Deep-Leakage~\cite{ZLH19}}
Deep-Leakage allows the adversary to access gradients (originally designed for federated learning systems) and aims to reconstruct training samples corresponding to the gradients.
Specifically, the adversary randomly creates ``dummy input'' and ``dummy label'' and computes gradients based on this input-label pair.
By optimizing this input-label pair to approximate true gradients, the ``dummy input'' and ``dummy label'' converges to target samples.

\subsubsection{Training-based Attack}

\mypara{Inv-Alignment~\cite{YZCL19}}
To overcome the limitation that data reconstruction requires white-box access to the target model, Yang et al.\ opt for a training-based approach that works with black-box access.
Briefly, they construct an autoencoder with the target model as the encoder part.
Once the autoencoder is well-trained, the decoder part can be leveraged to reconstruct inputs given corresponding posteriors.

\mypara{Updates-Leak~\cite{SBBFZ20}}
Updates-Leak considers the online-learning scenario where the adversary has access to different versions of the target model and tries to reconstruct samples used to update the model.
The adversary trains numerous shadow models to mimic the updating procedure and leverage the posterior difference to reconstruct target samples.

\subsubsection{Analysis-based Attack}

\mypara{Bias-Rec~\cite{HVYSI22}}
Haim et al.\ theoretically prove that the training data can be fully recovered given certain assumptions.
In detail, if the target model is a homogeneous ReLU network and trained on a binary dataset using gradient flow, then \textit{the parameters are linear combinations of the derivatives of the network at the training data points}~\cite{HVYSI22}.
According to this assertion, the adversary can derive training samples via optimization.

It is worth mentioning that the recent attack proposed by Balle et al.~\cite{BCH22} has shown promising results in reconstructing the missing sample in a dataset.
However, their attack assumes the adversary has the whole dataset except for the reconstructed one (in order to verify differential privacy properties), whose attack scenario significantly differs from common settings. 
Additionally, certain attacks~\cite{BGCDJ19,SME19,KD18} rely on invertible network architectures to execute their malicious actions. 
However, as these attack scenarios do not align with the scope of our investigation, we do not consider their attacks in this paper.

\subsection{Data Reconstruction vs.\ Membership Inference}

Membership inference attack (MIA) is another representative attack that aims to expose information about a training dataset, specifically by determining whether a target sample is part of it. 
There are two primary differences between data reconstruction and MIA, which necessitate the use of distinct attack methods.
First, MIA takes a sample as input to decide its membership status, whereas data reconstruction only has a target model.
This difference impels data reconstruction to employ an incompatible attack approach, as the state-of-the-art MIA regularly involves training samples in the attack process.
For example, Carlini et al.~\cite{CCNSTT22} train two sets of shadow models with datasets with/without the candidate sample. 
Such a training discrepancy is a crucial factor in enabling successful attacks.

From a different view, MIA can be framed as a decision problem while data reconstruction is a search problem.
This distinction implies that data reconstruction is a more challenging and sophisticated attack. 
In an ideal scenario, perfect data reconstruction would reveal all membership statuses, as the entire training dataset would be exposed. 
MIA could serve as the foundation for data reconstruction, given the assumption that the adversary has a dataset that includes all training samples. 
However, in reality, this assumption is not feasible, and the adversary must try all possible pixel combinations iteratively, which is impractical due to the enormity of the search space. 
Additionally, no existing MIA model can provide perfect accuracy in determining membership status, necessitating the development of alternative approaches for conducting data reconstruction attacks.

\subsection{Memorization}
In the realm of attacks targeting the disclosure of information from training datasets, memorization serves as a key concept closely linked to attack performance. 
Within machine learning, memorization refers to a model's unintentional retention of intricate details from its training data, which is particularly evident in high-capacity models. 
Feldman~\cite{F20} provides a succinct definition of memorization for a target sample \((x_i,y_i)\) with index \(i\), describing it as the impact of removing a data point on the model's prediction for that particular point:
\begin{equation}
\label{eqn:mem}
\begin{gathered}
   \textit{mem}(\mathcal{A},\mathcal{D},i)=
   \underset{f_\theta{\sim\mathcal{A}(\mathcal{D})}}{\text{Pr}}[f_\theta(x_i)=y_i]-\underset{f_\theta{\sim\mathcal{A}(\mathcal{D}^{\symbol{92} i})}}{\text{Pr}}[f_\theta(x_i)=y_i]
\end{gathered}
\end{equation}
where $\mathcal{D}^{\symbol{92} i}$ denotes the dataset with the sample \((x_i,y_i)\) removed.

While memorization can be beneficial, and even indispensable~\cite{F20}, for achieving high model performance, it simultaneously poses risks that adversaries can exploit to expose sensitive information.

Tram{\`e}r et al.~\cite{TSJLJHC22} demonstrates that enhancing a model's memorization through data poisoning attacks significantly increases the effectiveness of various privacy attacks, including membership inference, attribute inference, and data extraction. 
Further investigations by Carlini et al.~\cite{CJZPTT22} illustrate the ``privacy onion effect'', highlighting that vulnerabilities arising from memorization cannot be easily mitigated by merely removing outliers.

Conversely, reducing a model’s memorization of the training data can mitigate its vulnerability to privacy attacks, as seen in approaches like differential privacy~\cite{ACGMMTZ16,DR14,PSMRTE18}. 
However, this often comes at the expense of model performance. 
Despite its importance, the relationship between memorization and data reconstruction remains surprisingly underexplored. 
We aim to address this gap by benchmarking existing data reconstruction attacks from the perspective of memorization.

\begin{table*}[th]
\centering
\caption{We group ten reconstruction attacks from three dimensions, which indicate the necessary information for the attack, including training type, model access, and dataset access.
Note that attacks requiring more information about the model or dataset can be extended from attacks that require less information.}
\scalebox{0.92}{
\begin{tabular}{@{}ccccc@{}}
\toprule
\multirow{2}{*}{Training Type} & \multirow{2}{*}{Model Access} & \multicolumn{3}{c}{Dataset Access}           \\ \cmidrule(l){3-5} 
                               &                               & No Data       & Similar Distribution & Same Distribution     \\ \midrule
\multirow{5}{*}{Static}        & Black-Box                    &               &    Inv-Alignment          & Inv-Alignment \\ \cmidrule(l){2-5} 
                               & \multirow{4}{*}{White-Box}    & MI-Face       &    Revealer    & Revealer      \\
                               &                               & DeepDream     &    KEDMI       &   KEDMI            \\
                               &                               & DeepInversion &    PLGMI          &     PLGMI          \\
                               &                               & Bias-Rec      &              &               \\ \midrule
\multirow{2}{*}{Dynamic}       & Black-Box                    &               &              & Updates-Leak  \\ \cmidrule(l){2-5} 
                               & White-Box                    & Deep-Leakage  &              &               \\ \bottomrule
\end{tabular}
}
\label{table:taxonomy}
\end{table*}

\section{Defining Data Reconstruction}
\label{section:definition}

\subsection{Reconstruction Taxonomy}

To establish a rigorous definition of data reconstruction attacks, it is necessary to take into account various aspects such as training type, model access, and dataset access.
To this end, we present a taxonomy that captures these aspects and use it to formulate a formal definition of data reconstruction. 
We are motivated by two key questions that arise in this context.

\noindent\textbf{1) What data does the adversary aim to reconstruct?} 
The diversity of attack scenarios poses challenges to developing a unified definition of data reconstruction attacks.
Some attacks focus on the standard setting where the model is trained on one fixed dataset, while others focus on settings that entail dataset change during the training process, e.g., online learning.
Additionally, some attacks only consider data that contributes to certain updates.
This diversity makes it hard to incorporate all possibilities into a unified definition.

\noindent\textbf{2) What information does the adversary have?}
This problem also stems from the diverse attack scenarios where the adversary has varying levels of information. 
For example, when users release their models, the adversary has white-box access to the model, while if models only provide a query interface, the adversary may lack detailed information about the model's architecture and parameters. 
Furthermore, the adversary may or may not have access to the same distribution of the target dataset. 
Thus, it is crucial to consider all possible situations to provide a comprehensive definition of data reconstruction attacks.

In the following, we categorize reconstruction attacks from three dimensions: training type, model access, and dataset access:

\mypara{Training Type}
We categorize training into two types: \textit{static} and \textit{dynamic}, based on the target dataset's role. 
In static, the target dataset remains fixed during training, and the attack aims to recover it. 
In dynamic, the target dataset is introduced during training, causing model changes, as the updating dataset in online learning.

\mypara{Model Access}
The attacker may have one of the following access to the target model:
    \textit{Black-box Access}, which means the attacker could only query the model in an API manner.
    \textit{White-box Access}, which means the attacker could get full information about the target model, including the model architecture, parameters, and even gradients of the target sample calculated on the target model.
We also acknowledge that real-world scenarios may involve intermediate access, representing a hybrid of black-box and white-box approaches. 
For example, an encoder might only be queryable via an API (black-box), while subsequent classification layers are directly accessible (white-box). 
Our framework's analysis of these two extremes, pure black-box and full white-box, effectively bounds the attack performance for all such intermediate cases.

\mypara{Dataset Access}
The attacker may have one of the following types of knowledge about the target dataset: \textit{No Data}, meaning no access to any dataset information; \textit{Same Distribution}, meaning the attacker can sample data from the same distribution as the target dataset; \textit{Similar Distribution}, meaning the attacker can sample from distributions similar to the target. 
The key difference between ``same'' and ``similar'' distributions lies in the level of specificity of the distribution information that the adversary possesses: ``same'' refers to detailed information, e.g., images of a specific person, while ``similar'' refers to more general knowledge, e.g., knowing the dataset contains human faces.

In the following, we denote such information as extra knowledge (\knowledge), defined as below:

\begin{definition}[Extra Knowledge \knowledge]
Extra knowledge \knowledge provides the necessary information required for the reconstruction attack.
A standard extra knowledge should contain information from three dimensions:
\begin{enumerate}
    \item \textbf{Training Type:} static or dynamic
    \item \textbf{Model Access:} black-box access or white-box access
    \item \textbf{Dataset Access:} no data or similar distribution or same distribution
\end{enumerate}
\end{definition}

It should be emphasized that while our present investigation concerns the realm of vision, the attack taxonomy we have formulated has broader applicability to reconstruction attacks in diverse domains.

We categorize the ten attacks in~\autoref{table:taxonomy}. 
Current reconstruction attacks most focus on the scenario where the adversary has white-box access to the target model, as this access provides crucial information that aids in dataset reconstruction.
For attacks without model information, they tend to rely on information from a dataset that shares similar characteristics with the target dataset.  
For example, Inv-Alignment and Updates-Leak do not require white-box access but require data from the same distribution as a substitute.
Furthermore, attacks with model information can be further improved with the help of dataset information.
For instance, PLGMI leverages information from both the model and the dataset, resulting in improved performance compared to attacks that rely solely on model or dataset information, as demonstrated in the evaluation part.

Theoretically, attacks with black-box access can be easily extended to white-box attacks, while the inverse transition is not straightforward.
One potential solution to transfer white-box attacks to black-box attacks is through the use of model stealing to extract their internal parameters. 
Additionally, for attacks that leverage information about the same distribution, we also investigate the feasibility of using similar distribution. 
These aspects are examined in~\autoref{section:mem_utilization}.

\subsection{Reconstruction Definition}
\label{section:recdef}

Given the extra information, another remaining question is the number of samples to reconstruct. 
Existing attacks generate a surplus of samples and designate those that best align with the target dataset as the reconstruction outcome.
However, this approach is deemed unsuitable as it imparts specific information about the target dataset that is unavailable to the adversary. 
For the same reason, permitting the adversary to generate an infinite number of samples is inappropriate as generating every conceivable pixel combination can subsume the target dataset, yet such a reconstruction does not furnish any useful information.

Therefore, we explicitly stipulate that the reconstructed dataset has the same size as the target model.
Generating a greater number of samples than the target samples and selecting high-quality reconstructed samples is allowed. 
But it is crucial that the selection procedure does not involve any information about the target dataset.

It should be emphasized that the reconstruction size is a requisite for evaluation and not for the adversary. 
Furthermore, our definition encompasses scenarios where the focus is on reconstructing a subset of the training dataset. 
A detailed description of this is presented in~\autoref{section:ourmetrics}.

Provided the reconstruction size, the target model, and the necessary information indicated in extra knowledge, we formally define the reconstruction algorithm as follows:

\begin{definition}[Reconstruction Algorithm]

Given a target model $m\in\mathcal{M}$ and extra knowledge $k\in\knowledge$, reconstruction algorithm \algo could reconstruct a dataset $\drec=\mathcal{A}^k(m)\in\mathbb{D}$, i.e., \algo: $\mathcal{M}\times$ \knowledge $\rightarrow\mathbb{D}$, with the same size as the target dataset \dtar.
\end{definition}
In the static setting, the target dataset \dtar is the whole training dataset of the target model; in the dynamic setting, the target dataset refers to the subset of data that directly contributes to the model change.

\section{Evaluation Metric for Data Reconstruction}
\label{section:metrics}

A robust evaluation metric is crucial for the development of data reconstruction attacks, analogous to the role of loss functions in guiding model optimization.
This section reviews existing metrics, outlines the desired properties of ideal metrics, and introduces our proposed metrics in~\autoref{section:ourmetrics}.

\subsection{Existing Common Metrics}
\label{section:existingmetrics}

\mypara{Visualization}
Visualization is a useful tool for understanding the quality of reconstructions and has been widely used in previous work~\cite{FJR15,DeepDream,YMALMHJK20,ZJPWLS20,YZCL19,SBBFZ20,ZLH19,YWLXL22,SHCAK22}.
While visualization provides the most direct and intuitive impression of reconstruction quality, its non-quantitative nature and reliance on subjective human judgment limit its effectiveness as an evaluation metric. 
Consequently, it is crucial to also utilize quantitative metrics in order to accurately assess reconstruction quality.

\mypara{MSE/PSNR/SSIM}
Mean Squared Error (MSE) and other similarity measures, such as Peak Signal-to-Noise Ratio (PSNR) and Structural Similarity Index (SSIM), are commonly utilized to provide quantitative evaluation results~\cite{ZJPWLS20,YZCL19,SBBFZ20,ZLH19}.
However, such metrics only measure the similarity between individual samples rather than the overall dataset, which poses two issues. 
First, the choice of sample pairs for comparison is not fixed, causing evaluation results to vary based on the selected pairs. 
Second, sample-level metrics can't capture the diversity of the reconstructed dataset. 
For example, if all reconstructed samples are similar to \textit{one} target sample, the measured distance may be small, but the reconstruction may still be unsatisfactory.

\mypara{Feature Distance}
Feature distance measures the similarity in the feature space and has been used in previous works~\cite{ZJPWLS20,CKJQ21}. 
Concretely, for each reconstructed sample, the distance to the centroid of its class in the feature space is calculated. 
However, as the feature space is determined by an evaluation network, the evaluation results can be inconsistent because different evaluation networks use different feature spaces with varying class centroids. 
Additionally, like other sample-level metrics such as MSE, this measure doesn't capture the diversity of the reconstruction.

\mypara{Accuracy (Train)}
To incorporate the macro similarity of the reconstructed dataset, one method uses reconstructed samples to train a model for the same task~\cite{YMALMHJK20,ZJPWLS20,YWLXL22}.
The model's testing accuracy reflects the reconstruction quality, with higher accuracy indicating better reconstruction.
We point out that this metric overlooks the precision of the reconstruction, as demonstrated by a counterexample in~\autoref{subfigure:metric_trainacc}.
Concretely, we utilize the data-free model extraction method~\cite{TMWP21} to generate a dataset, which is then used to steal the target model. 
Although the resulting stolen model exhibits high accuracy, the generated dataset, which serves as the training dataset of the stolen model, is vastly dissimilar from the target dataset.

\mypara{Accuracy (Test)}
Instead of training a model on the reconstructed dataset, one can train an evaluation model on a dataset from the same distribution as the target model and assess whether the evaluation model can accurately classify each reconstructed sample~\cite{ZJPWLS20,YWLXL22,SHCAK22,CKJQ21,WFLKZM21}.
The idea is that high-quality reconstructions should contain recognizable patterns that the evaluation model can capture. 
However, we argue that this metric is ineffective, as shown by examples that resemble random noise but are still classified with high confidence by the evaluation model, as illustrated in~\autoref{subfigure:metric_testacc}.
These counterexamples, generated using MI-Face~\cite{FJR15}, achieve prediction confidences above 0.9 for their corresponding classes.

\subsection{Design Desiderata}
\label{section:desiderata}

To propose metrics that can reflect the quality of a reconstruction, we need to consider four questions: 1) can such metrics \textit{quantitatively} measure the quality? 2) can such metrics provide a \textit{consistent} result for a fixed reconstruction? 3) can such metrics embody the quality in a \textit{micro} aspect? 4) can such metrics incorporate the quality in a \textit{macro} aspect?

In response to the drawbacks of existing metrics, we propose a set of properties that suitable evaluation metrics should possess, including quantifiability, consistency, precision, and diversity:
\begin{itemize}
    \item \textbf{Quantifiability.}
    The evaluation metric should provide quantitative results and eliminate the influence of subjective factors.
    \item \textbf{Consistency. }
    The evaluation metric should be consistent, that is, given a pair of the reconstructed dataset and the target dataset, the evaluated result should be determined, regardless of the testing time, place, and order.
    \item \textbf{Precision.}
    The evaluation metric should be aware of the reconstruction precision. 
    Specifically, it should capture the sample-level similarity of reconstructed samples to target samples.
    \item \textbf{Diversity.}
    The evaluation metric should be aware of the reconstruction diversity.
    Concretely, the metric should reflect the percentage of data being reconstructed.
    An attack that can only recover a few samples accurately is not regarded as a successful reconstruction attack.
\end{itemize}

\subsection{Definition of Our Quantitative Metrics}
\label{section:ourmetrics}

To capture both precision and diversity aspects, we introduce two metrics from different perspectives. 
The first is a dataset-level metric that measures the distribution similarity between the reconstructed and target datasets.

\begin{definition}[Dataset-level Metric]
The dataset-level metric $\mu:\mathbb{D}\times\mathbb{D}\rightarrow \mathbb{R}$ is defined as a mapping that takes two datasets and produces a single real number $\text{D-Dis}$, i.e., $\text{D-Dis}=\mu(\mathcal{A}^k(m),\dtar)$.

\end{definition}
We leverage Fr\'echet Inception Distance (FID) to incorporate the macro quality of the reconstructed dataset, which measures the distance between different data distributions.
Specifically, we approximate the two dataset distributions by Gaussian distributions as follows:
    \[
    \dtar\sim\mathcal{N}(\nu_{\textit{tar}},\Sigma_{\textit{ori}})\text{, and } \mathcal{A}^k(m)\sim\mathcal{N}(\nu_{\textit{rec}},\Sigma_{\textit{rec}})
    \]
We compute the D-Dis value as:
    \[
    ||\nu_{\textit{tar}}-\nu_{\textit{rec}}||_2^2+\textit{tr}\left(\Sigma_{\textit{tar}}+\Sigma_{\textit{rec}}-2(\Sigma_{\textit{tar}}^{\frac{1}{2}}\cdot\Sigma_{\textit{rec}}\cdot\Sigma_{\textit{tar}}^{\frac{1}{2}})^{\frac{1}{2}}\right)
    \]
    where $||\cdot||_2$ denotes the Euclidean distance, and $\textit{tr}(\cdot)$ denotes the trace of a matrix. 
    $\nu$ and variance $\Sigma$ denote the parameters of the corresponding Gaussian distributions.
    
We choose Fr\'echet Inception Distance (FID) to measure the dataset-level distance due to its established use as a metric for evaluating the quality of generated distributions. 
Furthermore, FID is model-agnostic and can be calculated using any feature extractor, but common practice involves using the pre-trained InceptionV3 regardless of the dataset\footnote{https://github.com/mseitzer/pytorch-fid}. 
Using FID to measure dataset-level distance also resolves concerns related to consistency, as feature distances are calculated using the same model across different tasks.
We further introduce sample-level metrics to improve our evaluation by taking precision into account.
Specifically, we concentrate on how similar the reconstructed sample is to the target sample in the original data space.

\begin{definition}[Sample-level Metric]
\label{def:samplelevel}
The sample-level metric $\mu:\mathbb{D}\times\mathbb{D}\rightarrow \mathbb{R}^2$ maps two datasets into two real numbers $\text{S-Dis}$ and $\alpha$, i.e., $(\text{S-Dis},\alpha)=\mu(\mathcal{A}^k(m),\dtar)$, where the first value $\text{S-Dis}$ calculates the averaged minimal distance between reconstructed dataset and the target dataset, and the second value $\alpha$ denotes the coverage of reconstructed part. 
Formally:
\[
\text{S-Dis}=\frac{1}{|\mathcal{A}^k(m)|}\sum_{x_i\in\mathcal{A}^k(m)}d(x_i,f(x_i))
\]
where $f:\mathcal{A}^k(m)\rightarrow \dtar$, such that, $\forall x_i\in\mathcal{A}^k(m)$, $f$ maps $x_i$ to the sample $f(x_i)\in\dtar$ with the minimal distance to $x_i$.

For coverage $\alpha$, it indicates the reconstructed diversity:
\[
\alpha=\frac{|f(\mathcal{A}^k(m))|}{|\dtar|}
\]
Here, $f(\mathcal{A}^k(m))$ denotes the image of $f$, and $\alpha\in(0,1]$, larger coverage indicates better diversity.
\end{definition}
Note that the choice of $d$ may vary depending on the aspect of reconstruction quality to be assessed.
Both metrics proposed above provide deterministic evaluation results, only based on the content of the reconstructed dataset, which satisfies the property of quantifiability and consistency.
Furthermore, the coverage reflects the diversity of the reconstruction. 
The precision of the metrics is partially illustrated in~\autoref{section:results}, and we discuss it in more detail in~\autoref{section:deep_inves_metric}.

We can see that the reconstruction of a subset is encompassed by our reconstruction definition. 
To clarify, we can replicate such samples, and the outcome will remain unaffected as the corresponding image in the target dataset is fixed, resulting in an unchanged coverage and distance.
In the event that accurately reconstructing a subset, it is conceivable that it will be characterized by a small distance, yet it may have limited coverage.
A small distance indicates an accurate reconstruction, while the small coverage indicates that the reconstructed samples constitute only a minor fraction.

Provided the evaluation metric, we can parameterize the attack to characterize the ability of data reconstruction attack.
We first formulate the ideal case where the attack exactly reconstructs the target dataset under the measurement of $\mu$.

\begin{definition}[$\mu$-Exact Reconstruction]
We say a reconstruction algorithm \algo achieves $\mu$-exact reconstruction if $\mu(\mathcal{A}^k(m),\dtar)=\boldsymbol{0}$.
\end{definition}

$\mu$ denotes the distance measure that specifies the quality of the reconstruction, where the output could be a tuple if the measure has more than one evaluation metric.
Note that $\mu$-Exact Reconstruction is a necessary and insufficient condition for the perfect reconstruction, depending on the choice of measure $\mu$, which further highlights the importance of selecting appropriate metrics to evaluate the reconstruction performance.

The relaxed version allows the reconstructed dataset to have a small distance $\boldsymbol{\varepsilon}$ with the target dataset, we refer to this version as $(\boldsymbol{\varepsilon},\mu)$-Approximate Reconstruction:
\begin{definition}[$(\boldsymbol{\varepsilon},\mu)$-Approximate Reconstruction]
We say a reconstruction algorithm \algo achieves $(\boldsymbol{\varepsilon},\mu)$-approximate reconstruction if $\mu(\mathcal{A}^k(m),\dtar)\leq\boldsymbol{\varepsilon}$.
\end{definition}
\begin{remark}
$(\boldsymbol{\varepsilon},\mu)$-Approximate Reconstruction corresponds to the specific case of $\mu$-Exact Reconstruction when $\boldsymbol{\varepsilon}=\boldsymbol{0}$.
\end{remark}

Our framework's core metrics effectively address a broad spectrum of attack goals. 
For instance, reconstructing a single missing sample~\cite{SCEKPSTB23} translates within our definition to a low coverage score, indicating its narrow scope, combined with a reconstruction-distance metric to assess fidelity. 
In contrast, an attack targeting an entire class aims for high coverage~\cite{FJR15}, signifying the retrieval of numerous distinct dataset items. 
Overall, the two metrics work together: distance measures reconstruction accuracy, while coverage scales with the breadth of the attacker’s objective, from single-sample inference to full-dataset recovery.

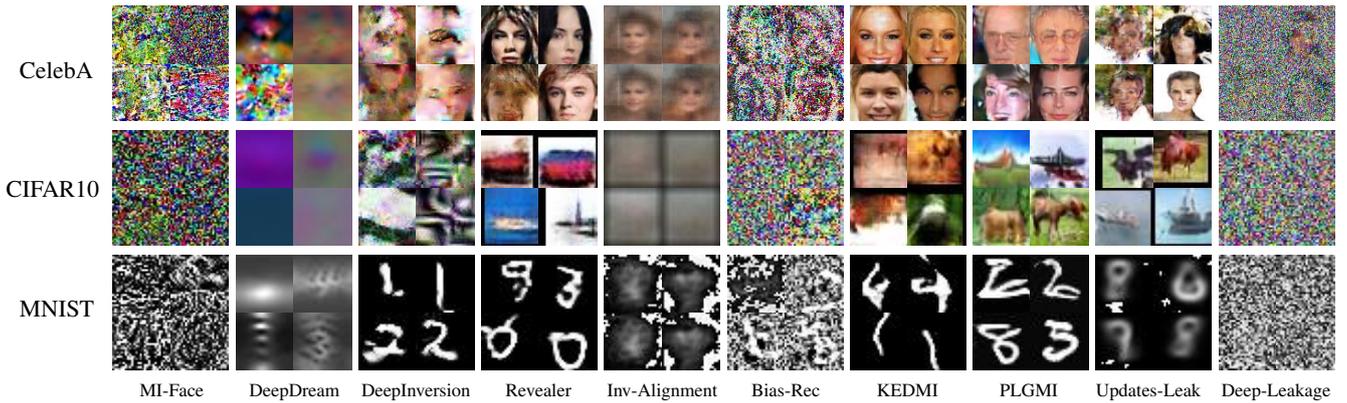
\begin{figure*}
\centering
\begin{picture}(500,165)
    \put(40,10){\includegraphics[width=0.09\columnwidth]{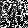}}
    \put(61.5,10){\includegraphics[width=0.09\columnwidth]{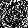}}
    \put(86,10){\includegraphics[width=0.09\columnwidth]{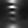}}
    \put(107.5,10){\includegraphics[width=0.09\columnwidth]{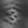}}
    \put(132,10){\includegraphics[width=0.09\columnwidth]{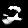}}
    \put(153.5,10){\includegraphics[width=0.09\columnwidth]{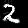}}
    \put(178,10){\includegraphics[width=0.09\columnwidth]{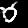}}
    \put(199.5,10){\includegraphics[width=0.09\columnwidth]{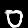}}
    \put(224,10){\includegraphics[width=0.09\columnwidth]{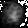}}
    \put(245.5,10){\includegraphics[width=0.09\columnwidth]{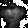}}
    \put(270,10){\includegraphics[width=0.09\columnwidth]{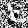}}
    \put(291.5,10){\includegraphics[width=0.09\columnwidth]{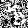}}
    \put(316,10){\includegraphics[width=0.09\columnwidth]{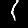}}
    \put(337.5,10){\includegraphics[width=0.09\columnwidth]{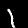}}
    \put(362,10){\includegraphics[width=0.09\columnwidth]{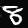}}
    \put(383.5,10){\includegraphics[width=0.09\columnwidth]{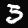}}
    \put(408,10){\includegraphics[width=0.09\columnwidth]{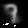}}
    \put(429.5,10){\includegraphics[width=0.09\columnwidth]{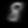}}
    \put(454,10){\includegraphics[width=0.09\columnwidth]{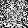}}
    \put(475.5,10){\includegraphics[width=0.09\columnwidth]{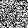}}
       
    \put(40,31.5){\includegraphics[width=0.09\columnwidth]{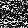}}
    \put(61.5,31.5){\includegraphics[width=0.09\columnwidth]{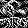}}
    \put(86,31.5){\includegraphics[width=0.09\columnwidth]{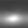}}
    \put(107.5,31.5){\includegraphics[width=0.09\columnwidth]{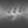}}
    \put(132,31.5){\includegraphics[width=0.09\columnwidth]{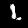}}
    \put(153.5,31.5){\includegraphics[width=0.09\columnwidth]{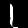}}
    \put(178,31.5){\includegraphics[width=0.09\columnwidth]{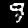}}
    \put(199.5,31.5){\includegraphics[width=0.09\columnwidth]{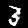}}
    \put(224,31.5){\includegraphics[width=0.09\columnwidth]{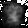}}
    \put(245.5,31.5){\includegraphics[width=0.09\columnwidth]{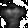}}
    \put(270,31.5){\includegraphics[width=0.09\columnwidth]{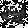}}
    \put(291.5,31.5){\includegraphics[width=0.09\columnwidth]{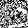}}
    \put(316,31.5){\includegraphics[width=0.09\columnwidth]{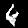}}
    \put(337.5,31.5){\includegraphics[width=0.09\columnwidth]{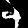}}
    \put(362,31.5){\includegraphics[width=0.09\columnwidth]{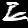}}
    \put(383.5,31.5){\includegraphics[width=0.09\columnwidth]{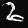}}
    \put(408,31.5){\includegraphics[width=0.09\columnwidth]{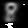}}
    \put(429.5,31.5){\includegraphics[width=0.09\columnwidth]{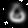}}
    \put(454,31.5){\includegraphics[width=0.09\columnwidth]{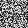}}
    \put(475.5,31.5){\includegraphics[width=0.09\columnwidth]{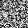}}
    
    \put(40,57){\includegraphics[width=0.09\columnwidth]{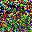}}
    \put(61.5,57){\includegraphics[width=0.09\columnwidth]{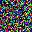}}
    \put(86,57){\includegraphics[width=0.09\columnwidth]{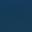}}
    \put(107.5,57){\includegraphics[width=0.09\columnwidth]{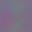}}
    \put(132,57){\includegraphics[width=0.09\columnwidth]{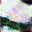}}
    \put(153.5,57){\includegraphics[width=0.09\columnwidth]{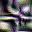}}
    \put(178,57){\includegraphics[width=0.09\columnwidth]{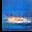}}
    \put(199.5,57){\includegraphics[width=0.09\columnwidth]{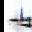}}
    \put(224,57){\includegraphics[width=0.09\columnwidth]{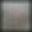}}
    \put(245.5,57){\includegraphics[width=0.09\columnwidth]{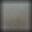}}
    \put(270,57){\includegraphics[width=0.09\columnwidth]{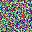}}
    \put(291.5,57){\includegraphics[width=0.09\columnwidth]{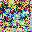}}
    \put(316,57){\includegraphics[width=0.09\columnwidth]{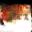}}
    \put(337.5,57){\includegraphics[width=0.09\columnwidth]{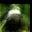}}
    \put(362,57){\includegraphics[width=0.09\columnwidth]{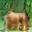}}
    \put(383.5,57){\includegraphics[width=0.09\columnwidth]{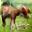}}
    \put(408,57){\includegraphics[width=0.09\columnwidth]{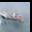}}
    \put(429.5,57){\includegraphics[width=0.09\columnwidth]{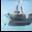}}
    \put(454,57){\includegraphics[width=0.09\columnwidth]{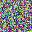}}
    \put(475.5,57){\includegraphics[width=0.09\columnwidth]{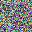}}    
    
    \put(40,78.5){\includegraphics[width=0.09\columnwidth]{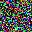}}
    \put(61.5,78.5){\includegraphics[width=0.09\columnwidth]{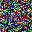}}
    \put(86,78.5){\includegraphics[width=0.09\columnwidth]{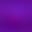}}
    \put(107.5,78.5){\includegraphics[width=0.09\columnwidth]{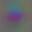}}
    \put(132,78.5){\includegraphics[width=0.09\columnwidth]{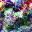}}
    \put(153.5,78.5){\includegraphics[width=0.09\columnwidth]{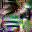}}
    \put(178,78.5){\includegraphics[width=0.09\columnwidth]{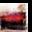}}
    \put(199.5,78.5){\includegraphics[width=0.09\columnwidth]{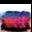}}
    \put(224,78.5){\includegraphics[width=0.09\columnwidth]{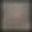}}
    \put(245.5,78.5){\includegraphics[width=0.09\columnwidth]{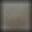}}
    \put(270,78.5){\includegraphics[width=0.09\columnwidth]{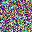}}
    \put(291.5,78.5){\includegraphics[width=0.09\columnwidth]{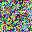}}
    \put(316,78.5){\includegraphics[width=0.09\columnwidth]{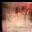}}
    \put(337.5,78.5){\includegraphics[width=0.09\columnwidth]{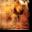}}
    \put(362,78.5){\includegraphics[width=0.09\columnwidth]{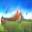}}
    \put(383.5,78.5){\includegraphics[width=0.09\columnwidth]{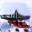}}
    \put(408,78.5){\includegraphics[width=0.09\columnwidth]{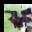}}
    \put(429.5,78.5){\includegraphics[width=0.09\columnwidth]{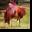}}
    \put(454,78.5){\includegraphics[width=0.09\columnwidth]{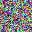}}
    \put(475.5,78.5){\includegraphics[width=0.09\columnwidth]{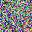}}

    \put(40,104){\includegraphics[width=0.09\columnwidth]{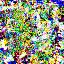}}
    \put(61.5,104){\includegraphics[width=0.09\columnwidth]{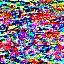}}
    \put(86,104){\includegraphics[width=0.09\columnwidth]{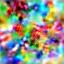}}
    \put(107.5,104){\includegraphics[width=0.09\columnwidth]{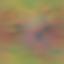}}
    \put(132,104){\includegraphics[width=0.09\columnwidth]{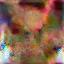}}
    \put(153.5,104){\includegraphics[width=0.09\columnwidth]{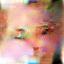}}
    \put(178,104){\includegraphics[width=0.09\columnwidth]{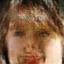}}
    \put(199.5,104){\includegraphics[width=0.09\columnwidth]{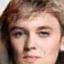}}
    \put(224,104){\includegraphics[width=0.09\columnwidth]{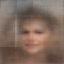}}
    \put(245.5,104){\includegraphics[width=0.09\columnwidth]{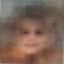}}
    \put(270,104){\includegraphics[width=0.09\columnwidth]{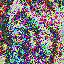}}
    \put(291.5,104){\includegraphics[width=0.09\columnwidth]{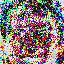}}
    \put(316,104){\includegraphics[width=0.09\columnwidth]{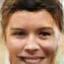}}
    \put(337.5,104){\includegraphics[width=0.09\columnwidth]{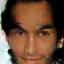}}
    \put(362,104){\includegraphics[width=0.09\columnwidth]{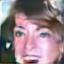}}
    \put(383.5,104){\includegraphics[width=0.09\columnwidth]{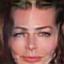}}
    \put(408,104){\includegraphics[width=0.09\columnwidth]{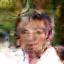}}
    \put(429.5,104){\includegraphics[width=0.09\columnwidth]{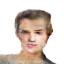}}
    \put(454,104){\includegraphics[width=0.09\columnwidth]{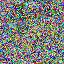}}
    \put(475.5,104){\includegraphics[width=0.09\columnwidth]{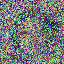}}
    
    \put(40,125.5){\includegraphics[width=0.09\columnwidth]{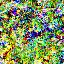}}
    \put(61.5,125.5){\includegraphics[width=0.09\columnwidth]{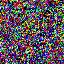}}
    \put(86,125.5){\includegraphics[width=0.09\columnwidth]{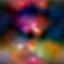}}
    \put(107.5,125.5){\includegraphics[width=0.09\columnwidth]{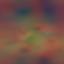}}
    \put(132,125.5){\includegraphics[width=0.09\columnwidth]{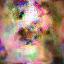}}
    \put(153.5,125.5){\includegraphics[width=0.09\columnwidth]{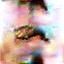}}
    \put(178,125.5){\includegraphics[width=0.09\columnwidth]{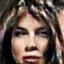}}
    \put(199.5,125.5){\includegraphics[width=0.09\columnwidth]{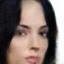}}
    \put(224,125.5){\includegraphics[width=0.09\columnwidth]{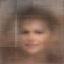}}
    \put(245.5,125.5){\includegraphics[width=0.09\columnwidth]{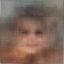}}
    \put(270,125.5){\includegraphics[width=0.09\columnwidth]{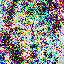}}
    \put(291.5,125.5){\includegraphics[width=0.09\columnwidth]{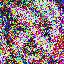}}
    \put(316,125.5){\includegraphics[width=0.09\columnwidth]{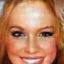}}
    \put(337.5,125.5){\includegraphics[width=0.09\columnwidth]{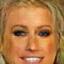}}
    \put(362,125.5){\includegraphics[width=0.09\columnwidth]{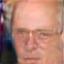}}
    \put(383.5,125.5){\includegraphics[width=0.09\columnwidth]{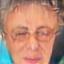}}
    \put(408,125.5){\includegraphics[width=0.09\columnwidth]{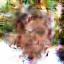}}
    \put(429.5,125.5){\includegraphics[width=0.09\columnwidth]{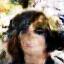}}
    \put(454,125.5){\includegraphics[width=0.09\columnwidth]{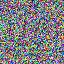}}
    \put(475.5,125.5){\includegraphics[width=0.09\columnwidth]{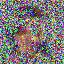}}
    
    \put(50,0){\scriptsize MI-Face}
    \put(91,0){\scriptsize DeepDream}
    \put(133,0){\scriptsize DeepInversion}
    \put(187,0){\scriptsize Revealer}
    \put(225,0){\scriptsize Inv-Alignment}
    \put(279,0){\scriptsize Bias-Rec}
    \put(326,0){\scriptsize KEDMI}
    \put(372,0){\scriptsize PLGMI}
    \put(408,0){\scriptsize Updates-Leak}
    \put(455,0){\scriptsize Deep-Leakage}
    
    \put(5,120){\small CelebA}
    \put(0,75){\small CIFAR10}
    \put(5,30){\small MNIST}
\end{picture}
\caption{Visualization of existing reconstruction attacks. 
For each attack, the left two images are reconstructed from the target model with a smaller training size ($1,000$ for CelebA and $100$ for CIFAR10 and MNIST), and the right two images are from the larger one ($20,000$).
}
\label{figure:reconstruction_visual}
\end{figure*}

\section{Evaluation}
\label{section:eval}

Due to space constraints, we defer the description of the experimental setup to~\autoref{section:setup}, where we provide details on the dataset, attack configurations, and evaluation metrics.

\begin{table*}[!t]
\centering
\caption{Evaluation results of existing reconstruction attacks. 
The target model is VGG16 trained on CelebA with 6 different sizes.
Attacks with a gray background belong to the dynamic training type.
For FID and MSE, a lower score indicates better reconstruction quality; while for SSIM and PSNR, a higher score indicates better performance. 
Experimental results for other model architectures and datasets can be found from~\autoref{table:eval_results_celeba_mobilenetv2} to~\autoref{table:eval_results_mnist_resnet18} in the appendix.
}
\scalebox{0.75}{
\begin{tabular}{@{}cccccccccc@{}}
\toprule
\multirow{2}{*}{Attack}& \multicolumn{2}{c}{\multirow{2}{*}{Metrics}} &\multicolumn{6}{c}{Target Data Size}\\ \cmidrule(l){4-9}& & & $1,000$ & $2,000$ & $5,000$ & $10,000$ & $15,000$ & $20,000$\\ \midrule
& \multicolumn{2}{c}{Memorization} & $\textbf{1.000}$ & $\textbf{0.981}$ & $\textbf{0.862}$ & $\textbf{0.539}$ & $\textbf{0.386}$ & $\textbf{0.301}$ \\ \midrule
\multirow{4}{*}{MI-Face} & Dataset-level & FID $\downarrow$  & $362.361$ &$365.570$ &$361.517$ &$340.959$ &$334.520$ &$336.906$ \\ \cmidrule(l){3-9}
& \multirow{3}{*}{Sample-level} & SSIM $\uparrow$ & $0.014 (100.00\%)$ & $0.019 (72.15\%)$ & $0.028 (57.88\%)$ & $0.024 (57.75\%)$ & $0.024 (57.28\%)$ & $0.023 (54.80\%)$ \\
& & PSNR $\uparrow$ & $7.978 (100.00\%)$ & $8.453 (53.80\%)$ & $8.872 (30.74\%)$ & $8.851 (17.63\%)$ & $9.005 (13.61\%)$ & $9.037 (10.08\%)$ \\
& & MSE $\downarrow$  & $0.163 (100.00\%)$ & $0.145 (53.80\%)$ & $0.132 (30.74\%)$ & $0.131 (17.63\%)$ & $0.127 (13.61\%)$ & $0.126 (10.08\%)$ \\ \cmidrule(l){1-9}
\multirow{4}{*}{DeepDream} & Dataset-level & FID $\downarrow$  & $337.139$ &$306.974$ &$297.590$ &$315.676$ &$345.472$ &$370.396$ \\ \cmidrule(l){3-9}
& \multirow{3}{*}{Sample-level} & SSIM $\uparrow$ & $0.079 (100.00\%)$ & $0.140 (57.65\%)$ & $0.234 (26.44\%)$ & $0.290 (14.38\%)$ & $0.321 (8.95\%)$ & $0.346 (6.81\%)$ \\
& & PSNR $\uparrow$ & $9.162 (100.00\%)$ & $10.304 (53.70\%)$ & $11.957 (26.60\%)$ & $13.006 (14.43\%)$ & $13.980 (9.29\%)$ & $14.027 (7.63\%)$ \\
& & MSE $\downarrow$  & $0.128 (100.00\%)$ & $0.100 (53.70\%)$ & $0.069 (26.60\%)$ & $0.053 (14.43\%)$ & $0.042 (9.29\%)$ & $0.041 (7.63\%)$ \\ \cmidrule(l){1-9}
\multirow{4}{*}{DeepInversion} & Dataset-level & FID $\downarrow$  & $287.497$ &$283.429$ &$273.183$ &$273.415$ &$245.736$ &$234.672$ \\ \cmidrule(l){3-9}
& \multirow{3}{*}{Sample-level} & SSIM $\uparrow$ & $0.100 (100.00\%)$ & $0.110 (61.70\%)$ & $0.118 (42.44\%)$ & $0.140 (28.97\%)$ & $0.150 (27.07\%)$ & $0.153 (22.47\%)$ \\
& & PSNR $\uparrow$ & $9.676 (100.00\%)$ & $10.094 (56.10\%)$ & $10.550 (31.68\%)$ & $11.143 (21.07\%)$ & $11.388 (19.09\%)$ & $11.343 (15.73\%)$ \\
& & MSE $\downarrow$  & $0.119 (100.00\%)$ & $0.105 (56.10\%)$ & $0.094 (31.68\%)$ & $0.081 (21.07\%)$ & $0.076 (19.09\%)$ & $0.077 (15.73\%)$ \\ \cmidrule(l){1-9}
\multirow{4}{*}{Revealer} & Dataset-level & FID $\downarrow$  & $116.712$ &$103.428$ &$94.899$ &$93.961$ &$93.781$ &$92.982$ \\ \cmidrule(l){3-9}
& \multirow{3}{*}{Sample-level} & SSIM $\uparrow$ & $0.101 (100.00\%)$ & $0.116 (66.75\%)$ & $0.135 (50.50\%)$ & $0.150 (44.05\%)$ & $0.157 (40.47\%)$ & $0.162 (38.33\%)$ \\
& & PSNR $\uparrow$ & $9.144 (100.00\%)$ & $9.720 (60.90\%)$ & $10.087 (43.68\%)$ & $10.449 (37.20\%)$ & $10.622 (33.66\%)$ & $10.733 (32.17\%)$ \\
& & MSE $\downarrow$  & $0.123 (100.00\%)$ & $0.113 (60.90\%)$ & $0.103 (43.68\%)$ & $0.094 (37.20\%)$ & $0.091 (33.66\%)$ & $0.088 (32.17\%)$ \\ \cmidrule(l){1-9}
\multirow{4}{*}{Inv-Alignment} & Dataset-level & FID $\downarrow$  & $344.049$ &$243.849$ &$359.419$ &$229.609$ &$361.569$ &$357.910$ \\ \cmidrule(l){3-9}
& \multirow{3}{*}{Sample-level} & SSIM $\uparrow$ & $0.255 (100.00\%)$ & $0.312 (51.45\%)$ & $0.285 (22.56\%)$ & $0.353 (13.09\%)$ & $0.336 (9.46\%)$ & $0.328 (7.89\%)$ \\
& & PSNR $\uparrow$ & $11.292 (100.00\%)$ & $12.463 (52.40\%)$ & $13.023 (23.10\%)$ & $13.858 (13.93\%)$ & $14.007 (9.74\%)$ & $14.253 (8.02\%)$ \\
& & MSE $\downarrow$  & $0.081 (100.00\%)$ & $0.061 (52.40\%)$ & $0.052 (23.10\%)$ & $0.043 (13.93\%)$ & $0.041 (9.74\%)$ & $0.039 (8.02\%)$ \\ \cmidrule(l){1-9}
\multirow{4}{*}{Bias-Rec} & Dataset-level & FID $\downarrow$ & $327.883$ &$323.892$ &$316.452$ &$319.774$ &$318.895$ &$315.227$ \\ \cmidrule(l){3-9}
& \multirow{3}{*}{Sample-level} & SSIM $\uparrow$ & $0.039 (38.20\%)$ & $0.040 (35.55\%)$ & $0.043 (30.88\%)$ & $0.044 (27.50\%)$ & $0.045 (25.44\%)$ & $0.047 (24.28\%)$ \\
& & PSNR $\uparrow$ & $9.783 (3.20\%)$ & $10.211 (0.75\%)$ & $10.249 (0.80\%)$ & $10.311 (0.53\%)$ & $10.222 (0.60\%)$ & $10.354 (0.57\%)$ \\
& & MSE $\downarrow$ & $0.105 (3.20\%)$ & $0.096 (0.75\%)$ & $0.095 (0.80\%)$ & $0.093 (0.53\%)$ & $0.095 (0.60\%)$ & $0.093 (0.57\%)$ \\ \cmidrule(l){1-9}
\multirow{4}{*}{KEDMI} & Dataset-level & FID $\downarrow$  & $121.689$ &$110.088$ &$101.730$ &$94.341$ &$95.852$ &$97.667$ \\ \cmidrule(l){3-9}
& \multirow{3}{*}{Sample-level} & SSIM $\uparrow$ & $0.111 (100.00\%)$ & $0.124 (57.30\%)$ & $0.144 (32.60\%)$ & $0.157 (24.29\%)$ & $0.160 (21.10\%)$ & $0.167 (18.93\%)$  \\
& & PSNR $\uparrow$ & $9.167 (100.00\%)$ & $9.715 (54.50\%)$ & $10.605 (30.28\%)$ & $11.126 (20.60\%)$ & $11.428 (17.12\%)$ & $11.560 (15.36\%)$ \\
& & MSE $\downarrow$  & $0.133 (100.00\%)$ & $0.115 (54.50\%)$ & $0.092 (30.28\%)$ & $0.081 (20.60\%)$ & $0.075 (17.12\%)$ & $0.073 (15.36\%)$ \\ \cmidrule(l){1-9}
\multirow{4}{*}{PLGMI} & Dataset-level & FID $\downarrow$  & $127.722$ &$107.883$ &$104.842$ &$97.730$ &$84.836$ &$85.143$ \\ \cmidrule(l){3-9}
& \multirow{3}{*}{Sample-level} & SSIM $\uparrow$ & $0.110 (100.00\%)$ & $0.132 (60.40\%)$ & $0.136 (37.02\%)$ & $0.149 (26.15\%)$ & $0.154 (21.83\%)$ & $0.161 (18.54\%)$ \\
& & PSNR $\uparrow$ & $9.448 (100.00\%)$ & $10.164 (56.10\%)$ & $10.588 (31.58\%)$ & $10.921 (20.55\%)$ & $11.299 (16.44\%)$ & $11.513 (13.54\%)$ \\
& & MSE $\downarrow$  & $0.122 (100.00\%)$ & $0.102 (56.10\%)$ & $0.092 (31.58\%)$ & $0.085 (20.55\%)$ & $0.078 (16.44\%)$ & $0.074 (13.54\%)$ \\ \cmidrule(l){1-9}
\multirow{4}{*}{Updates-Leak} & Dataset-level & FID $\downarrow$  & \cellcolor{mygray} $192.446$ & \cellcolor{mygray} $259.231$ & \cellcolor{mygray} $263.899$ & \cellcolor{mygray} $275.845$ & \cellcolor{mygray} $312.011$ & \cellcolor{mygray} $260.100$ \\ \cmidrule(l){3-9}
& \multirow{3}{*}{Sample-level} & SSIM $\uparrow$ & \cellcolor{mygray} $0.203 (18.00\%)$ & \cellcolor{mygray} $0.184 (11.00\%)$ & \cellcolor{mygray} $0.194 (9.00\%)$ & \cellcolor{mygray} $0.155 (22.00\%)$ & \cellcolor{mygray} $0.170 (18.00\%)$ & \cellcolor{mygray} $0.187 (5.00\%)$ \\
& & PSNR $\uparrow$ & \cellcolor{mygray} $13.173 (14.00\%)$ & \cellcolor{mygray} $13.165 (4.00\%)$ & \cellcolor{mygray} $12.693 (6.00\%)$ & \cellcolor{mygray} $12.554 (14.00\%)$ & \cellcolor{mygray} $12.537 (10.00\%)$ & \cellcolor{mygray} $12.509 (6.00\%)$ \\
& & MSE $\downarrow$  & \cellcolor{mygray} $0.049 (14.00\%)$ & \cellcolor{mygray} $0.050 (4.00\%)$ & \cellcolor{mygray} $0.056 (6.00\%)$ & \cellcolor{mygray} $0.058 (14.00\%)$ & \cellcolor{mygray} $0.058 (10.00\%)$ & \cellcolor{mygray} $0.058 (6.00\%)$ \\ \cmidrule(l){1-9}
\multirow{4}{*}{Deep-Leakage} & Dataset-level & FID $\downarrow$  & \cellcolor{mygray} $376.852$ & \cellcolor{mygray} $383.272$ & \cellcolor{mygray} $382.227$ & \cellcolor{mygray} $384.841$ & \cellcolor{mygray} $384.023$ & \cellcolor{mygray} $383.338$ \\ \cmidrule(l){3-9}
& \multirow{3}{*}{Sample-level} & SSIM $\uparrow$ & \cellcolor{mygray} $0.031 (49.00\%)$ & \cellcolor{mygray} $0.030 (48.00\%)$ & \cellcolor{mygray} $0.034 (43.00\%)$ & \cellcolor{mygray} $0.033 (45.00\%)$ & \cellcolor{mygray} $0.037 (43.00\%)$ & \cellcolor{mygray} $0.040 (52.00\%)$ \\
& & PSNR $\uparrow$ & \cellcolor{mygray} $10.957 (3.00\%)$ & \cellcolor{mygray} $10.965 (2.00\%)$ & \cellcolor{mygray} $11.014 (4.00\%)$ & \cellcolor{mygray} $11.076 (2.00\%)$ & \cellcolor{mygray} $11.152 (2.00\%)$ & \cellcolor{mygray} $11.199 (2.00\%)$ \\
& & MSE $\downarrow$  & \cellcolor{mygray} $0.080 (3.00\%)$ & \cellcolor{mygray} $0.080 (2.00\%)$ & \cellcolor{mygray} $0.079 (4.00\%)$ & \cellcolor{mygray} $0.078 (2.00\%)$ & \cellcolor{mygray} $0.077 (2.00\%)$ & \cellcolor{mygray} $0.076 (2.00\%)$ \\ \bottomrule
\end{tabular}
}
\label{table:eval_results_celeba_vgg}
\end{table*}

\begin{figure*}[!t]
\centering
\begin{subfigure}{0.49\columnwidth}
\includegraphics[width=\columnwidth]{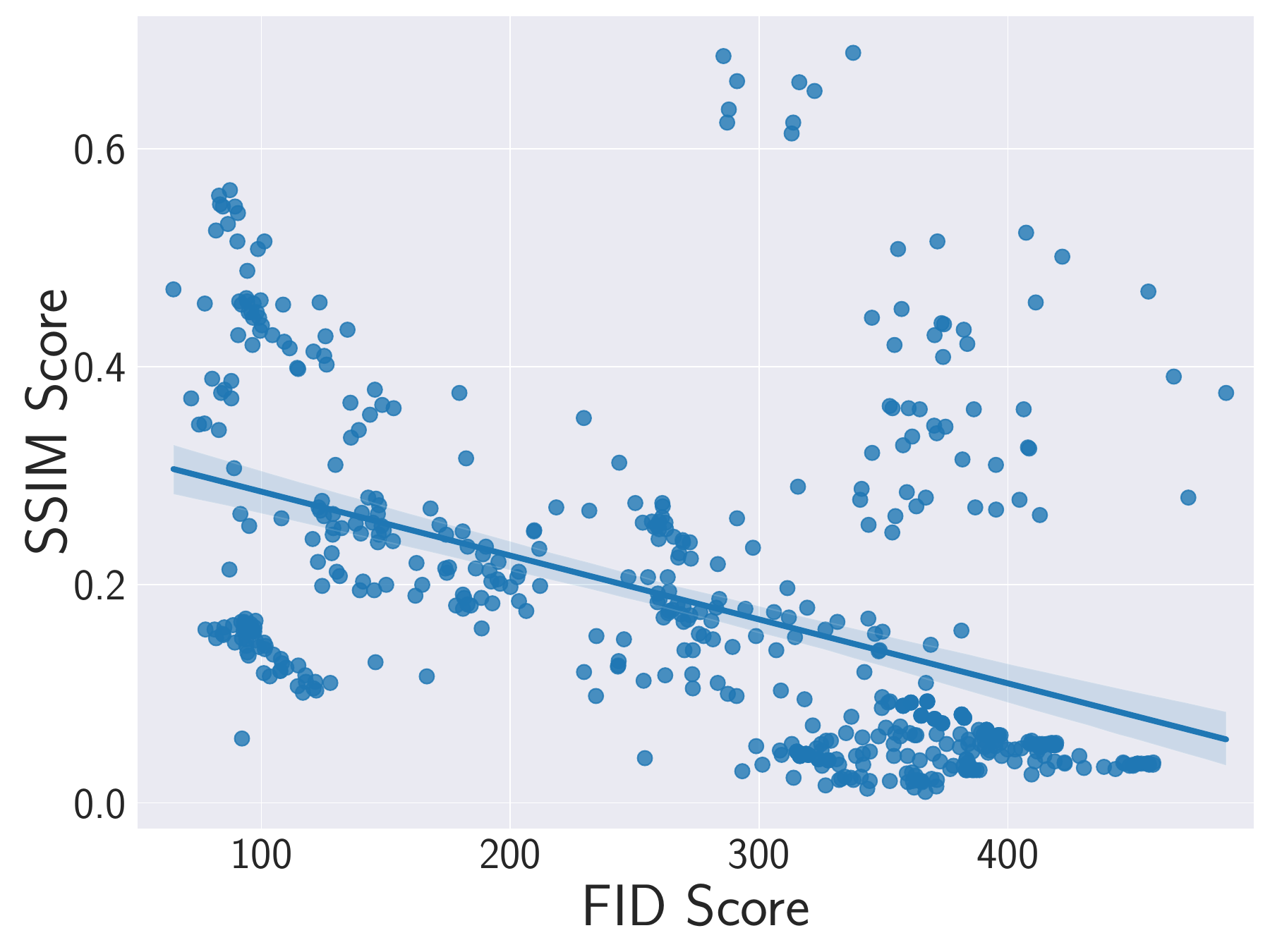}
\caption{FID vs. SSIM}
\label{fig:fidvsssim}
\end{subfigure}
\begin{subfigure}{0.49\columnwidth}
\includegraphics[width=\columnwidth]{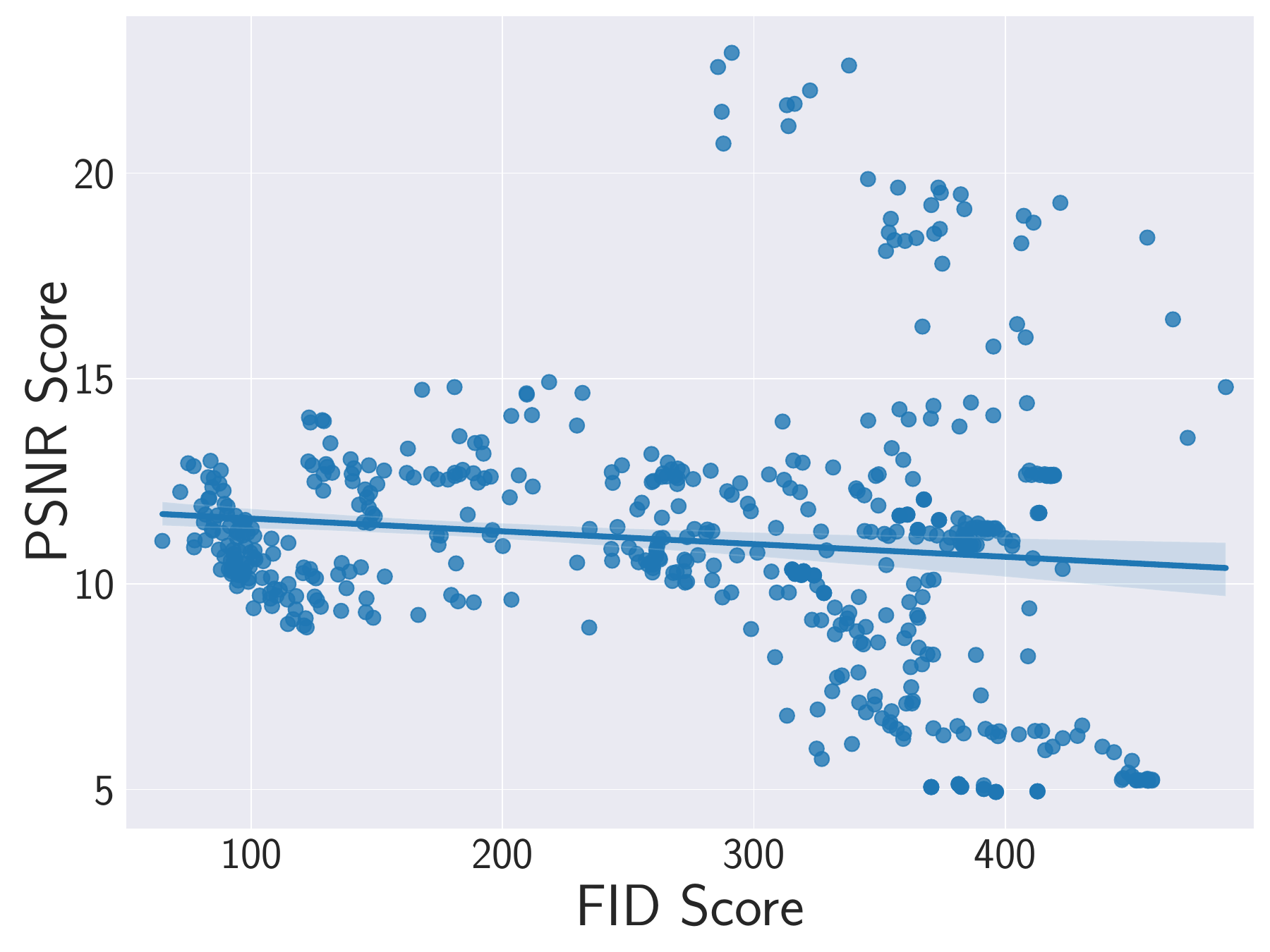}
\caption{FID vs. PSNR}
\label{fig:fidvspsnr}
\end{subfigure}
\begin{subfigure}{0.49\columnwidth}
\includegraphics[width=\columnwidth]{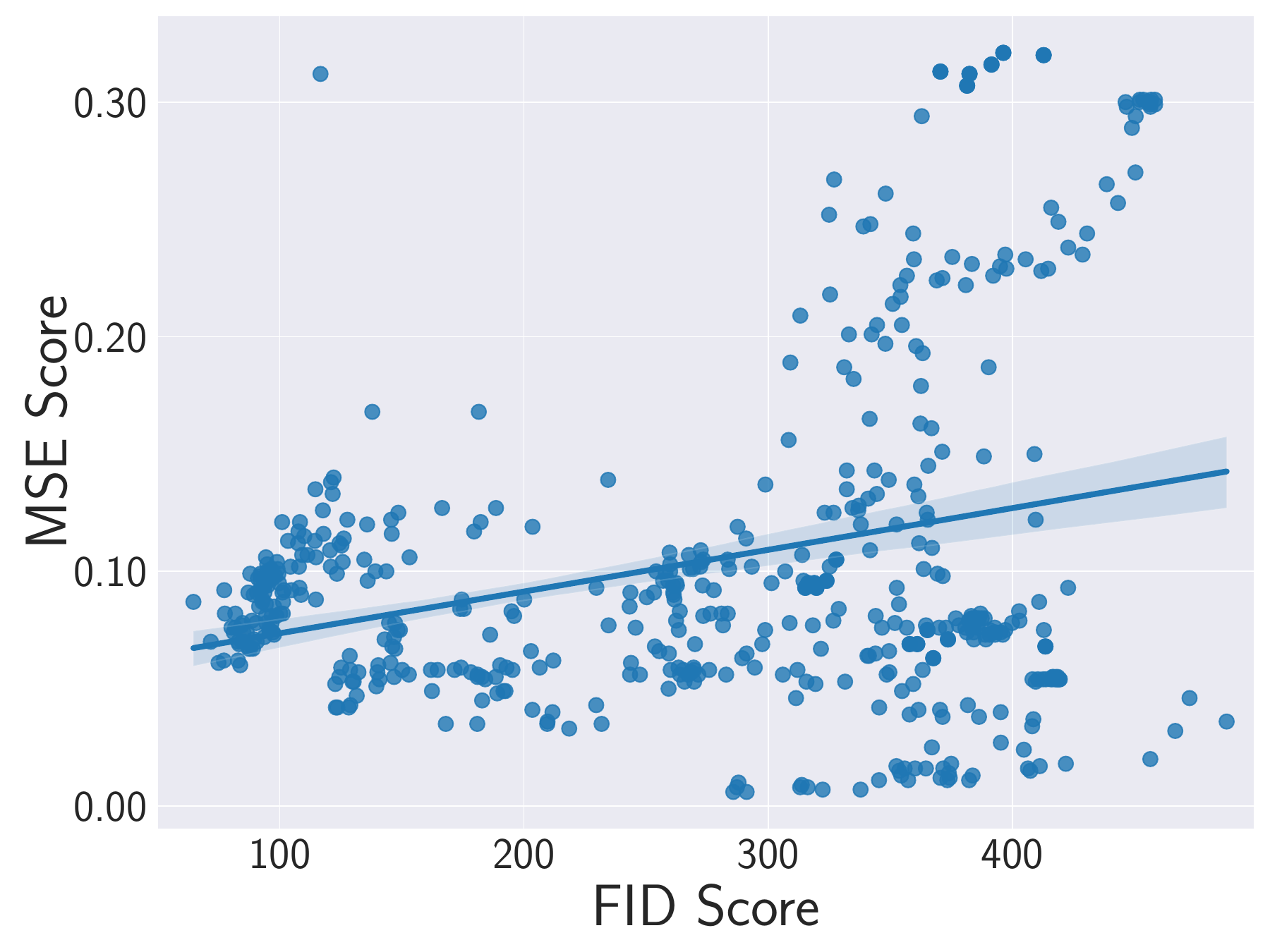}
\caption{FID vs. MSE}
\label{fig:fidvsmse}
\end{subfigure}
\begin{subfigure}{0.49\columnwidth}
\includegraphics[width=\columnwidth]{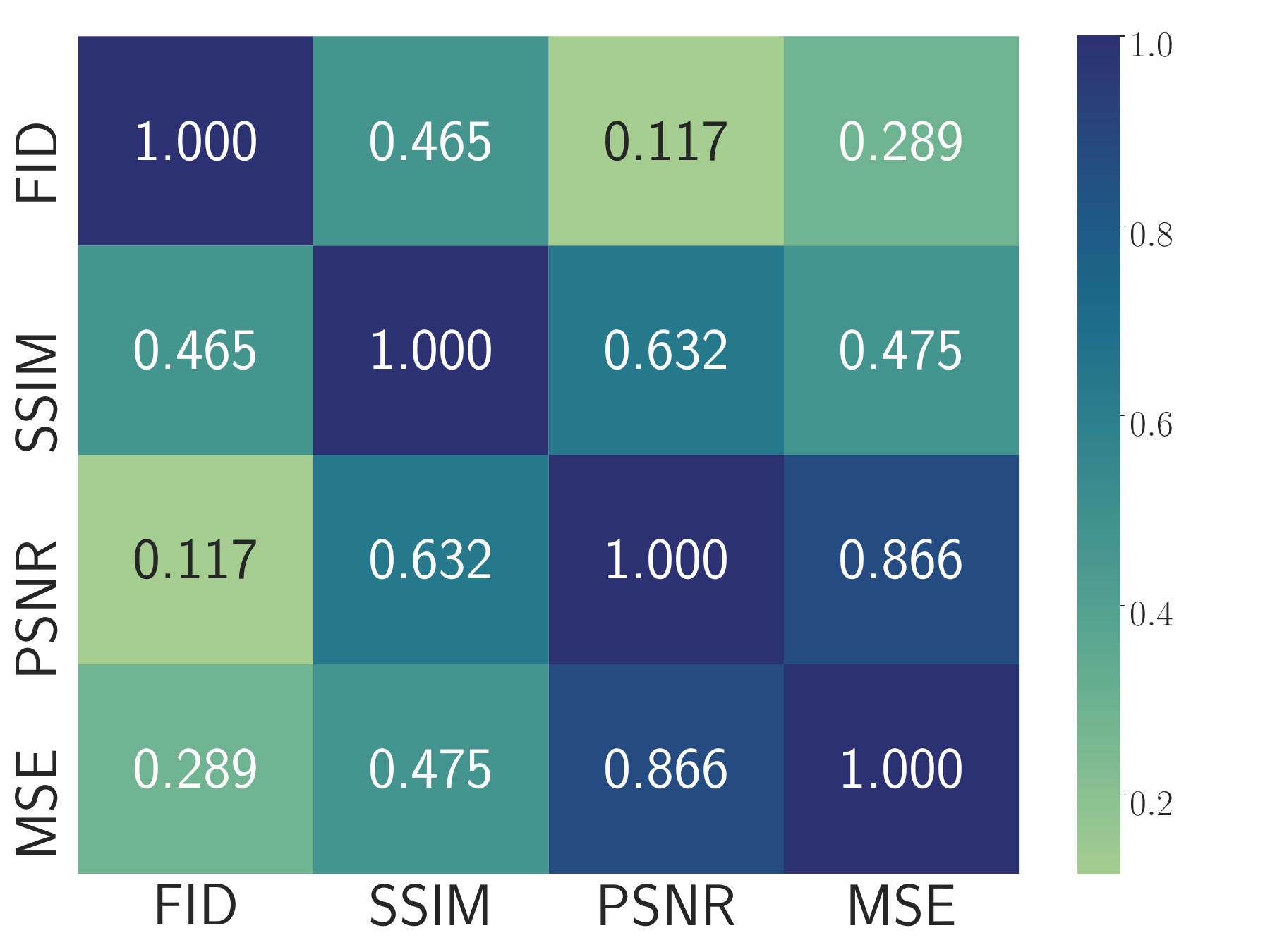}
\caption{Correlation}
\label{fig:metricscorr}
\end{subfigure}
\caption{Relationship between sample-level metrics and the dataset-level metric, we plot the correlation heatmap using the absolute value of correlation between different metrics.}
\label{figure:metrics_relation}
\end{figure*}

\subsection{Results under Quantitative Metrics}
\label{section:results}

We divide ten reconstruction attacks into two groups based on the training type. 
To provide an overview of the attack performance, we visualize the reconstruction results of all ten attacks in~\autoref{figure:reconstruction_visual}. 
Overall, none of the attacks can achieve exact reconstruction, and their quality varies significantly. 
To better analyze and compare these attacks, we apply the proposed metrics to evaluate the reconstruction datasets, and report the results in~\autoref{table:eval_results_celeba_vgg}.

\mypara{Dataset-level Metric}
For the CelebA dataset with the largest size, we visually confirm that three GAN-based attacks (Revealer, KEDMI, and PLGMI) outperform other methods, which is consistent with their lower FID scores. 
For dynamic training-type attacks, Updates-Leak shows a lower FID score than Deep-Leakage, indicating better reconstruction quality. 
This pattern holds for the other two datasets. 
We also observe that DeepInversion's performance worsens as dataset complexity\footnote{We follow the criteria used in~\cite{LWHSZBCFZ22} to determine the complexity of the dataset.
Briefly, gray-scale datasets are simpler than colored datasets.} increases, with FID scores rising from $64.678$ (MNIST) to $131.545$ (CIFAR10) to $234.672$ (CelebA). 
In contrast, GAN-based attacks, such as PLGMI, maintain clear semantic reconstruction, reflected in their low FID scores: $82.940$ (MNIST), $123.018$ (CIFAR10), and $85.143$ (CelebA).

In general, the FID score reflects reconstruction quality, with lower values indicating better reconstructions. 
However, we argue that relying solely on FID may not fully capture reconstruction quality, as shown with the CelebA dataset. 
As seen in~\autoref{figure:reconstruction_visual}, Inv-Alignment generates reconstructions that clearly show the semantic meaning of the target dataset, i.e., human faces, whereas some DeepInversion reconstructions lack sufficient information, despite having a lower FID score ($234.672$) compared to Inv-Alignment ($357.610$). This highlights the need to incorporate sample-level metrics like SSIM, PSNR, and MSE to more comprehensively evaluate reconstruction quality.

\mypara{Sample-level Metric}
Sample-level metrics complement dataset-level metrics by providing a micro-scale evaluation of reconstructions, which, in certain contexts, may better align with human perception. 
As previously noted, Inv-Alignment surpasses DeepInversion on the CelebA dataset; however, this superiority is not reflected by the FID score. 
In contrast, all three sample-level metrics deliver results consistent with our expectations, as shown in~\autoref{figure:reconstruction_visual}.
Additionally, sample-level metrics and dataset-level metrics generally provide consistent outcomes, as demonstrated in~\autoref{figure:metrics_relation}. 
Specifically, attacks achieving higher performance at the dataset level also tend to perform well on sample-level metrics. 
For instance, \autoref{fig:fidvsssim} shows that reconstructions with higher FID scores typically exhibit lower SSIM values, with similar trends observed across other metrics.

Sample-level metrics also play a vital role in assessing the diversity of reconstructed samples (coverage), a key indicator of a reconstruction attack's success. 
Diversity is challenging to evaluate using dataset-level metrics, as these rely on a few representative samples to approximate the distribution. 
Diversity is crucial for determining reconstruction quality, as generating identical data similar to a few samples in the training dataset might achieve high performance but does not constitute an effective reconstruction attack, as it limits the exposed information to only a few data points. 
To measure coverage, we identify the nearest pair for each reconstructed sample in the target dataset within the same class. 
The proportion of such pairs in the target dataset indicates the reconstruction's diversity. 
Given that coverage is influenced by both the number of classes and reconstructed samples, we present the results for CIFAR10 and MNIST together in~\autoref{figure:coverage_relation_cifar_mnist}, and separately for CelebA. The findings reveal an intriguing observation: high-quality reconstructions do not necessarily correspond to greater diversity in the reconstructed samples. 
Furthermore, results in~\autoref{table:eval_results_celeba_vgg} show a decline in coverage as the size of the training dataset increases. 
This trend aligns with our hypothesis that current reconstruction attacks tend to focus on capturing general information about the training dataset rather than individual sample details. 
Consequently, the ability to reconstruct diverse, unique samples diminishes as the training dataset grows, leading to lower coverage metrics for larger datasets.

\mypara{Summary}
While there is a relationship between dataset-level and sample-level metrics, their correlation is relatively weak, especially when evaluating low-quality reconstructions, as illustrated in~\autoref{fig:metricscorr}. 
For example, both MI-Face and DeepDream perform poorly on the CelebA task, which is reflected in their FID scores ($336.906$ vs. $370.396$). 
However, their SSIM scores vary significantly ($0.023$ vs. $0.346$), highlighting the inconsistency between metrics.

To examine the relationship between metrics in the context of high-quality reconstructions, we focus on three GAN-based attacks that generate higher-quality outputs. 
In these cases, the correlation coefficients improve: the correlation between FID and SSIM increases from $0.465$ to $0.597$, FID and PSNR from $0.117$ to $0.721$, and FID and MSE from $0.289$ to $0.511$.

Furthermore, low-quality reconstructions introduce discrepancies in coverage, even within the same attack. 
For instance, in the case of Bias-Rec on CelebA, coverage varies significantly between SSIM ($24.28\%$) and PSNR ($0.57\%$). 
When low-quality reconstructions are excluded, coverage across sample-level metrics aligns more closely, with the correlation coefficient between SSIM and PSNR, as well as between SSIM and MSE, increasing dramatically from $0.429$ to $0.993$.

Despite these improvements, the correlation between dataset-level and sample-level metrics remains insufficiently strong to ignore their distinctions. 
Dataset-level metrics provide a broad perspective on reconstruction quality, whereas sample-level metrics offer detailed insights into the fidelity of individual samples. 
Therefore, it is essential to utilize multiple metrics during evaluation. 
Although a single score can be derived using the minimum of normalized metrics to represent reconstruction quality, it is generally advisable to consider all available metrics. 
Ultimately, the decision between prioritizing usability and ensuring accuracy or effectiveness involves a careful trade-off.

\section{Memorization in Data Reconstruction}
\label{section:memorization}

We observe that existing attacks show varying performance, and even the same attack may exploit different levels of vulnerabilities when target models are trained with datasets of varying sizes. 
From the perspective of model owners, understanding which models are more vulnerable to data reconstruction attacks is essential. 
In this section, we examine model vulnerability to data reconstruction attacks through the lens of memorization. 
This choice is motivated by the close connection between memorization and membership inference attacks. 
Previous work~\cite{TSJLJHC22,CJZPTT22} suggests that strongly memorized samples are more vulnerable to membership inference. 
Given that data reconstruction can be framed as a search problem within the context of membership inference, we investigate whether models with higher memorization scores are likewise more prone to data reconstruction attacks.

As shown in~\autoref{table:eval_results_celeba_vgg}, we trained models using six datasets of varying sizes, which resulted in different levels of memorization for individual samples. 
To quantify memorization at the model level, we extend the sample-based memorization definition (\autoref{eqn:mem}) to cover the entire model. 
Specifically, we use the average memorization score of the first 1,000 samples in the training dataset as a proxy for the model’s overall memorization score:
\begin{equation}
\label{eqn:model_mem}
\begin{gathered}
   \textit{model-mem}(\mathcal{A},\mathcal{D})=
   \mathop{\mathbb{E}}_{x_i\in \mathcal{D}}[\underset{f_\theta{\sim\mathcal{A}(\mathcal{D})}}{\text{Pr}}[f_\theta(x_i)=y_i]- \\
   \underset{f_\theta{\sim\mathcal{A}(\mathcal{D}^{\symbol{92} i})}}{\text{Pr}}[f_\theta(x_i)=y_i]]
\end{gathered}
\end{equation}
As expected, we observe that the model memorization score decreases from $1.000$ to $0.301$ as the training size increases from $1,000$ to $20,000$. 
However, the performance of different attacks is inconsistent. 
For instance, with the DeepDream attack, the FID distance to the training dataset is $337.139$ when the memorization score is $1.000$, and it decreases as the memorization score drops to $0.301$. 
In contrast, for attacks such as Revealer, reconstruction quality (as measured by FID) improves as the memorization score decreases.

We attribute this discrepancy to two potential factors. 
First, the attack methods may lack the necessary capacity to accurately capture underlying vulnerabilities, particularly in the case of non-generative model-based attacks. 
This further raises the question of whether current methods are genuinely extracting private information from the model or merely imputing plausible samples, we provide some initial discussion in~\autoref{section:mem_utilization}. 
Second, the evaluation metrics used may not effectively capture the true performance of these attacks. 
As discussed in~\autoref{section:results}, while existing metrics provide a rough estimate of reconstruction quality, there remains a gap between the numerical results and human perception. 
In some cases, measurements do not align with visual evaluations. 
We discuss this challenge in the next section.

\begin{table}[!t]
\centering
\setlength{\tabcolsep}{3.0pt}
\caption{Evaluation with GPT-4o.}
\label{table:eval_gpt4o}
\scalebox{0.7}{
\begin{tabular}{@{}cccccccc@{}}
\toprule
\multirow{2}{*}{Attack}& \multirow{2}{*}{Metrics} &\multicolumn{6}{c}{Target Data Size}\\ 
\cmidrule(l){3-8}& & $1,000$ & $2,000$ & $5,000$ & $10,000$ & $15,000$ & $20,000$\\ \midrule
& Memorization & $\textbf{1.000}$ & $\textbf{0.981}$ & $\textbf{0.862}$ & $\textbf{0.539}$ & $\textbf{0.386}$ & $\textbf{0.301}$ \\ 
\midrule
\multirow{3}{*}{Revealer} & \# of Major & $349$ &$182$ &$188$ &$133$ &$94$ &$54$ \\
& \# of All & $166$ &$35$ &$45$ &$46$ &$26$ &$15$ \\
& Pred Rate & $0.335$ &$0.185$ &$0.194$ &$0.137$ &$0.096$ &$0.053$ \\
\midrule
\multirow{3}{*}{KEDMI} & \# of Major & $303$ &$157$ &$227$ &$188$ &$90$ &$35$ \\
& \# of All & $115$ &$14$ &$37$ &$37$ &$14$ &$6$ \\
& Pred Rate & $0.281$ &$0.169$ &$0.217$ &$0.189$ &$0.097$ &$0.047$ \\
\midrule
\multirow{3}{*}{PLGMI} & \# of Major & $325$ &$133$ &$162$ &$192$ &$120$ &$68$ \\
& \# of All & $116$ &$12$ &$12$ &$44$ &$15$ &$6$ \\
& Pred Rate & $0.283$ &$0.142$ &$0.174$ &$0.200$ &$0.125$ &$0.076$ \\
\midrule
\multirow{3}{*}{PLGMI (Pre)} & \# of Major & $484$ &$146$ &$186$ &$126$ &$38$ &$20$ \\
& \# of All & $246$ &$15$ &$29$ &$21$ &$8$ &$0$ \\
& Pred Rate & $0.446$ &$0.160$ &$0.188$ &$0.125$ &$0.054$ &$0.026$ \\
\bottomrule
\end{tabular}
}
\end{table}

\begin{table}[!t]
\centering
\setlength{\tabcolsep}{3.0pt}
\caption{Evaluation with InternVL 2.5 and Claude 3.7 for PLGMI (pre) on VGG16 trained on CelebA.}
\label{table:eval_internvl_claude}
\setlength{\tabcolsep}{3.pt}
\scalebox{0.68}{
\begin{tabular}{@{}cccccccc@{}}
\toprule
\multirow{2}{*}{LLMs}& \multirow{2}{*}{Metrics} &\multicolumn{6}{c}{Target Data Size}\\ 
\cmidrule(l){3-8}& & $1,000$ & $2,000$ & $5,000$ & $10,000$ & $15,000$ & $20,000$\\
\midrule
\multirow{3}{*}{InternVL 2.5}& \# of Major & $531$ &$134$ &$92$ &$149$ &$65$ &$29$ \\
& \# of All & $69$ &$0$ &$0$ &$2$ &$0$ &$0$ \\
& Pred Rate & $0.410$ &$0.157$ &$0.130$ &$0.165$ &$0.096$ &$0.042$ \\
\midrule
\multirow{3}{*}{Claude 3.7}& \# of Major & $479$ & $188$ & $41$ & $125$ & $130$ & $37$ \\
& \# of All & $143$ & $31$ & $0$ & $23$ & $17$ & $0$ \\
& Pred Rate & $0.434$ & $0.179$ & $0.056$ & $0.134$ & $0.158$ & $0.039$ \\
\bottomrule
\end{tabular}
}
\end{table}

\subsection{Linkage between Model Memorization and Dataset Leakage}
\label{section:deep_inves_metric}

To evaluate the effectiveness of data reconstruction, we have introduced a set of metrics designed to quantitatively assess reconstruction quality. 
While these metrics offer a coarse-grained view of reconstruction success, previous findings suggest that high quantitative scores do not necessarily correlate with high-quality reconstructions. 
Although human inspection remains the most direct and intuitive method for quality assessment, it is often prohibitive in terms of cost and scale for extensive datasets. 
Therefore, to achieve an efficient and scalable evaluation, we propose utilizing GPT-4o. 
This model is recognized for its robust performance and has been further refined during the Reinforcement Learning from Human Feedback (RLHF) phase to align with human preferences.

We evaluate the attack performance using 1,000 randomly chosen CelebA images as targets. 
Each target appears in six training sets of increasing size. 
For every set, we run the reconstruction algorithm and keep the 1,000 results with the highest PSNR score relative to their targets. 
Thus, each target image has six candidate reconstructions—one from each training size.
To decide which reconstruction looks closest to the ground truth, we prompt GPT-4o: ``\textit{From the second to the seventh image, which image is more similar to the first one? Please make sure your response must be the index of that image and don’t say any other words.}''
We repeat this query five times and tally how often each training size is chosen. We report three summary measures: ``\# of major'' (how often a setting wins the majority vote), ``\# of all'' (how often the vote is unanimous), and ``pred rate'' (its overall selection frequency across all queries).

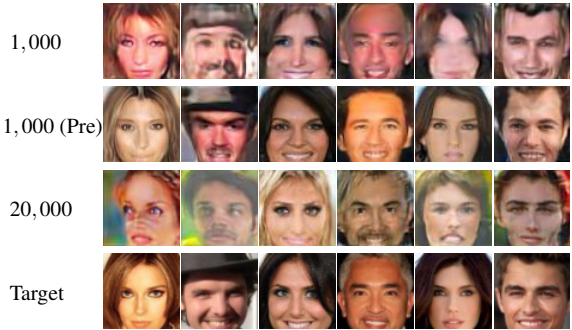
\begin{figure}
\centering
\scalebox{0.9}{
\begin{picture}(250,140)
    \put(42,0){\includegraphics[width=0.13\columnwidth]{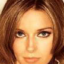}}
    \put(74.5,0){\includegraphics[width=0.13\columnwidth]{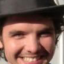}}
    \put(107,0){\includegraphics[width=0.13\columnwidth]{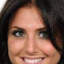}}
    \put(139.5,0){\includegraphics[width=0.13\columnwidth]{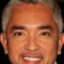}}
    \put(172,0){\includegraphics[width=0.13\columnwidth]{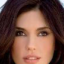}}
    \put(204.5,0){\includegraphics[width=0.13\columnwidth]{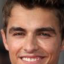}}

    \put(42,35){\includegraphics[width=0.13\columnwidth]{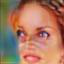}}
    \put(74.5,35){\includegraphics[width=0.13\columnwidth]{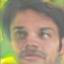}}
    \put(107,35){\includegraphics[width=0.13\columnwidth]{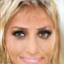}}
    \put(139.5,35){\includegraphics[width=0.13\columnwidth]{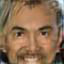}}
    \put(172,35){\includegraphics[width=0.13\columnwidth]{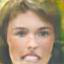}}
    \put(204.5,35){\includegraphics[width=0.13\columnwidth]{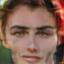}}

    \put(42,70){\includegraphics[width=0.13\columnwidth]{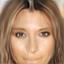}}
    \put(74.5,70){\includegraphics[width=0.13\columnwidth]{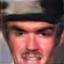}}
    \put(107,70){\includegraphics[width=0.13\columnwidth]{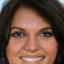}}
    \put(139.5,70){\includegraphics[width=0.13\columnwidth]{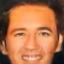}}
    \put(172,70){\includegraphics[width=0.13\columnwidth]{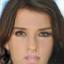}}
    \put(204.5,70){\includegraphics[width=0.13\columnwidth]{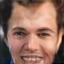}}

    \put(42,105){\includegraphics[width=0.13\columnwidth]{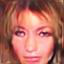}}
    \put(74.5,105){\includegraphics[width=0.13\columnwidth]{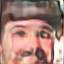}}
    \put(107,105){\includegraphics[width=0.13\columnwidth]{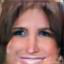}}
    \put(139.5,105){\includegraphics[width=0.13\columnwidth]{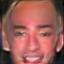}}
    \put(172,105){\includegraphics[width=0.13\columnwidth]{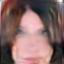}}
    \put(204.5,105){\includegraphics[width=0.13\columnwidth]{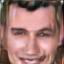}}
    
    \put(3,117){\small $1,000$}
    \put(0,82){\small $1,000$ (Pre)}
    \put(3,47){\small $20,000$}
    \put(3,12){\small Target}
\end{picture}
}
\caption{Visualization of PLGMI on different target models.}
\label{figure:visual_plgmi}
\end{figure}

Our analysis concentrates on three GAN-based approaches, which consistently yield high-quality reconstructions. As indicated in~\autoref{table:eval_gpt4o}, despite some variability, all three metrics across the $1,000$ identities generally exhibit a trend where reconstruction quality diminishes as the memorization score decreases. 
These results can also be generalized to other leading LLMs, such as InternVL 2.5 and Claude 3.7, as demonstrated in~\autoref{table:eval_internvl_claude}. 
Visual evidence in~\autoref{figure:visual_plgmi} further supports the observation that reconstructions from the smallest target dataset size ($1,000$ samples) resemble the target images more closely than those from the largest size ($20,000$ samples). 
This observation aligns with our expectations but contrasts with results from previous metrics.

We interpret these findings in two ways. 
From the perspective of GAN-based attacks, these methods reconstruct data at a class level and aim to closely approximate the distribution of the targets. 
Consequently, when a target dataset includes multiple samples within a class, achieving a close match for any specific target sample becomes unlikely. 
Conversely, smaller target datasets result in less generalized feature learning, manifesting as blurred areas in reconstructions, like the eyes and cheeks. 
While this blurriness negatively impacts performance as measured by both dataset-level and sample-level metrics, its effect on visual assessments is relatively modest, provided the blurring is not extensive.

This second interpretation brings forth another noteworthy observation. 
For each attack method, the optimal performance is achieved at a dataset size of $1000$. 
Interestingly, the second-best performance does not occur at a slightly larger size, such as $2000$, but frequently occurs at medium dataset sizes, such as $5,000$ or $10,000$ samples. 
This suggests that the reconstruction process benefits from the model's ability to learn additional features. 
However, this advantage is counterbalanced by the challenges posed by larger datasets, as a more extensive training dataset complicates the accurate recovery of individual samples, consistent with our earlier findings. 

To further explore the benefits of feature learning, we trained a model pre-trained on disjoint datasets and fine-tuned it with the target data. 
This approach enables the model to learn additional features, as evidenced by the improved testing accuracy. 
We evaluate the attack on this fine-tuned model in~\autoref{table:eval_gpt4o}. 
The results indicate that pre-training enhances the advantage of using a smaller dataset size, as the pre-trained model has already developed a degree of generalizability and is able to learn sample-specific features more efficiently.

To better understand the effect of pre-training, we compare attack performance between models trained from scratch and those fine-tuned from pre-trained weights, as presented in~\autoref{table:plgmi_vs_plgmi_pre}. 
Fine-tuning a pre-trained model enables it to learn additional features beyond those required for generalization, leading to improved reconstruction performance.

Together, these findings demonstrate that learning additional features enhances reconstruction quality. 
Notably, reconstructions from pre-trained models exhibit finer structural details and less blurring, as shown in~\autoref{figure:visual_plgmi}.

We further validate the role of learning additional features from the opposite perspective by reducing the bottleneck size (i.e., the width of the final layer after convolutions), thereby limiting the model’s capacity to retain features~\cite{TZ15}. 
As shown in~\autoref{table:eval_feature_size}, this reduction in learned feature information leads to a degradation in reconstruction performance.

\begin{table}[!t]
\centering
\setlength{\tabcolsep}{5.0pt}
\caption{PLGMI vs. PLGMI (Pre).}
\scalebox{0.7}{
\begin{tabular}{@{}ccccccc@{}}
\toprule
\multirow{2}{*}{Metrics} &\multicolumn{6}{c}{Target Data Size}\\ 
\cmidrule(l){2-7}& $1,000$ & $2,000$ & $5,000$ & $10,000$ & $15,000$ & $20,000$\\
\midrule
\# of Major & $11:989$ & $17:983$ & $6:994$ & $6:994$ & $3:997$ & $6:994$ \\
\# of All & $1:956$ & $2:939$ & $1:964$ & $0:961$ & $0:979$ & $0:952$ \\
Pred Rate & \(\dfrac{0.015}{0.985}\) & \(\dfrac{0.023}{0.977}\) & \(\dfrac{0.012}{0.988}\) & \(\dfrac{0.013}{0.987}\) & \(\dfrac{0.006}{0.994}\) & \(\dfrac{0.014}{0.986}\) \\
\bottomrule
\end{tabular}
}
\label{table:plgmi_vs_plgmi_pre}
\end{table}

\begin{table}[!t]
\centering
\setlength{\tabcolsep}{2.0pt}
\caption{GPT-4o evaluation with different sizes of feature embeddings on CelebA with data size of 1,000 and 20,000.}
\scalebox{0.8}{
\begin{tabular}{@{}cccccccc@{}}
\toprule
\multirow{2}{*}{Attack}& \multirow{2}{*}{Metrics} &\multicolumn{3}{c}{Feature size (1000)} & \multicolumn{3}{c}{Feature size (20000)}\\ 
\cmidrule(l){3-5}\cmidrule(l){6-8}& & $2,048$ (Ori) & $512$ & $128$ & $2,048$ (Ori) & $512$ & $128$ \\
\midrule
\multirow{3}{*}{PLGMI}& \# of Major & $917$ & $75$ & $8$ & $742$ & $251$ & $7$ \\
& \# of All & $719$ & $12$ & $0$ & $465$ & $39$ & $0$ \\
& Pred Rate & $0.884$ & $0.105$ & $0.011$ & $0.725$ & $0.263$ & $0.012$ \\
\bottomrule
\end{tabular}
}
\label{table:eval_feature_size}
\end{table}

From this study, we derive two key insights. 
First, from an attack perspective, for GAN-based methods that produce high-quality reconstructions, we observe a trade-off between the generalizability of the target model and the quality of the reconstructions. 
This insight could inspire further advancements in data reconstruction attacks. 
Second, from an evaluation perspective, the quantitative metrics employed reliably assess attack performance when the differences between methods are pronounced (e.g., comparing GAN-based methods to techniques producing less semantically meaningful reconstructions, such as MI-Face). 
However, since current reconstruction attacks are far from perfectly recovering target datasets, existing dataset-level and sample-level metrics often fail to align precisely with human perception, particularly when differences are subtle. 

With the advancements in large language models (LLMs), some limitations of visualization as an evaluation metric can be mitigated. 
We encourage future research to explore integrating LLMs with quantitative metrics for a more comprehensive evaluation of data reconstruction attacks.

\section{Utilization of Attack Knowledge}
\label{section:mem_utilization}

This section investigates whether data reconstruction reveals sensitive information or just replicates attack knowledge, and how different levels of attack information affect performance.

\begin{figure}[!t]
\centering
\begin{subfigure}{0.45\columnwidth}
\includegraphics[width=\columnwidth]{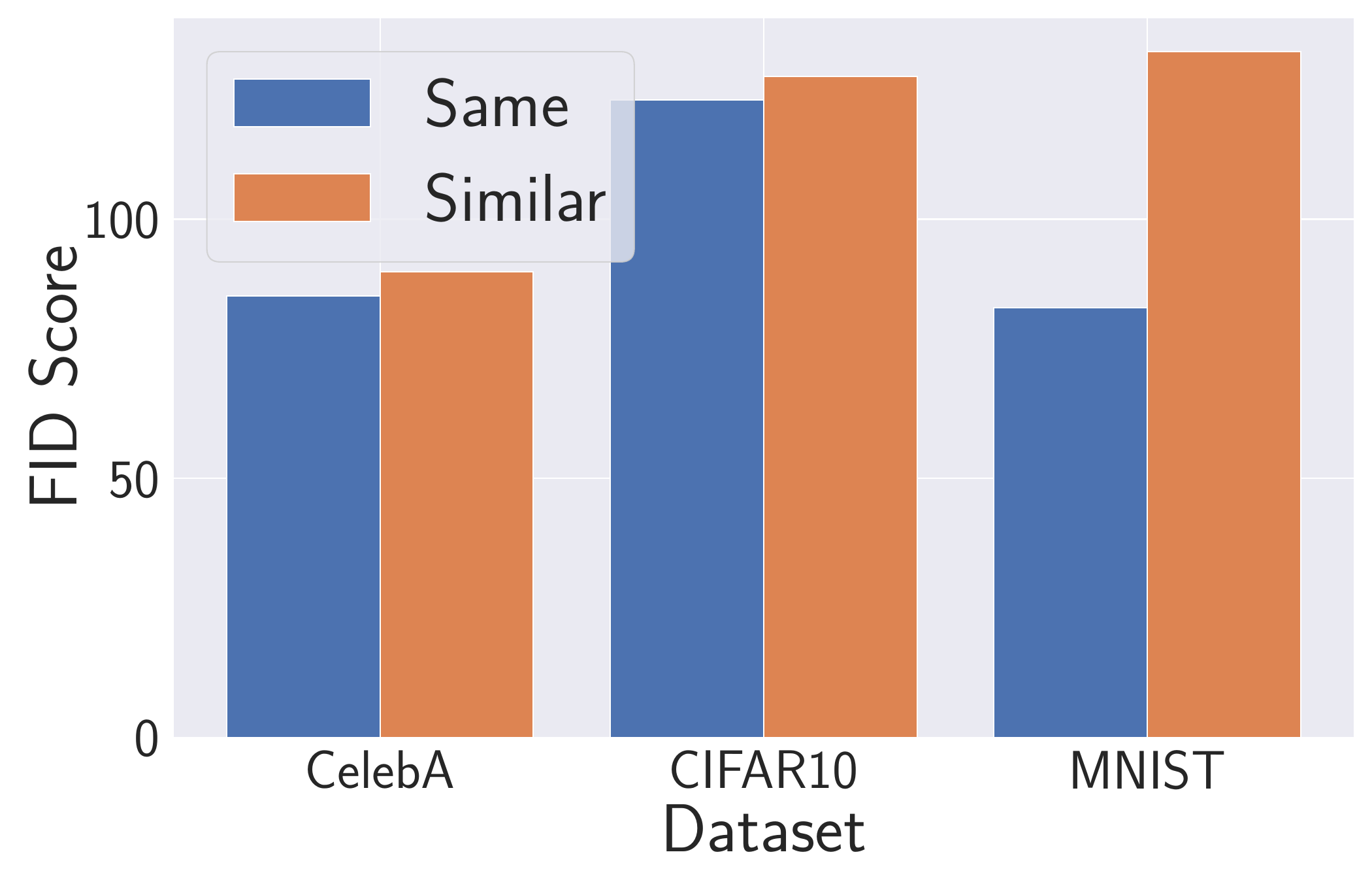}
\caption{Influence of data access}
\label{figure:influence_of_data_access_plgmi}
\end{subfigure}
\begin{subfigure}{0.45\columnwidth}
\includegraphics[width=\columnwidth]{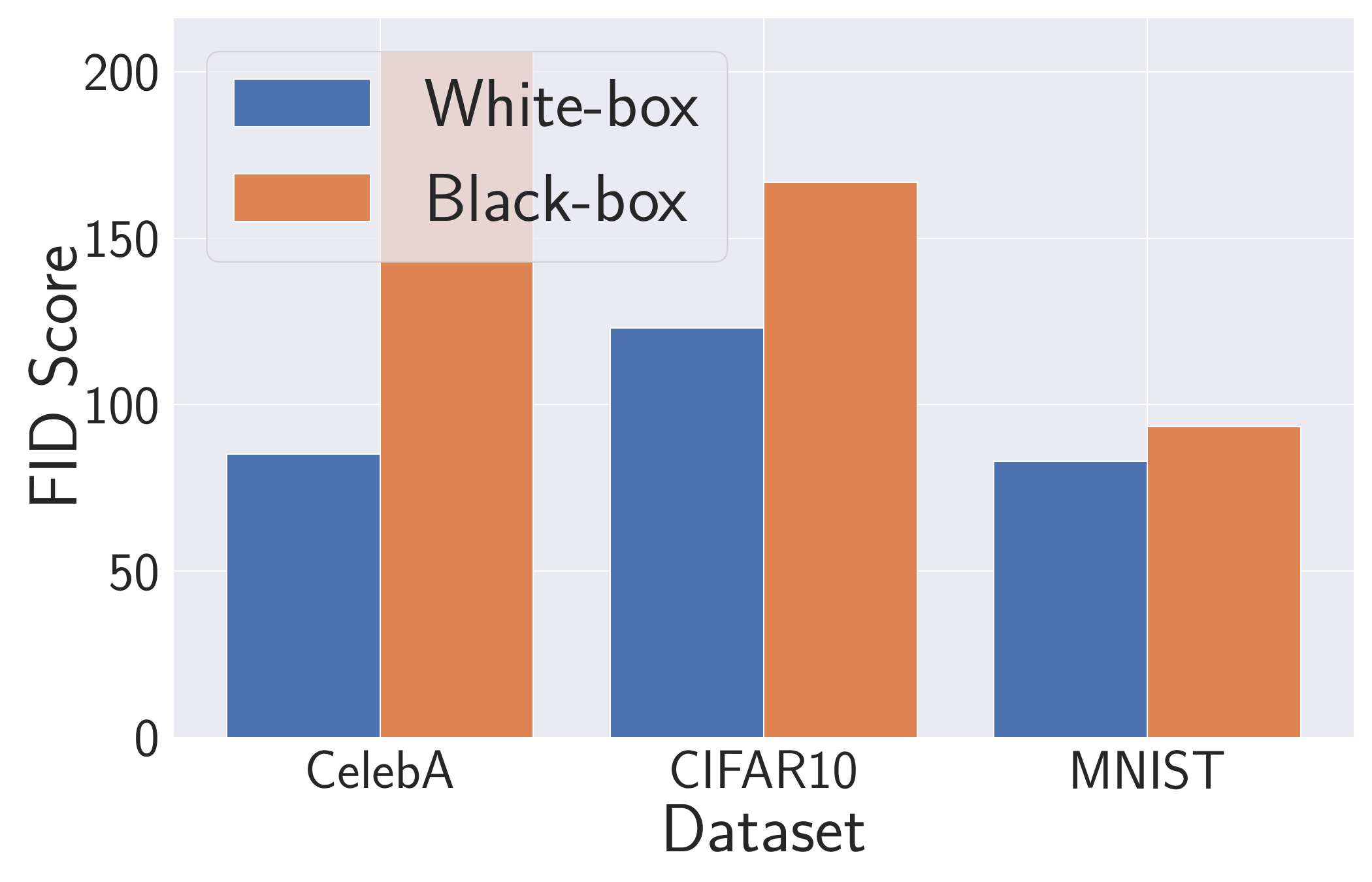}
\caption{Influence of model access}
\label{figure:influence_of_model_access_plgmi}
\end{subfigure}
\caption{Influence of auxiliary information for PLGMI. Experimental results for other attacks can be found in~\autoref{figure:influence_of_knowledge_more}.}
\label{figure:influence_of_knowledge}
\end{figure}

\mypara{Influence of Data Access}
We first assess how current methods use auxiliary datasets to enhance attacks, focusing on whether reconstruction merely involves imputing data from a similarly distributed dataset. By replacing auxiliary datasets with similar but distinct ones, that is, Kuzushiji-MNIST~\cite{CBKLYH18} for MNIST, CIFAR100~\cite{CIFAR} for CIFAR10, and FFHQ~\cite{KLA19} for CelebA, we observe moderate performance drops in GAN-based attacks (see \autoref{figure:influence_of_data_access_plgmi} and \autoref{figure:influence_of_knowledge_more}). 
This suggests reconstruction extends beyond mere imputation from auxiliary data.

\begin{figure}
\centering
\begin{picture}(200, 70)
    \put(40,10){\includegraphics[width=0.12\columnwidth]{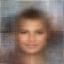}}
    \put(69,10){\includegraphics[width=0.12\columnwidth]{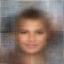}}
    \put(102,10){\includegraphics[width=0.12\columnwidth]{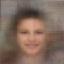}}
    \put(131,10){\includegraphics[width=0.12\columnwidth]{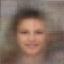}}
    
    \put(40,39){\includegraphics[width=0.12\columnwidth]{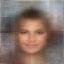}}
    \put(69,39){\includegraphics[width=0.12\columnwidth]{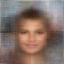}}
    \put(102,39){\includegraphics[width=0.12\columnwidth]{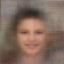}}
    \put(131,39){\includegraphics[width=0.12\columnwidth]{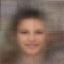}}
    
    \put(60,0){\scriptsize Clean}
    \put(120,0){\scriptsize Backdoor}
\end{picture}
\caption{Effect of additional data to the attack performance of Inv-Alignment. 
}
\label{figure:reconstruction_bd_inv}
\end{figure}

The picture changes for low-fidelity methods such as Inv-Alignment. 
Supplying auxiliary data drawn from a different distribution can even yield better results than providing data from the original distribution. 
For instance, providing CIFAR100 as the auxiliary set yields a FID of $330.31$, whereas supplying data from the same distribution gives $357.91$.

To further investigate this limitation, we conducted a backdoor experiment. 
A model was trained on a mixed dataset that included both clean and backdoored images, where the backdoored images contained a 16 × 16 black square in the bottom-right corner. 
We then reconstructed the training set while giving Inv-Alignment both clean and backdoored samples. 
Although the trigger should be a strong cue, it never appears in the reconstructions (see~\autoref{figure:reconstruction_bd_inv}). 
These findings suggest that low-quality reconstruction methods fail to exploit critical attack information.

\begin{table}[!t]
\centering
\setlength{\tabcolsep}{3.0pt}
\caption{Effect of batch normalization for DeepInversion on VGG16 trained on CelebA.}
\scalebox{0.8}{
\begin{tabular}{@{}ccccccc@{}}
\toprule
\multirow{2}{*}{BatchNorm} &\multicolumn{6}{c}{Target Data Size}\\ 
\cmidrule(l){2-7}& $1,000$ & $2,000$ & $5,000$ & $10,000$ & $15,000$ & $20,000$\\
\midrule
updated & $287.497$ & $283.429$ & $273.183$ & $273.415$ & $245.736$ & $234.672$ \\
fixed & $372.803$ & $348.387$ & $346.099$ & $346.149$ & $325.639$ & $311.791$ \\
\bottomrule
\end{tabular}
}
\label{table:deepinversion_bn}
\end{table}

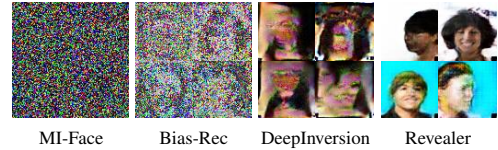
\begin{figure}
\centering
\begin{picture}(150,60)
    \put(-15,10){\includegraphics[width=0.09\columnwidth]{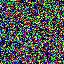}}
    \put(6.5,10){\includegraphics[width=0.09\columnwidth]{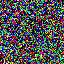}}
    \put(31,10){\includegraphics[width=0.09\columnwidth]{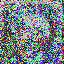}}
    \put(52.5,10){\includegraphics[width=0.09\columnwidth]{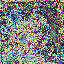}}
    \put(77,10){\includegraphics[width=0.09\columnwidth]{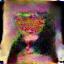}}
    \put(98.5,10){\includegraphics[width=0.09\columnwidth]{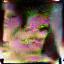}}
    \put(123,10){\includegraphics[width=0.09\columnwidth]{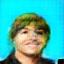}}
    \put(144.5,10){\includegraphics[width=0.09\columnwidth]{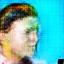}}
       
    \put(-15,31.5){\includegraphics[width=0.09\columnwidth]{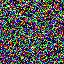}}
    \put(6.5,31.5){\includegraphics[width=0.09\columnwidth]{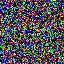}}
    \put(31,31.5){\includegraphics[width=0.09\columnwidth]{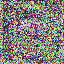}}
    \put(52.5,31.5){\includegraphics[width=0.09\columnwidth]{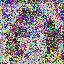}}
    \put(77,31.5){\includegraphics[width=0.09\columnwidth]{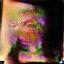}}
    \put(98.5,31.5){\includegraphics[width=0.09\columnwidth]{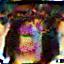}}
    \put(123,31.5){\includegraphics[width=0.09\columnwidth]{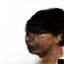}}
    \put(144.5,31.5){\includegraphics[width=0.09\columnwidth]{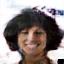}}
        
    \put(-5,0){\scriptsize MI-Face}
    \put(40,0){\scriptsize Bias-Rec}
    \put(78,0){\scriptsize DeepInversion}
    \put(132,0){\scriptsize Revealer}
\end{picture}
\caption{Visualization of attacks on VGG16 trained on backdoored CelebA with size 20,000. 
}
\label{figure:reconstruction_bd_all}
\end{figure}

\mypara{Influence of Model Access}
We next investigate whether access to model parameters increases information leakage by comparing attack performance under black-box and white-box settings.

For black-box evaluations, we first apply state-of-the-art model stealing techniques~\cite{TMWP21} to create a white-box surrogate model from the black-box target model. 
We then execute standard attacks on this surrogate model.

The results presented in~\autoref{figure:influence_of_model_access_plgmi} and~\autoref{figure:influence_of_knowledge_more} demonstrate that several attack methods—particularly those with stronger baseline performance—benefit from white-box access, as expected. 
The performance gap widens for more complex reconstruction tasks, indicating that these methods effectively exploit model parameters to improve their performance. 
To illustrate this point, we examine two specific attacks: DeepDream and DeepInversion. 
The key distinction between them is that DeepInversion explicitly leverages the statistical information encoded in BatchNorm layers. 
As shown in~\autoref{table:eval_results_celeba_vgg}, DeepInversion consistently outperforms DeepDream, highlighting the benefit of incorporating internal model statistics.

To further investigate the role of statistical information stored in model parameters, we conduct an experiment where a target model is trained with BatchNorm parameters fixed, allowing only the remaining parameters to be updated. 
We then apply DeepInversion to both the standard and modified models. 
As shown in~\autoref{table:deepinversion_bn}, the reconstruction performance significantly degrades in the modified setting, despite using the same attack method. 
This performance drop underscores the importance of access to BatchNorm statistics, which appear critical for high-quality reconstruction.

However, it is important to note that not all attack methods exhibit improved performance under white-box conditions. 
In some cases, the lack of performance gain suggests that these methods fail to effectively utilize the information memorized by the target model. 
To further validate this, we evaluate attacks on a backdoored model, where a predefined trigger, known to be strongly memorized, is embedded during training. 
As shown in~\autoref{figure:reconstruction_bd_all}, among the selected methods, DeepInversion and Revealer are able to clearly reconstruct the square trigger, and they also produce the highest-quality reconstructions overall. 
This finding aligns with earlier observations and further emphasizes the limitations of certain attack methods in fully leveraging model-internal information.

\section{Discussion and Conclusion}
\label{section:discussion}

\begin{table}[!t]
\centering
\setlength{\tabcolsep}{3.0pt}
\caption{Evaluation on Swin with GPT-4o.}
\label{table:eval_swin_gpt4o}
\scalebox{0.78}{
\begin{tabular}{@{}cccccccc@{}}
\toprule
\multirow{2}{*}{Attack}& \multirow{2}{*}{Metrics} &\multicolumn{6}{c}{Target Data Size}\\ 
\cmidrule(l){3-8}& & $1,000$ & $2,000$ & $5,000$ & $10,000$ & $15,000$ & $20,000$\\
\midrule
\multirow{3}{*}{PLGMI}& \# of Major & $101$ &$435$ &$229$ &$185$ &$46$ &$4$ \\
& \# of All & $10$ &$43$ &$11$ &$10$ &$3$ &$0$ \\
& Pred Rate & $0.120$ &$0.369$ &$0.251$ &$0.197$ &$0.056$ &$0.007$ \\
\bottomrule
\end{tabular}
}
\end{table}

\begin{table}[!t]
\centering
\caption{Evaluation of the Vec2Text~\cite{MKSR23} attack performance across varying target data sizes on different datasets, measured by BLEU, Sim. (Similarity), and R-L (ROUGE-L).}
\setlength{\tabcolsep}{3.pt} 
\scalebox{0.78}{
\begin{tabular}{@{}llcccccc@{}}
\toprule
\multirow{2}{*}{Dataset} & \multirow{2}{*}{Metrics} & \multicolumn{6}{c}{Target Data Size} \\
\cmidrule(l){3-8}
& & $1,000$ & $2,000$ & $5,000$ & $10,000$ & $15,000$ & $20,000$ \\
\midrule
\multirow{3}{*}{SST2} & BLEU & $0.169$ & $0.172$ & $0.152$ & $0.083$ & $0.066$ & $0.058$ \\
& Sim. & $0.707$ & $0.706$ & $0.663$ & $0.514$ & $0.460$ & $0.440$ \\
& R-L & $0.457$ & $0.463$ & $0.432$ & $0.298$ & $0.251$ & $0.231$ \\
\midrule
\multirow{3}{*}{AGNews} & BLEU & $0.039$ & $0.042$ & $0.036$ & $0.028$ & $0.026$ & $0.025$ \\
& Sim. & $0.658$ & $0.657$ & $0.585$ & $0.503$ & $0.470$ & $0.453$ \\
& R-L & $0.203$ & $0.210$ & $0.195$ & $0.175$ & $0.166$ & $0.163$ \\
\midrule
\multirow{3}{*}{IMDB} & BLEU & $0.010$ & $0.010$ & $0.009$ & $0.009$ & $0.008$ & $0.008$ \\
& Sim. & $0.531$ & $0.501$ & $0.477$ & $0.468$ & $0.463$ & $0.465$ \\
& R-L & $0.136$ & $0.135$ & $0.135$ & $0.132$ & $0.132$ & $0.132$ \\
\bottomrule
\end{tabular}
}
\label{table:eval_results_vec2text}
\end{table}

Our findings in~\autoref{table:eval_gpt4o} reveal a clear trend: model memorization is strongly correlated with reconstruction performance. 
Specifically, models that exhibit higher levels of memorization toward individual training samples tend to be more vulnerable to reconstruction attacks.

We extend this analysis to larger, contemporary model architectures—particularly transformer-based models, which are widely adopted in current practice. 
In our experiments with Swin~\cite{LLCHWZLG21} and MAE~\cite{HCXLDG22} architectures, results presented in~\autoref{table:eval_swin_gpt4o} support similar observations: models trained on smaller datasets exhibit greater vulnerability compared to those trained on larger datasets.

We further broaden our analysis to encompass additional data modalities, with a focus on the text domain, motivated by the rapid rise of Large Language Models. 
In this setting, we investigate two representative reconstruction attack strategies: Vec2Text~\cite{MKSR23}, which reconstructs input text from its embeddings, and Complete~\cite{CTWJHLRBSEOR21,WLBZ24}, which attempts to recover the original prompt provided to an LLM. 
Detailed descriptions of these methods and the experimental setup are provided in~\autoref{app:nlp_setup}. 
To quantitatively measure attack efficacy, we utilized three metrics. 
Semantic similarity served as a macro-level metric, assessing the overall likeness between the reconstructed and original texts. 
For micro-level evaluation, analogous to pixel-level comparisons in image reconstruction, we employed BLEU and ROUGE-L scores to capture finer-grained textual similarities.

As shown in~\autoref{table:eval_results_vec2text}, despite the difference in modality, we observe a consistent pattern: training on larger datasets reduces the model’s memorization of individual samples, which in turn leads to poorer reconstruction quality. 
This suggests that memorization plays a critical role in determining vulnerability to reconstruction.

Building on these observations, a natural defense strategy emerges: reducing memorization may decrease susceptibility to reconstruction attacks. 
This insight helps explain why Differential Privacy (DP) is effective—by limiting the model’s exposure to individual samples, DP inherently reduces memorization. 
Beyond DP, we also explore model pruning as an alternative. 
As shown in~\autoref{table:eval_results_celeba_prune_pretrain}, pruning a substantial number of neuron connections discards stored information, which can mitigate unnecessary memorization and thus reduce reconstruction attack success. 
Notably, pruning results in minimal performance degradation—typically under 1\%—whereas DP often incurs a much larger accuracy drop~\cite{HPTD15}. 
This suggests that pruning may be a more practical defense method in some cases and highlights the potential of developing defenses that target sample-specific memorization.

\begin{table}[!t]
\centering
\caption{Effect of the features learned by the target model to the attack performance for VGG16 trained on CelebA.
Lower FID indicates better attack performance.}
\setlength{\tabcolsep}{3.pt}
\scalebox{0.8}{
\begin{tabular}{@{}ccccccc@{}}
\toprule
\multirow{2}{*}{Attack} &\multicolumn{6}{c}{Target Data Size}\\ 
\cmidrule(l){2-7}& $1,000$ & $2,000$ & $5,000$ & $10,000$ & $15,000$ & $20,000$\\
\midrule
PLGMI (DP) & $173.735$ & $156.141$ & $140.480$ & $127.090$ & $122.218$ & $119.856$  \\
PLGMI (Prune) & $154.435$ & $153.764$ & $138.074$ & $126.206$ & $122.993$ & $121.437$ \\
PLGMI & $127.722$ & $107.833$ & $104.842$ & $97.730$ & $84.836$ & $85.143$ \\
PLGMI (Pre) & $94.603$ & $82.571$ & $86.860$ & $78.429$ & $82.768$ & $80.814$ \\
\bottomrule
\end{tabular}
}
\label{table:eval_results_celeba_prune_pretrain}
\end{table}

This intuition is further supported by the observation that datasets used for fine-tuning pre-trained models are often more susceptible to reconstruction. 
During fine-tuning, the model, having already learned general data distributions from pre-training, tends to develop stronger memorization of the specific features unique to the fine-tuning samples. 
This underscores the need for caution when fine-tuning pre-trained models, as this process can heighten data exposure risks. 
This finding also resonates with existing research on attacks that manipulate pre-trained models to render fine-tuned versions more vulnerable to membership inference~\cite{WMHGGC24,LWCX24}. 
Such research indicates that adversaries might modify pre-trained models to make the fine-tuning dataset easier to reconstruct, potentially by encouraging the model to memorize more sample-specific features.

Furthermore, caution should be exercised when releasing model parameters, particularly those that encapsulate statistical information about the training data, as demonstrated in~\autoref{table:deepinversion_bn}. 
Such parameters can be exploited in reconstruction efforts.

Finally, for the advancement of reconstruction attack methodologies, researchers should aim to optimally utilize all available information. 
This includes strategically leveraging knowledge of the dataset distribution or statistical details embedded within model parameters. 
Concurrently, a complementary and important research avenue involves developing robust reconstruction techniques that can succeed even without access to such auxiliary information.

\section*{Acknowledgments}

We would like to thank the anonymous reviewers for their insightful comments and constructive suggestions.
We are especially grateful to Tianhao Wang for the in-depth discussions and valuable feedback throughout the development of this work.
This work is partially funded by the European Health
and Digital Executive Agency (HADEA) within the project
``Understanding the individual host response against Hepatitis D Virus to develop a personalized approach for
the management of hepatitis D'' (DSolve, grant agreement number 101057917) and the BMBF with the project
``Repr\"asentative, synthetische Gesundheitsdaten mit starken
Privatsph \"arengarantien'' (PriSyn, 16KISAO29K).

\section*{Ethics Considerations}

All experiments in this study were conducted using publicly available open-source datasets, ensuring that no private or sensitive data was involved. 
The target models were trained exclusively on open-source benchmark datasets, which were utilized solely for the purposes of this research. 
Furthermore, the reconstruction attacks were designed to reconstruct data from these open-source datasets only, and no attempt was made to access or infer any sensitive or personal information. 
This approach aligns with ethical research practices and ensures that the work complies with privacy standards and community guidelines.

\section*{Open Science}

In compliance with the open science policy, we are committed to promoting transparency and reproducibility in our research. 
To this end, we share the artifacts associated with our work, including the data reconstruction attack framework and its evaluation code, available at~\url{https://doi.org/10.5281/zenodo.15603060}. 
These resources are provided to facilitate further research and the development of more efficient data reconstruction attack methods, enabling researchers to better evaluate privacy leakage in machine learning models. 
By making these tools publicly available, we aim to contribute meaningfully to the broader scientific community and uphold the principles of open science.

\clearpage

\bibliographystyle{plain}
\bibliography{normal_generated_py3}

\begin{thebibliography}{10}

\bibitem{DeepDream}
\url{https://ai.googleblog.com/2015/06/inceptionism-going-deeper-into-neural.html}.

\bibitem{CIFAR}
\url{https://www.cs.toronto.edu/~kriz/cifar.html}.

\bibitem{MNIST}
\url{http://yann.lecun.com/exdb/mnist/}.

\bibitem{ACGMMTZ16}
Martin Abadi, Andy Chu, Ian Goodfellow, Brendan McMahan, Ilya Mironov, Kunal Talwar, and Li~Zhang.
\newblock {Deep Learning with Differential Privacy}.
\newblock In {\em {ACM SIGSAC Conference on Computer and Communications Security (CCS)}}, pages 308--318. ACM, 2016.

\bibitem{AMSVVF15}
Giuseppe Ateniese, Luigi~V. Mancini, Angelo Spognardi, Antonio Villani, Domenico Vitali, and Giovanni Felici.
\newblock {Hacking smart machines with smarter ones: How to extract meaningful data from machine learning classifiers}.
\newblock {\em {Int. J. Secur. Networks}}, 2015.

\bibitem{BCH22}
Borja Balle, Giovanni Cherubin, and Jamie Hayes.
\newblock {Reconstructing Training Data with Informed Adversaries}.
\newblock In {\em {IEEE Symposium on Security and Privacy (S\&P)}}, pages 1138--1156. IEEE, 2022.

\bibitem{BTS13}
Martin B{\"{a}}uml, Makarand Tapaswi, and Rainer Stiefelhagen.
\newblock {Semi-supervised Learning with Constraints for Person Identification in Multimedia Data}.
\newblock In {\em {IEEE Conference on Computer Vision and Pattern Recognition (CVPR)}}, pages 3602--3609. IEEE, 2013.

\bibitem{BGCDJ19}
Jens Behrmann, Will Grathwohl, Ricky T.~Q. Chen, David Duvenaud, and J{\"{o}}rn{-}Henrik Jacobsen.
\newblock {Invertible Residual Networks}.
\newblock In {\em {International Conference on Machine Learning (ICML)}}, pages 573--582. PMLR, 2019.

\bibitem{CCNSTT22}
Nicholas Carlini, Steve Chien, Milad Nasr, Shuang Song, Andreas Terzis, and Florian Tram{\`{e}}r.
\newblock {Membership Inference Attacks From First Principles}.
\newblock In {\em {IEEE Symposium on Security and Privacy (S\&P)}}, pages 1897--1914. IEEE, 2022.

\bibitem{CJZPTT22}
Nicholas Carlini, Matthew Jagielski, Chiyuan Zhang, Nicolas Papernot, Andreas Terzis, and Florian Tram{\`{e}}r.
\newblock {The Privacy Onion Effect: Memorization is Relative}.
\newblock In {\em {Annual Conference on Neural Information Processing Systems (NeurIPS)}}. NeurIPS, 2022.

\bibitem{CTWJHLRBSEOR21}
Nicholas Carlini, Florian Tram{\`{e}}r, Eric Wallace, Matthew Jagielski, Ariel Herbert{-}Voss, Katherine Lee, Adam Roberts, Tom~B. Brown, Dawn Song, {\'{U}}lfar Erlingsson, Alina Oprea, and Colin Raffel.
\newblock {Extracting Training Data from Large Language Models}.
\newblock In {\em {USENIX Security Symposium (USENIX Security)}}, pages 2633--2650. USENIX, 2021.

\bibitem{CG162}
Paola Cerchiello and Paolo Giudici.
\newblock {Big Data Analysis for Financial Risk Management}.
\newblock {\em {Journal of Big Data}}, 2016.

\bibitem{CAOJTU23}
Harsh Chaudhari, John Abascal, Alina Oprea, Matthew Jagielski, Florian Tram{\`{e}}r, and Jonathan~R. Ullman.
\newblock {SNAP: Efficient Extraction of Private Properties with Poisoning}.
\newblock In {\em {IEEE Symposium on Security and Privacy (S\&P)}}, pages 1935--1952. IEEE, 2023.

\bibitem{CKJQ21}
Si~Chen, Mostafa Kahla, Ruoxi Jia, and Guo{-}Jun Qi.
\newblock {Knowledge-Enriched Distributional Model Inversion Attacks}.
\newblock In {\em {IEEE International Conference on Computer Vision (ICCV)}}, pages 16158--16167. IEEE, 2021.

\bibitem{CTCP21}
Christopher A.~Choquette Choo, Florian Tram{\`e}r, Nicholas Carlini, and Nicolas Papernot.
\newblock {Label-Only Membership Inference Attacks}.
\newblock In {\em {International Conference on Machine Learning (ICML)}}, pages 1964--1974. PMLR, 2021.

\bibitem{CBKLYH18}
Tarin Clanuwat, Mikel Bober{-}Irizar, Asanobu Kitamoto, Alex Lamb, Kazuaki Yamamoto, and David Ha.
\newblock {Deep Learning for Classical Japanese Literature}.
\newblock {\em {CoRR abs/1812.01718}}, 2018.

\bibitem{DDYPB23}
Haonan Duan, Adam Dziedzic, Mohammad Yaghini, Nicolas Papernot, and Franziska Boenisch.
\newblock {On the Privacy Risk of In-context Learning}.
\newblock In {\em {Workshop on Trustworthy Natural Language Processing (TrustNLP)}}, 2023.

\bibitem{DR14}
Cynthia Dwork and Aaron Roth.
\newblock {\em {The Algorithmic Foundations of Differential Privacy}}.
\newblock Now Publishers Inc., 2014.

\bibitem{F20}
Vitaly Feldman.
\newblock {Does Learning Require Memorization? A Short Tale about a Long Tail}.
\newblock In {\em {Annual ACM Symposium on Theory of Computing (STOC)}}, pages 954--959. ACM, 2020.

\bibitem{FJR15}
Matt Fredrikson, Somesh Jha, and Thomas Ristenpart.
\newblock {Model Inversion Attacks that Exploit Confidence Information and Basic Countermeasures}.
\newblock In {\em {ACM SIGSAC Conference on Computer and Communications Security (CCS)}}, pages 1322--1333. ACM, 2015.

\bibitem{GWYGB18}
Karan Ganju, Qi~Wang, Wei Yang, Carl~A. Gunter, and Nikita Borisov.
\newblock {Property Inference Attacks on Fully Connected Neural Networks using Permutation Invariant Representations}.
\newblock In {\em {ACM SIGSAC Conference on Computer and Communications Security (CCS)}}, pages 619--633. ACM, 2018.

\bibitem{GPMXWOCB14}
Ian Goodfellow, Jean Pouget-Abadie, Mehdi Mirza, Bing Xu, David Warde-Farley, Sherjil Ozair, Aaron Courville, and Yoshua Bengio.
\newblock {Generative Adversarial Nets}.
\newblock In {\em {Annual Conference on Neural Information Processing Systems (NIPS)}}, pages 2672--2680. NIPS, 2014.

\bibitem{HVYSI22}
Niv Haim, Gal Vardi, Gilad Yehudai, Ohad Shamir, and Michal Irani.
\newblock {Reconstructing Training Data from Trained Neural Networks}.
\newblock In {\em {Annual Conference on Neural Information Processing Systems (NeurIPS)}}. NeurIPS, 2022.

\bibitem{HPTD15}
Song Han, Jeff Pool, John Tran, and William~J. Dally.
\newblock {Learning both Weights and Connections for Efficient Neural Networks}.
\newblock {\em {CoRR abs/1506.02626}}, 2015.

\bibitem{HCXLDG22}
Kaiming He, Xinlei Chen, Saining Xie, Yanghao Li, Piotr Doll{\'{a}}r, and Ross~B. Girshick.
\newblock {Masked Autoencoders Are Scalable Vision Learners}.
\newblock In {\em {IEEE Conference on Computer Vision and Pattern Recognition (CVPR)}}, pages 15979--15988. IEEE, 2022.

\bibitem{HZRS16}
Kaiming He, Xiangyu Zhang, Shaoqing Ren, and Jian Sun.
\newblock {Deep Residual Learning for Image Recognition}.
\newblock In {\em {IEEE Conference on Computer Vision and Pattern Recognition (CVPR)}}, pages 770--778. IEEE, 2016.

\bibitem{HLXCZ22}
Xinlei He, Zheng Li, Weilin Xu, Cory Cornelius, and Yang Zhang.
\newblock {Membership-Doctor: Comprehensive Assessment of Membership Inference Against Machine Learning Models}.
\newblock {\em {CoRR abs/2208.10445}}, 2022.

\bibitem{HWWBSZ21}
Xinlei He, Rui Wen, Yixin Wu, Michael Backes, Yun Shen, and Yang Zhang.
\newblock {Node-Level Membership Inference Attacks Against Graph Neural Networks}.
\newblock {\em {CoRR abs/2102.05429}}, 2021.

\bibitem{KPQ21}
Sanjay Kariyappa, Atul Prakash, and Moinuddin~K. Qureshi.
\newblock {{MAZE:} Data-Free Model Stealing Attack Using Zeroth-Order Gradient Estimation}.
\newblock In {\em {IEEE Conference on Computer Vision and Pattern Recognition (CVPR)}}, pages 13814--13823. IEEE, 2021.

\bibitem{KLA19}
Tero Karras, Samuli Laine, and Timo Aila.
\newblock {A Style-Based Generator Architecture for Generative Adversarial Networks}.
\newblock In {\em {IEEE Conference on Computer Vision and Pattern Recognition (CVPR)}}, pages 4401--4410. IEEE, 2019.

\bibitem{KD18}
Diederik~P. Kingma and Prafulla Dhariwal.
\newblock {Glow: Generative Flow with Invertible 1x1 Convolutions}.
\newblock In {\em {Annual Conference on Neural Information Processing Systems (NeurIPS)}}, pages 10236--10245. NeurIPS, 2018.

\bibitem{KA18}
Pahulpreet~Singh Kohli and Shriya Arora.
\newblock {Application of Machine Learning in Disease Prediction}.
\newblock In {\em {International Conference on Computing Communication and Automation (ICCCA)}}, pages 1--4. IEEE, 2018.

\bibitem{LLWHYZFZ24}
Hao Li, Zheng Li, Siyuan Wu, Chengrui Hu, Yutong Ye, Min Zhang, Dengguo Feng, and Yang Zhang.
\newblock {SeqMIA: Sequential-Metric Based Membership Inference Attack}.
\newblock {\em {CoRR abs/2407.15098}}, 2024.

\bibitem{LHYZ24}
Zheng Li, Xinlei He, Ning Yu, and Yang Zhang.
\newblock {Membership Inference Attack Against Masked Image Modeling}.
\newblock {\em {CoRR abs/2408.06825}}, 2024.

\bibitem{LLHYBZ22}
Zheng Li, Yiyong Liu, Xinlei He, Ning Yu, Michael Backes, and Yang Zhang.
\newblock {Auditing Membership Leakages of Multi-Exit Networks}.
\newblock In {\em {ACM SIGSAC Conference on Computer and Communications Security (CCS)}}, pages 1917--1931. ACM, 2022.

\bibitem{LZ21}
Zheng Li and Yang Zhang.
\newblock {Membership Leakage in Label-Only Exposures}.
\newblock In {\em {ACM SIGSAC Conference on Computer and Communications Security (CCS)}}, pages 880--895. ACM, 2021.

\bibitem{LWCX24}
Ruixuan Liu, Tianhao Wang, Yang Cao, and Li~Xiong.
\newblock {PreCurious: How Innocent Pre-Trained Language Models Turn into Privacy Traps}.
\newblock In {\em {ACM SIGSAC Conference on Computer and Communications Security (CCS)}}, pages 3511--3524. ACM, 2024.

\bibitem{LZBZ22}
Yiyong Liu, Zhengyu Zhao, Michael Backes, and Yang Zhang.
\newblock {Membership Inference Attacks by Exploiting Loss Trajectory}.
\newblock In {\em {ACM SIGSAC Conference on Computer and Communications Security (CCS)}}, pages 2085--2098. ACM, 2022.

\bibitem{LWHSZBCFZ22}
Yugeng Liu, Rui Wen, Xinlei He, Ahmed Salem, Zhikun Zhang, Michael Backes, Emiliano~De Cristofaro, Mario Fritz, and Yang Zhang.
\newblock {ML-Doctor: Holistic Risk Assessment of Inference Attacks Against Machine Learning Models}.
\newblock In {\em {USENIX Security Symposium (USENIX Security)}}, pages 4525--4542. USENIX, 2022.

\bibitem{LLCHWZLG21}
Ze~Liu, Yutong Lin, Yue Cao, Han Hu, Yixuan Wei, Zheng Zhang, Stephen Lin, and Baining Guo.
\newblock {Swin Transformer: Hierarchical Vision Transformer using Shifted Windows}.
\newblock In {\em {IEEE International Conference on Computer Vision (ICCV)}}, pages 9992--10002. IEEE, 2021.

\bibitem{LLWT15}
Ziwei Liu, Ping Luo, Xiaogang Wang, and Xiaoou Tang.
\newblock {Deep Learning Face Attributes in the Wild}.
\newblock In {\em {IEEE International Conference on Computer Vision (ICCV)}}, pages 3730--3738. IEEE, 2015.

\bibitem{MDPHNP11}
Andrew~L. Maas, Raymond~E. Daly, Peter~T. Pham, Dan Huang, Andrew~Y. Ng, and Christopher Potts.
\newblock {Learning Word Vectors for Sentiment Analysis}.
\newblock In {\em {Annual Meeting of the Association for Computational Linguistics (ACL)}}, pages 142--150. ACL, 2011.

\bibitem{MGC22}
Saeed Mahloujifar, Esha Ghosh, and Melissa Chase.
\newblock {Property Inference from Poisoning}.
\newblock In {\em {IEEE Symposium on Security and Privacy (S\&P)}}, pages 1120--1137. IEEE, 2022.

\bibitem{MSCS19}
Luca Melis, Congzheng Song, Emiliano~De Cristofaro, and Vitaly Shmatikov.
\newblock {Exploiting Unintended Feature Leakage in Collaborative Learning}.
\newblock In {\em {IEEE Symposium on Security and Privacy (S\&P)}}, pages 497--512. IEEE, 2019.

\bibitem{MKSR23}
John~X. Morris, Volodymyr Kuleshov, Vitaly Shmatikov, and Alexander~M. Rush.
\newblock {Text Embeddings Reveal (Almost) As Much As Text}.
\newblock In {\em {Conference on Empirical Methods in Natural Language Processing (EMNLP)}}, pages 12448--12460. Association for Computational Linguistics, 2023.

\bibitem{PSMRTE18}
Nicolas Papernot, Shuang Song, Ilya Mironov, Ananth Raghunathan, Kunal Talwar, and {\'{U}}lfar Erlingsson.
\newblock {Scalable Private Learning with {PATE}}.
\newblock In {\em {International Conference on Learning Representations (ICLR)}}, 2018.

\bibitem{SBBFZ20}
Ahmed Salem, Apratim Bhattacharya, Michael Backes, Mario Fritz, and Yang Zhang.
\newblock {Updates-Leak: Data Set Inference and Reconstruction Attacks in Online Learning}.
\newblock In {\em {USENIX Security Symposium (USENIX Security)}}, pages 1291--1308. USENIX, 2020.

\bibitem{SCEKPSTB23}
Ahmed Salem, Giovanni Cherubin, David Evans, Boris K{\"{o}}pf, Andrew Paverd, Anshuman Suri, Shruti Tople, and Santiago~Zanella B{\'{e}}guelin.
\newblock {SoK: Let the Privacy Games Begin! A Unified Treatment of Data Inference Privacy in Machine Learning}.
\newblock In {\em {IEEE Symposium on Security and Privacy (S\&P)}}, pages 327--345. IEEE, 2023.

\bibitem{SZHBFB19}
Ahmed Salem, Yang Zhang, Mathias Humbert, Pascal Berrang, Mario Fritz, and Michael Backes.
\newblock {ML-Leaks: Model and Data Independent Membership Inference Attacks and Defenses on Machine Learning Models}.
\newblock In {\em {Network and Distributed System Security Symposium (NDSS)}}. Internet Society, 2019.

\bibitem{SHZZC18}
Mark Sandler, Andrew~G. Howard, Menglong Zhu, Andrey Zhmoginov, and Liang{-}Chieh Chen.
\newblock {MobileNetV2: Inverted Residuals and Linear Bottlenecks}.
\newblock In {\em {IEEE Conference on Computer Vision and Pattern Recognition (CVPR)}}, pages 4510--4520. IEEE, 2018.

\bibitem{SAB22}
Sunandini Sanyal, Sravanti Addepalli, and R.~Venkatesh Babu.
\newblock {Towards Data-Free Model Stealing in a Hard Label Setting}.
\newblock {\em {CoRR abs/2204.11022}}, 2022.

\bibitem{SSSS17}
Reza Shokri, Marco Stronati, Congzheng Song, and Vitaly Shmatikov.
\newblock {Membership Inference Attacks Against Machine Learning Models}.
\newblock In {\em {IEEE Symposium on Security and Privacy (S\&P)}}, pages 3--18. IEEE, 2017.

\bibitem{SZ15}
Karen Simonyan and Andrew Zisserman.
\newblock {Very Deep Convolutional Networks for Large-Scale Image Recognition}.
\newblock In {\em {International Conference on Learning Representations (ICLR)}}, 2015.

\bibitem{SPWCMNP13}
Richard Socher, Alex Perelygin, Jean Wu, Jason Chuang, Christopher~D. Manning, Andrew~Y. Ng, and Christopher Potts.
\newblock {Recursive Deep Models for Semantic Compositionality Over a Sentiment Treebank}.
\newblock In {\em {Conference on Empirical Methods in Natural Language Processing (EMNLP)}}, pages 1631--1642. ACL, 2013.

\bibitem{SRS17}
Congzheng Song, Thomas Ristenpart, and Vitaly Shmatikov.
\newblock {Machine Learning Models that Remember Too Much}.
\newblock In {\em {ACM SIGSAC Conference on Computer and Communications Security (CCS)}}, pages 587--601. ACM, 2017.

\bibitem{SS20}
Congzheng Song and Vitaly Shmatikov.
\newblock {Overlearning Reveals Sensitive Attributes}.
\newblock In {\em {International Conference on Learning Representations (ICLR)}}, 2020.

\bibitem{SME19}
Yang Song, Chenlin Meng, and Stefano Ermon.
\newblock {MintNet: Building Invertible Neural Networks with Masked Convolutions}.
\newblock In {\em {Annual Conference on Neural Information Processing Systems (NeurIPS)}}, pages 11002--11012. NeurIPS, 2019.

\bibitem{SHCAK22}
Lukas Struppek, Dominik Hintersdorf, Antonio De~Almeida Correia, Antonia Adler, and Kristian Kersting.
\newblock {Plug {\&} Play Attacks: Towards Robust and Flexible Model Inversion Attacks}.
\newblock In {\em {International Conference on Machine Learning (ICML)}}, pages 20522--20545. PMLR, 2022.

\bibitem{SE21}
Anshuman Suri and David Evans.
\newblock {Formalizing and Estimating Distribution Inference Risks}.
\newblock {\em {CoRR abs/2109.06024}}, 2021.

\bibitem{TZ15}
Naftali Tishby and Noga Zaslavsky.
\newblock {Deep Learning and the Information Bottleneck Principle}.
\newblock {\em {CoRR abs/1503.02406}}, 2015.

\bibitem{TSJLJHC22}
Florian Tram{\`e}r, Reza Shokri, Ayrton~San Joaquin, Hoang Le, Matthew Jagielski, Sanghyun Hong, and Nicholas Carlini.
\newblock {Truth Serum: Poisoning Machine Learning Models to Reveal Their Secrets}.
\newblock In {\em {ACM SIGSAC Conference on Computer and Communications Security (CCS)}}. ACM, 2022.

\bibitem{TZJRR16}
Florian Tram{\`e}r, Fan Zhang, Ari Juels, Michael~K. Reiter, and Thomas Ristenpart.
\newblock {Stealing Machine Learning Models via Prediction APIs}.
\newblock In {\em {USENIX Security Symposium (USENIX Security)}}, pages 601--618. USENIX, 2016.

\bibitem{TMWP21}
Jean{-}Baptiste Truong, Pratyush Maini, Robert~J. Walls, and Nicolas Papernot.
\newblock {Data-Free Model Extraction}.
\newblock In {\em {IEEE Conference on Computer Vision and Pattern Recognition (CVPR)}}, pages 4771--4780. IEEE, 2021.

\bibitem{WFLKZM21}
Kuan{-}Chieh Wang, Yan Fu, Ke~Li, Ashish Khisti, Richard~S. Zemel, and Alireza Makhzani.
\newblock {Variational Model Inversion Attacks}.
\newblock In {\em {Annual Conference on Neural Information Processing Systems (NeurIPS)}}, pages 9706--9719. NeurIPS, 2021.

\bibitem{WZJ21}
Tianhao Wang, Yuheng Zhang, and Ruoxi Jia.
\newblock {Improving Robustness to Model Inversion Attacks via Mutual Information Regularization}.
\newblock In {\em {AAAI Conference on Artificial Intelligence (AAAI)}}, pages 11666--11673. AAAI, 2021.

\bibitem{WBZ25}
Rui Wen, Michael Backes, and Yang Zhang.
\newblock {Understanding Data Importance in Machine Learning Attacks: Does Valuable Data Pose Greater Harm?}
\newblock In {\em {Network and Distributed System Security Symposium (NDSS)}}. Internet Society, 2025.

\bibitem{WLBZ24}
Rui Wen, Zheng Li, Michael Backes, and Yang Zhang.
\newblock {Membership Inference Attacks Against In-Context Learning}.
\newblock In {\em {ACM SIGSAC Conference on Computer and Communications Security (CCS)}}. ACM, 2024.

\bibitem{WWBZS23}
Rui Wen, Tianhao Wang, Michael Backes, Yang Zhang, and Ahmed Salem.
\newblock {Last One Standing: A Comparative Analysis of Security and Privacy of Soft Prompt Tuning, LoRA, and In-Context Learning}.
\newblock {\em {CoRR abs/2310.11397}}, 2023.

\bibitem{WMHGGC24}
Yuxin Wen, Leo Marchyok, Sanghyun Hong, Jonas Geiping, Tom Goldstein, and Nicholas Carlini.
\newblock {Privacy Backdoors: Enhancing Membership Inference through Poisoning Pre-trained Models}.
\newblock In {\em {Annual Conference on Neural Information Processing Systems (NeurIPS)}}. NeurIPS, 2024.

\bibitem{WWBBHSZ24}
Yixin Wu, Rui Wen, Michael Backes, Pascal Berrang, Mathias Humbert, Yun Shen, and Yang Zhang.
\newblock {Quantifying Privacy Risks of Prompts in Visual Prompt Learning}.
\newblock In {\em {USENIX Security Symposium (USENIX Security)}}. USENIX, 2024.

\bibitem{YZCL19}
Ziqi Yang, Jiyi Zhang, Ee-Chien Chang, and Zhenkai Liang.
\newblock {Neural Network Inversion in Adversarial Setting via Background Knowledge Alignment}.
\newblock In {\em {ACM SIGSAC Conference on Computer and Communications Security (CCS)}}, page 225–240. ACM, 2019.

\bibitem{YSLXSH22}
Mang Ye, Jianbing Shen, Gaojie Lin, Tao Xiang, Ling Shao, and Steven C.~H. Hoi.
\newblock {Deep Learning for Person Re-Identification: A Survey and Outlook}.
\newblock {\em {IEEE Transactions on Pattern Analysis and Machine Intelligence}}, 2022.

\bibitem{YMALMHJK20}
Hongxu Yin, Pavlo Molchanov, Jose~M. Alvarez, Zhizhong Li, Arun Mallya, Derek Hoiem, Niraj~K. Jha, and Jan Kautz.
\newblock {Dreaming to Distill: Data-Free Knowledge Transfer via DeepInversion}.
\newblock In {\em {IEEE Conference on Computer Vision and Pattern Recognition (CVPR)}}, pages 8712--8721. IEEE, 2020.

\bibitem{YCZZYZ23}
Xiaojian Yuan, Kejiang Chen, Jie Zhang, Weiming Zhang, Nenghai Yu, and Yang Zhang.
\newblock {Pseudo Label-Guided Model Inversion Attack via Conditional Generative Adversarial Network}.
\newblock In {\em {AAAI Conference on Artificial Intelligence (AAAI)}}. AAAI, 2023.

\bibitem{YWLXL22}
Zhuowen Yuan, Fan Wu, Yunhui Long, Chaowei Xiao, and Bo~Li.
\newblock {SecretGen: Privacy Recovery on Pre-trained Models via Distribution Discrimination}.
\newblock In {\em {European Conference on Computer Vision (ECCV)}}, pages 139--155. Springer, 2022.

\bibitem{ZYWBZ24}
Minxing Zhang, Ning Yu, Rui Wen, Michael Backes, and Yang Zhang.
\newblock {Generated Distributions Are All You Need for Membership Inference Attacks Against Generative Models}.
\newblock In {\em {Winter Conference on Applications of Computer Vision (WACV)}}, pages 4827--4837. IEEE, 2024.

\bibitem{ZZL15}
Xiang Zhang, Junbo Zhao, and Yann LeCun.
\newblock {Character-level Convolutional Networks for Text Classification}.
\newblock In {\em {Annual Conference on Neural Information Processing Systems (NIPS)}}, pages 649--657. NIPS, 2015.

\bibitem{ZJPWLS20}
Yuheng Zhang, Ruoxi Jia, Hengzhi Pei, Wenxiao Wang, Bo~Li, and Dawn Song.
\newblock {The Secret Revealer: Generative Model-Inversion Attacks Against Deep Neural Networks}.
\newblock In {\em {IEEE Conference on Computer Vision and Pattern Recognition (CVPR)}}, pages 250--258. IEEE, 2020.

\bibitem{ZLH19}
Ligeng Zhu, Zhijian Liu, and Song Han.
\newblock {Deep Leakage from Gradients}.
\newblock In {\em {Annual Conference on Neural Information Processing Systems (NeurIPS)}}, pages 14747--14756. NeurIPS, 2019.

\end{thebibliography}

\appendix

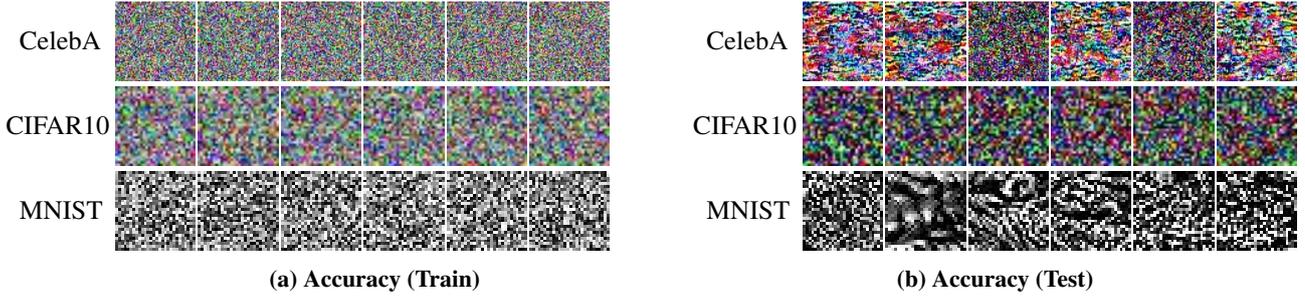
\begin{figure*}
    \centering
    \begin{subfigure}{0.95\columnwidth}
    \centering
       \begin{picture}(100,100)
            \put(-47,64){\includegraphics[width=0.13\columnwidth]{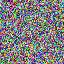}}
            \put(-16,64){\includegraphics[width=0.13\columnwidth]{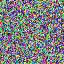}}
            \put(15,64){\includegraphics[width=0.13\columnwidth]{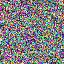}}
            \put(46,64){\includegraphics[width=0.13\columnwidth]{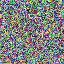}}
            \put(77,64){\includegraphics[width=0.13\columnwidth]{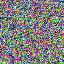}}
            \put(108,64){\includegraphics[width=0.13\columnwidth]{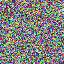}}
            \put(-83,76){CelebA}

            \put(-47,32){\includegraphics[width=0.13\columnwidth]{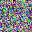}}
            \put(-16,32){\includegraphics[width=0.13\columnwidth]{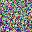}}
            \put(15,32){\includegraphics[width=0.13\columnwidth]{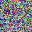}}
            \put(46,32){\includegraphics[width=0.13\columnwidth]{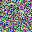}}
            \put(77,32){\includegraphics[width=0.13\columnwidth]{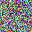}}
            \put(108,32){\includegraphics[width=0.13\columnwidth]{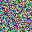}}
            \put(-88,44){CIFAR10}

            \put(-47,0){\includegraphics[width=0.13\columnwidth]{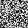}}
            \put(-16,0){\includegraphics[width=0.13\columnwidth]{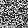}}
            \put(15,0){\includegraphics[width=0.13\columnwidth]{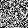}}
            \put(46,0){\includegraphics[width=0.13\columnwidth]{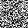}}
            \put(77,0){\includegraphics[width=0.13\columnwidth]{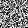}}
            \put(108,0){\includegraphics[width=0.13\columnwidth]{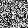}}
            \put(-83,12){MNIST}
        \end{picture}
        \caption{Accuracy (Train)}
        \label{subfigure:metric_trainacc}
     \end{subfigure}     
    \begin{subfigure}{0.95\columnwidth}

    \centering
        \begin{picture}(50,100)
            \put(-47,64){\includegraphics[width=0.13\columnwidth]{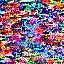}}
            \put(-16,64){\includegraphics[width=0.13\columnwidth]{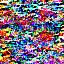}}
            \put(15,64){\includegraphics[width=0.13\columnwidth]{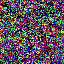}}
            \put(46,64){\includegraphics[width=0.13\columnwidth]{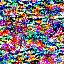}}
            \put(77,64){\includegraphics[width=0.13\columnwidth]{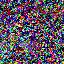}}
            \put(108,64){\includegraphics[width=0.13\columnwidth]{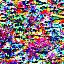}}
            \put(-83,76){CelebA}

            \put(-47,32){\includegraphics[width=0.13\columnwidth]{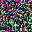}}
            \put(-16,32){\includegraphics[width=0.13\columnwidth]{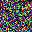}}
            \put(15,32){\includegraphics[width=0.13\columnwidth]{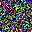}}
            \put(46,32){\includegraphics[width=0.13\columnwidth]{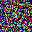}}
            \put(77,32){\includegraphics[width=0.13\columnwidth]{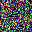}}
            \put(108,32){\includegraphics[width=0.13\columnwidth]{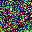}}
            \put(-88,44){CIFAR10}

            \put(-47,0){\includegraphics[width=0.13\columnwidth]{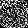}}
            \put(-16,0){\includegraphics[width=0.13\columnwidth]{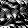}}
            \put(15,0){\includegraphics[width=0.13\columnwidth]{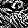}}
            \put(46,0){\includegraphics[width=0.13\columnwidth]{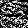}}
            \put(77,0){\includegraphics[width=0.13\columnwidth]{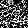}}
            \put(108,0){\includegraphics[width=0.13\columnwidth]{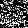}}
            \put(-83,12){MNIST}
        \end{picture}
        \caption{Accuracy (Test)}
        \label{subfigure:metric_testacc}
    \end{subfigure}
\caption{
Counterexamples of accuracy-based metrics.
The left examples show that using random-like samples can train a well-generalized model, and the right examples demonstrate that random-like samples can be predicted with high confidence.}
\label{figure:vetometrics}
\end{figure*}

\begin{figure*}[!t]
\centering
\begin{subfigure}{0.4\columnwidth}
\includegraphics[width=\columnwidth]{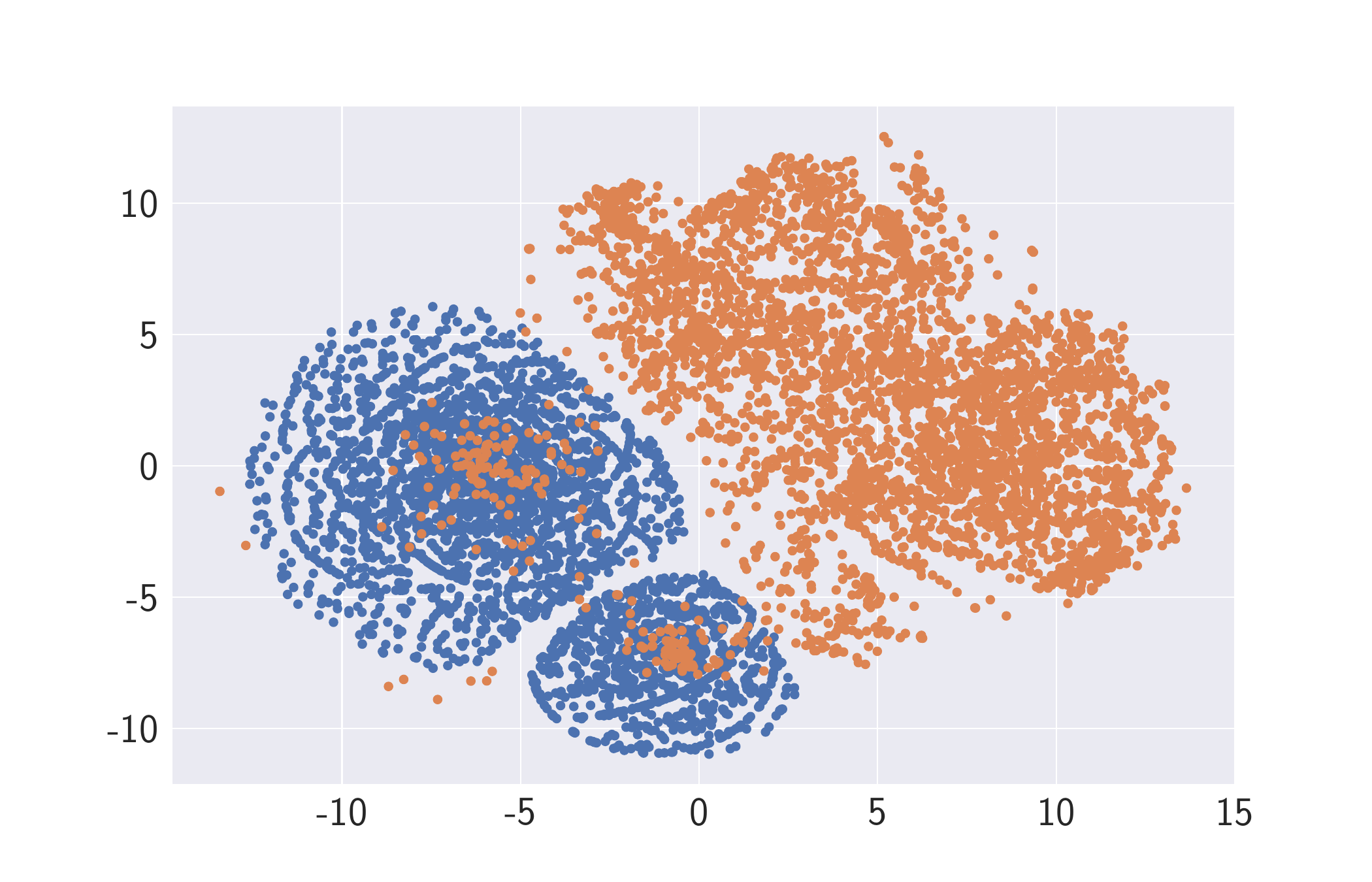}
\caption{ MI-Face }
\label{fig:tsne_mi_ce}
\end{subfigure}
\begin{subfigure}{0.4\columnwidth}
\includegraphics[width=\columnwidth]{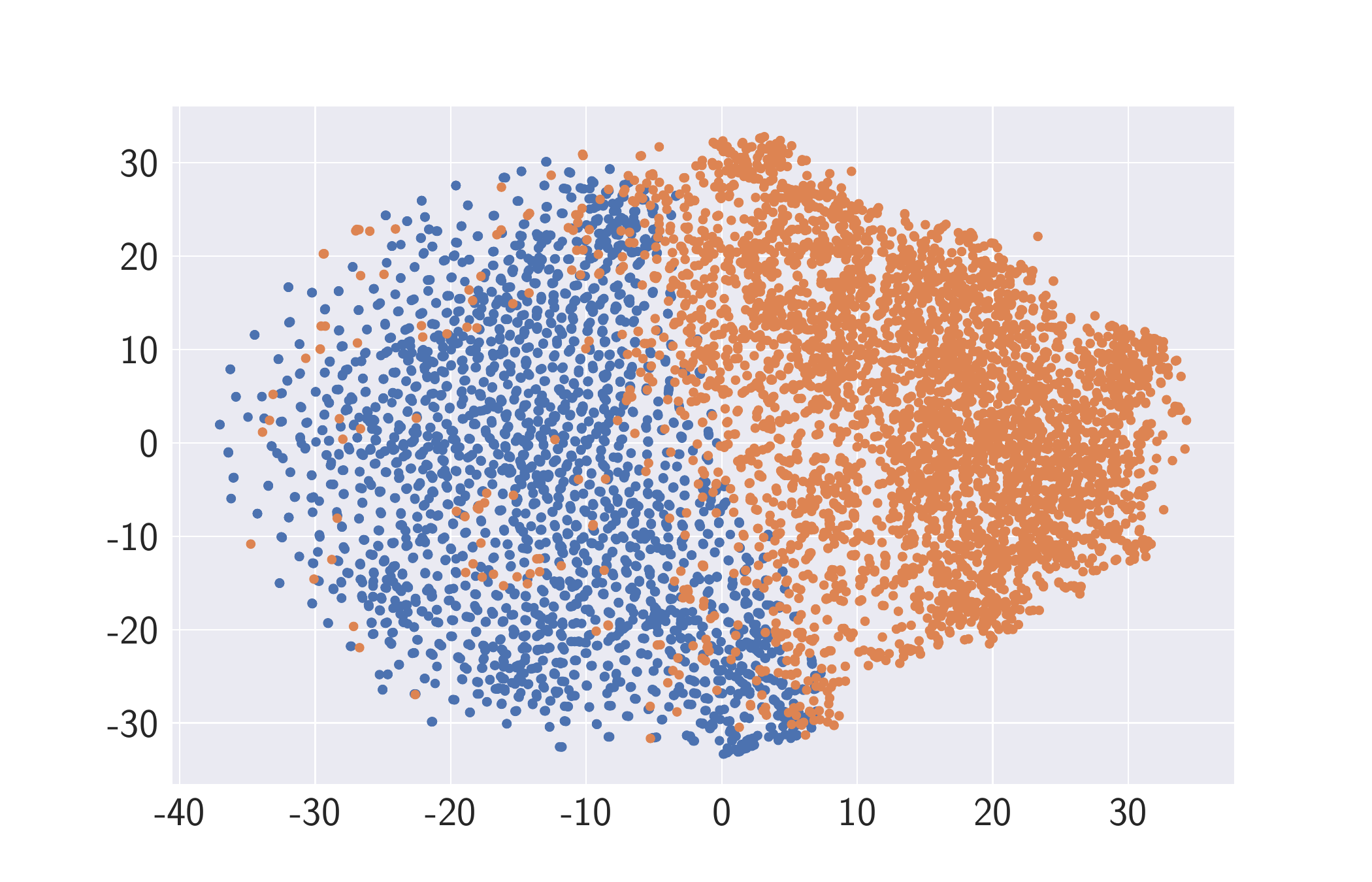}
\caption{ Deepdream }
\label{fig:tsne_deepdream_ce}
\end{subfigure}
\begin{subfigure}{0.4\columnwidth}
\includegraphics[width=\columnwidth]{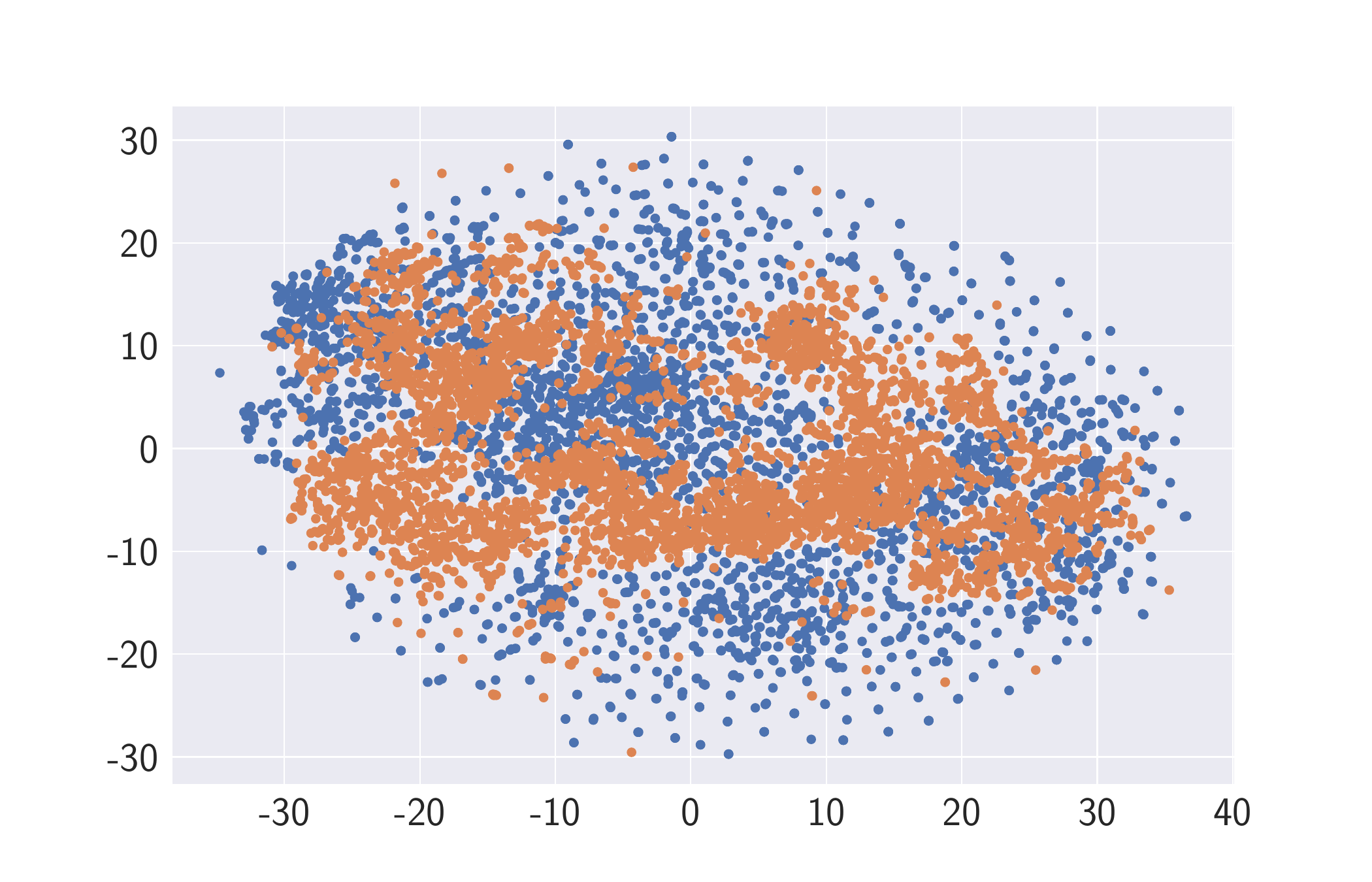}
\caption{ DeepInversion }
\label{fig:tsne_deepinversion_ce}
\end{subfigure}
\begin{subfigure}{0.4\columnwidth}
\includegraphics[width=\columnwidth]{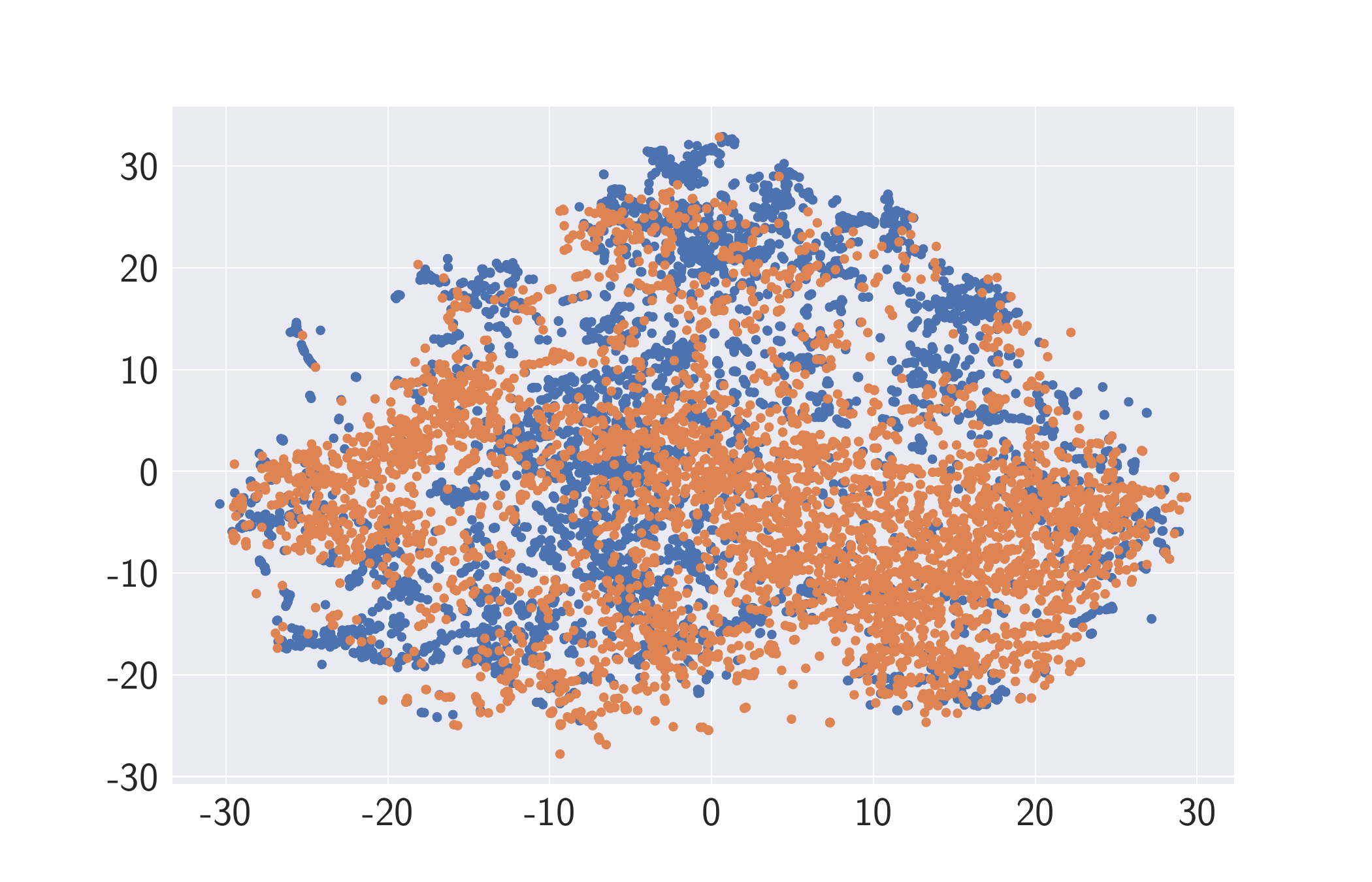}
\caption{ Revealer }
\label{fig:tsne_revealer_ce}
\end{subfigure}
\begin{subfigure}{0.4\columnwidth}
\includegraphics[width=\columnwidth]{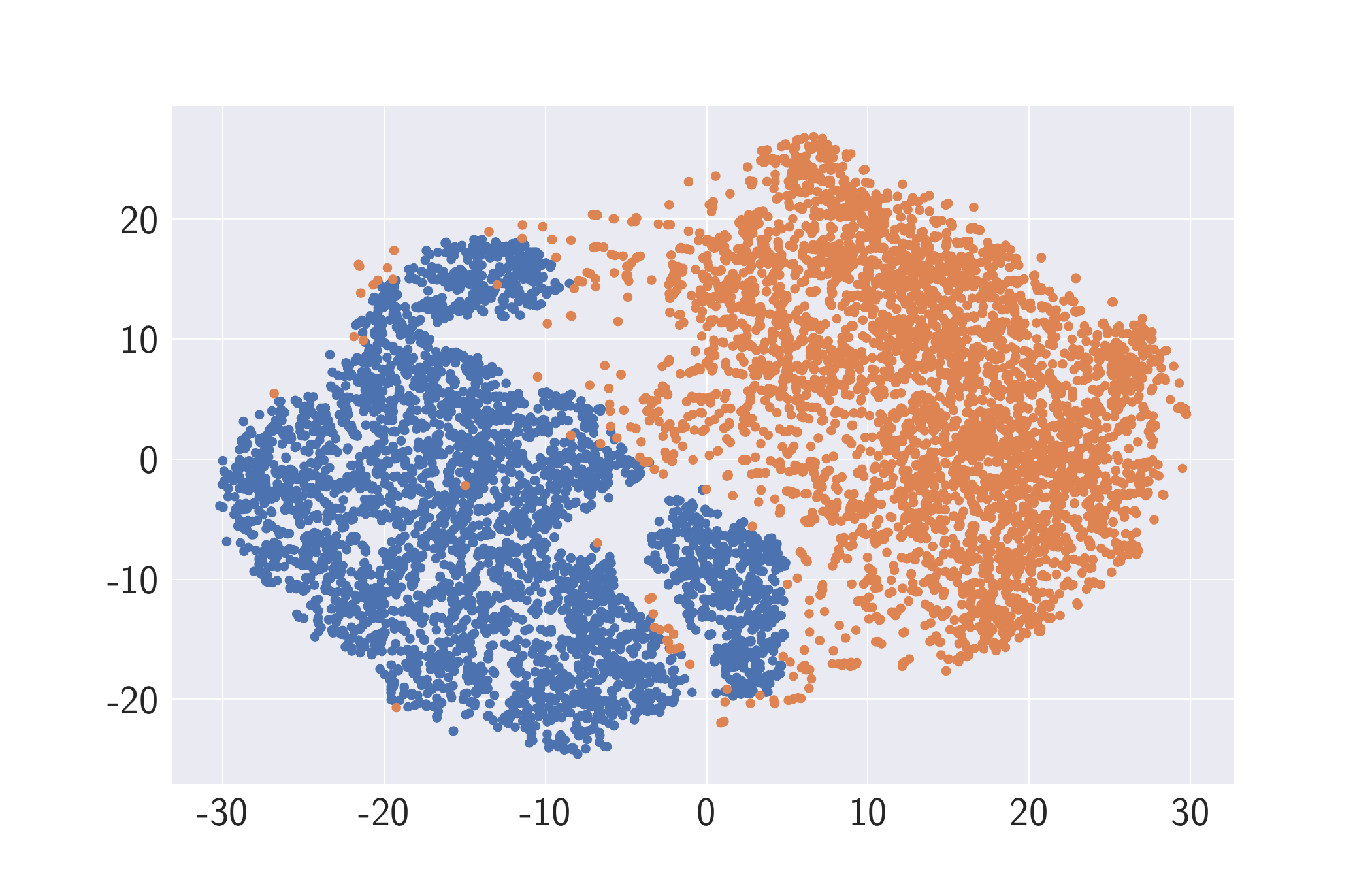}
\caption{ Inv-Alignment }
\label{fig:tsne_inv_ce}
\end{subfigure}
\begin{subfigure}{0.4\columnwidth}
\includegraphics[width=\columnwidth]{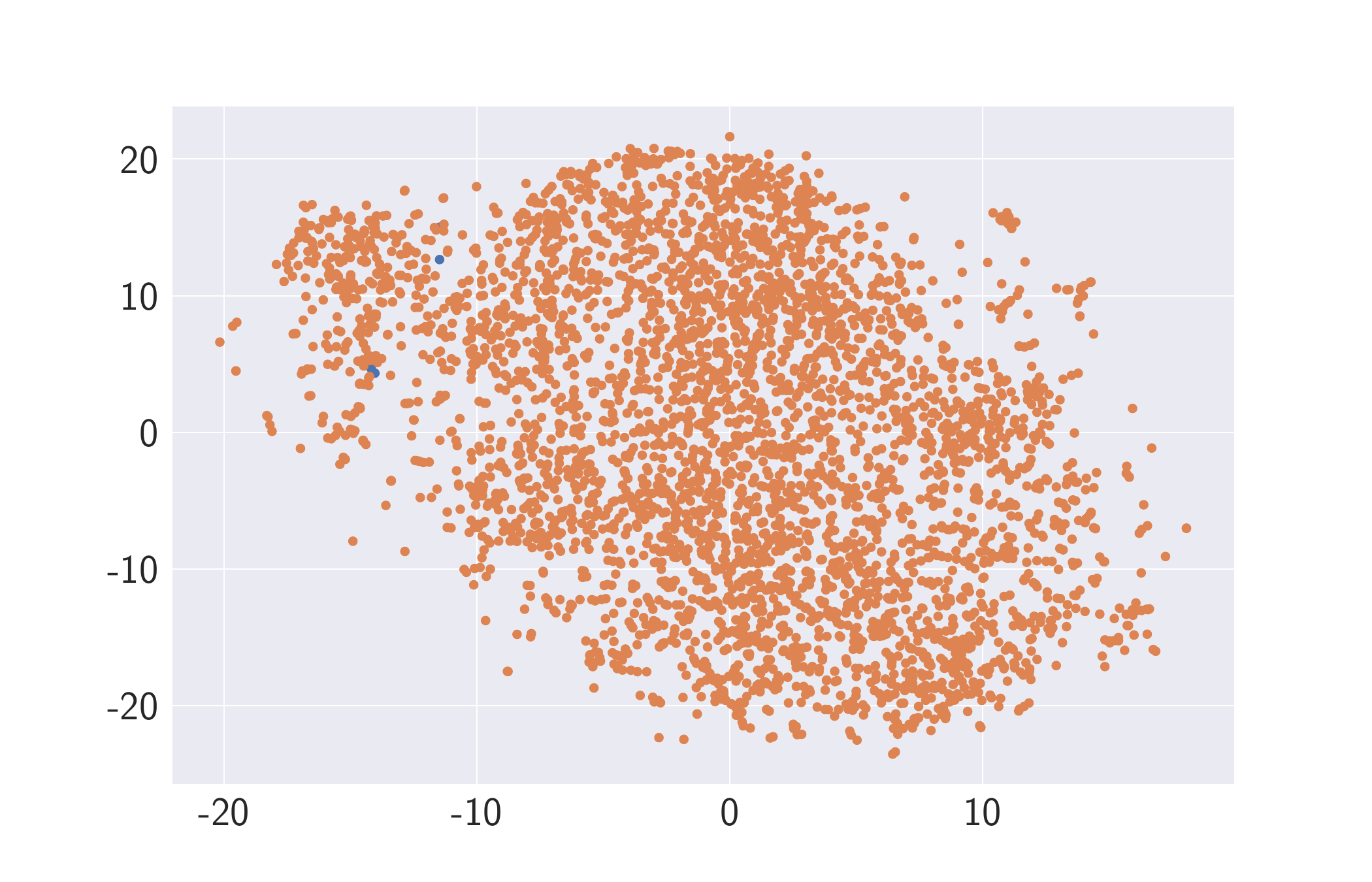}
\caption{ Bias-Rec }
\label{fig:tsne_impbias_ce}
\end{subfigure}
\begin{subfigure}{0.4\columnwidth}
\includegraphics[width=\columnwidth]{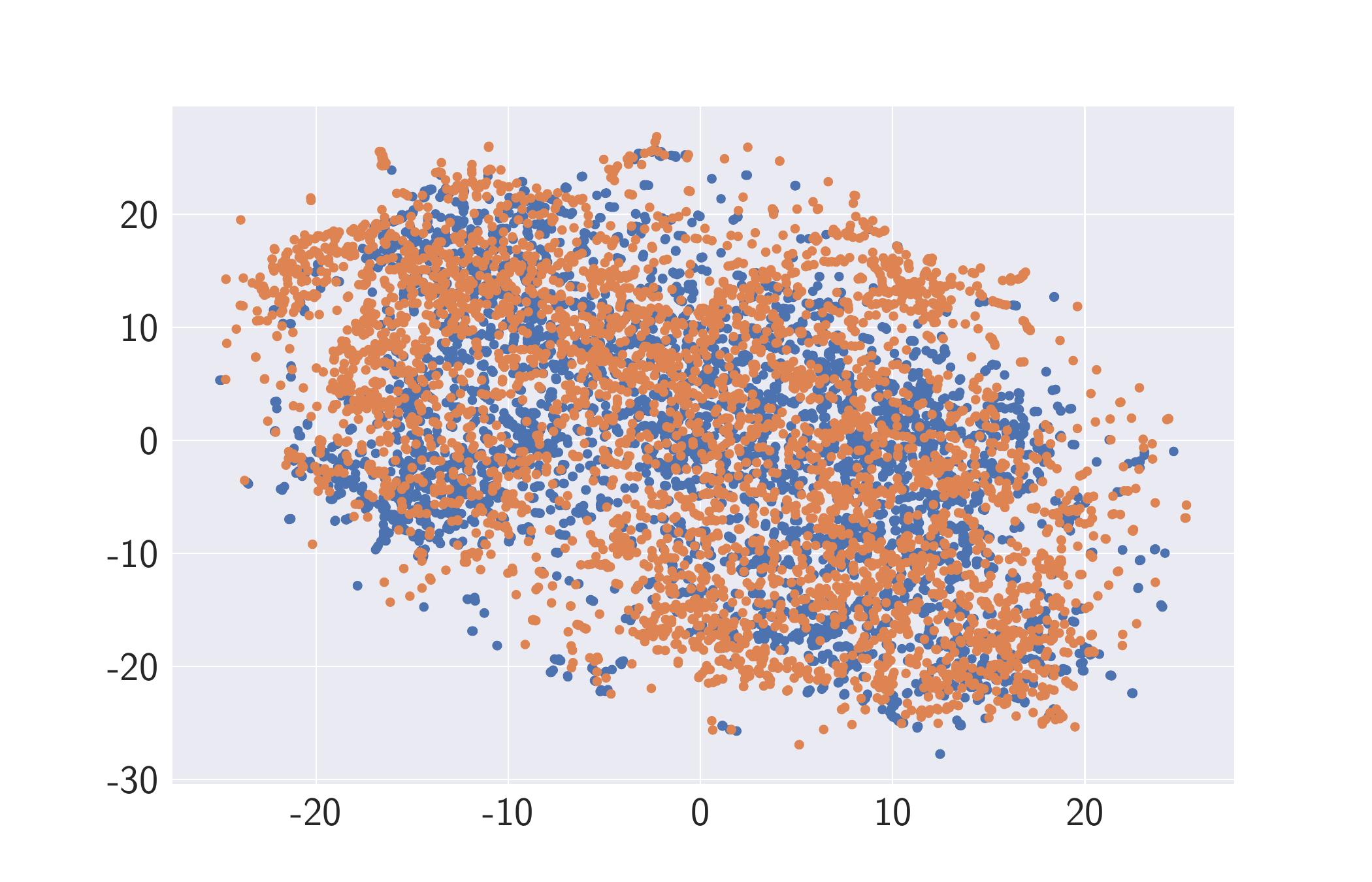}
\caption{ KEDMI }
\label{fig:tsne_kedmi_ce}
\end{subfigure}
\begin{subfigure}{0.4\columnwidth}
\includegraphics[width=\columnwidth]{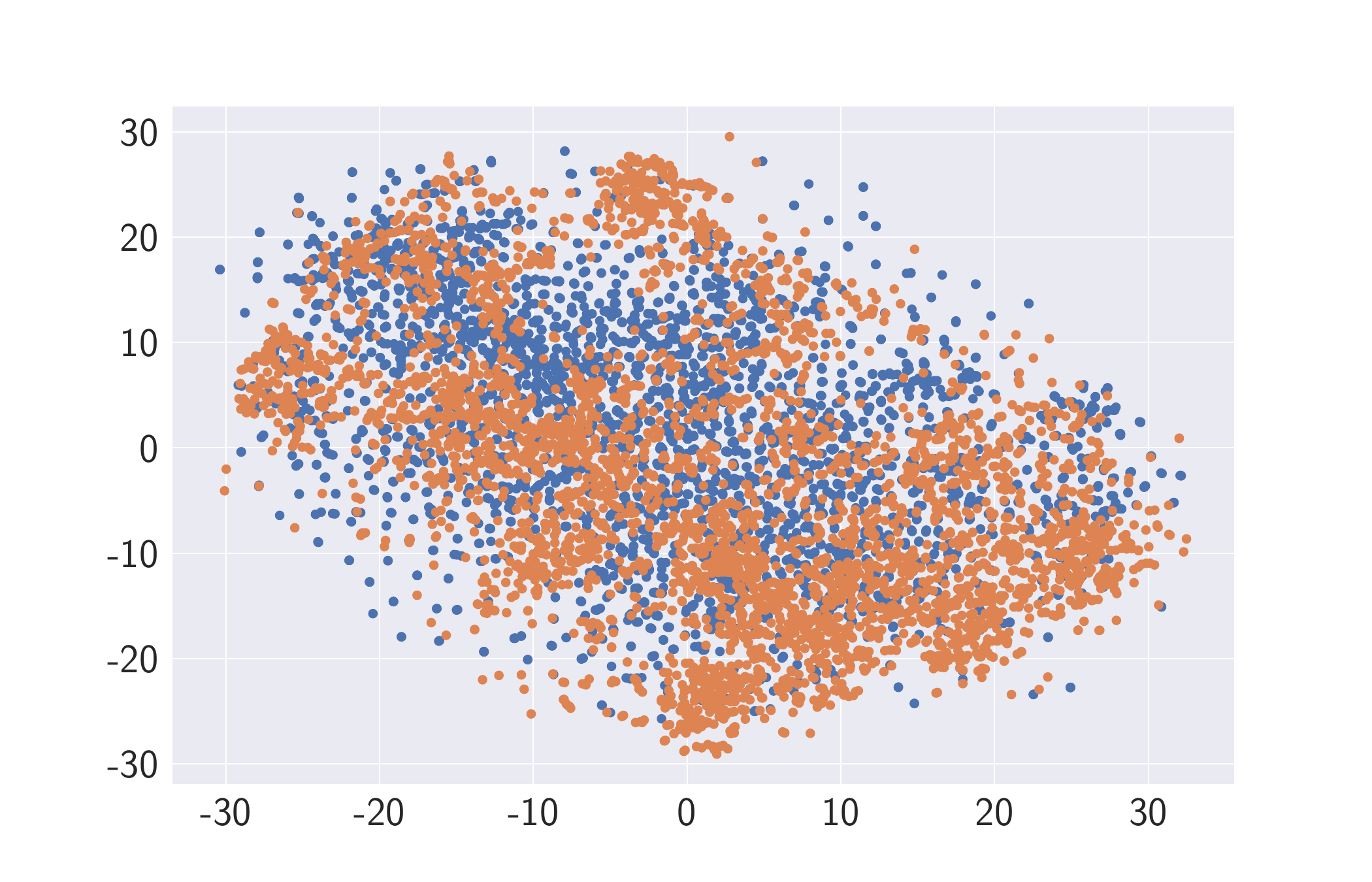}
\caption{ PLGMI }
\label{fig:tsne_plgmi_celeba}
\end{subfigure}
\begin{subfigure}{0.4\columnwidth}
\includegraphics[width=\columnwidth]{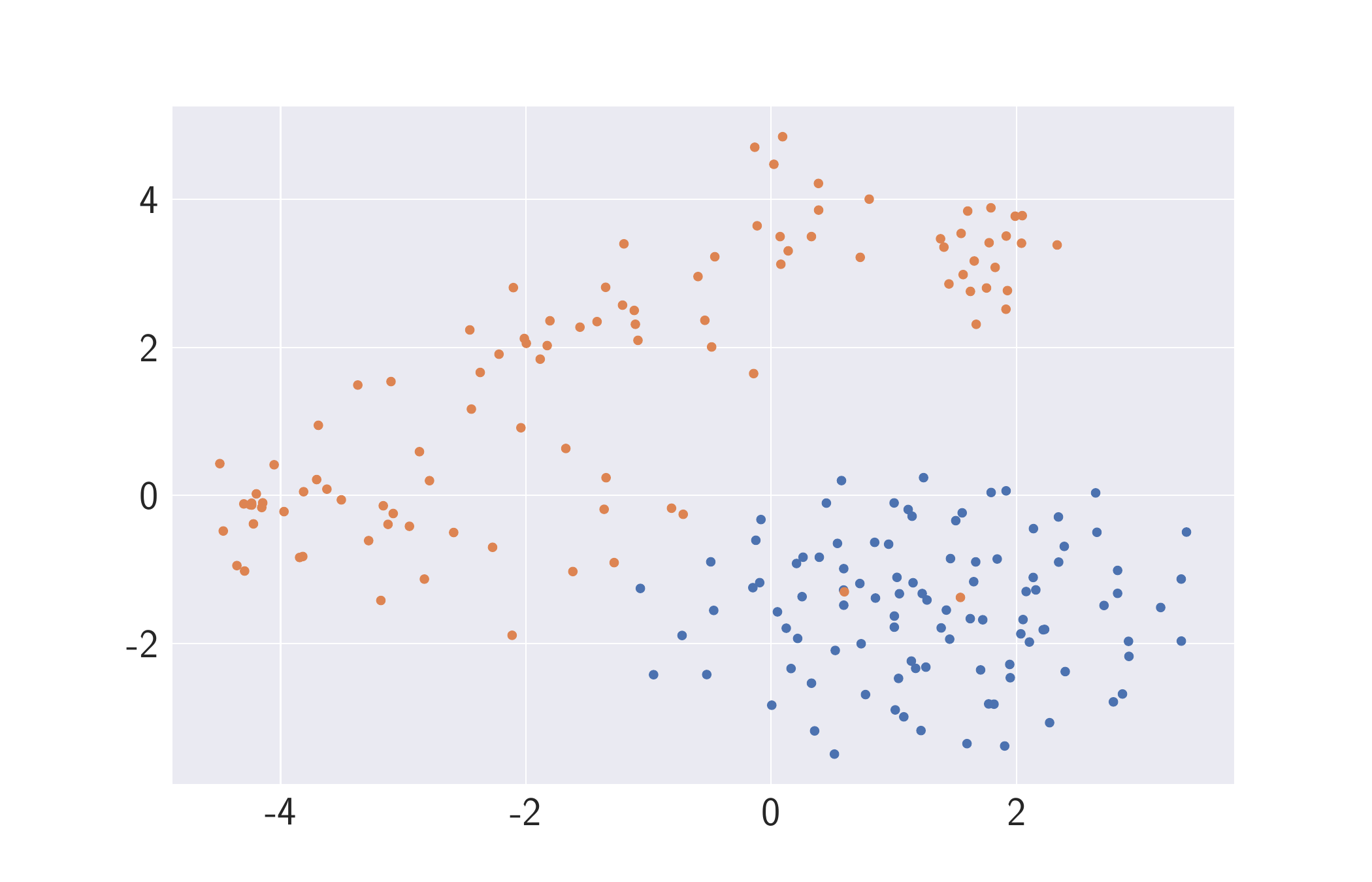}
\caption{ Updates-Leak }
\label{fig:tsne_udleak_ce}
\end{subfigure}
\begin{subfigure}{0.4\columnwidth}
\includegraphics[width=\columnwidth]{distribution_tsne_new/tsne_celeba_gradleak_vgg_20000.pdf}
\caption{ Deep-Leakage }
\label{fig:tsne_gradleak_ce}
\end{subfigure}
\caption{t-SNE visualization of reconstructions and corresponding target datasets for CelebA.}
\label{figure:tsne_reconstruct_celeba}
\end{figure*}

\begin{figure*}[!t]
\centering
\begin{subfigure}{0.4\columnwidth}
\includegraphics[width=\columnwidth]{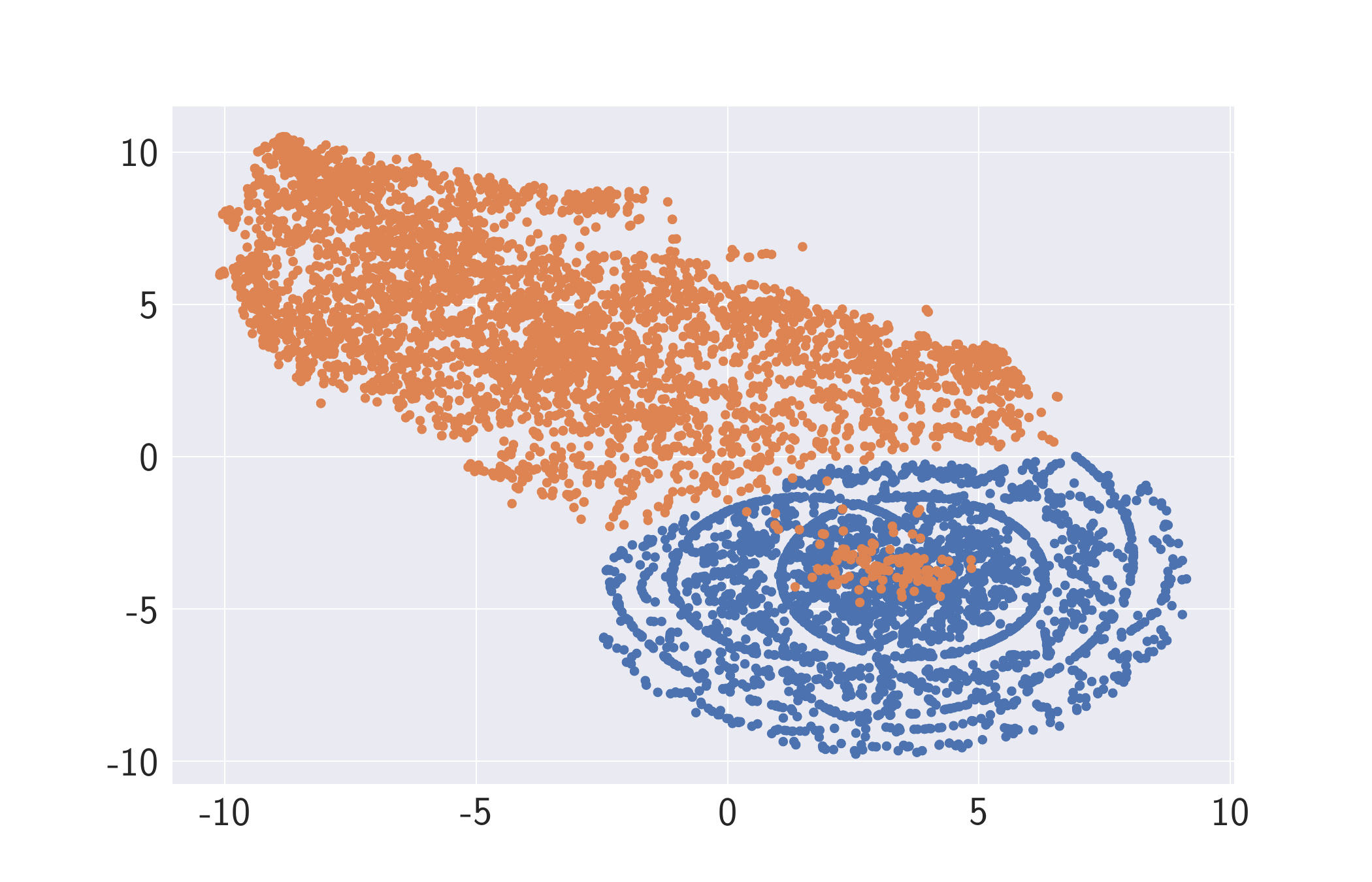}
\caption{ MI-Face }
\label{fig:tsne_mi_cifar}
\end{subfigure}
\begin{subfigure}{0.4\columnwidth}
\includegraphics[width=\columnwidth]{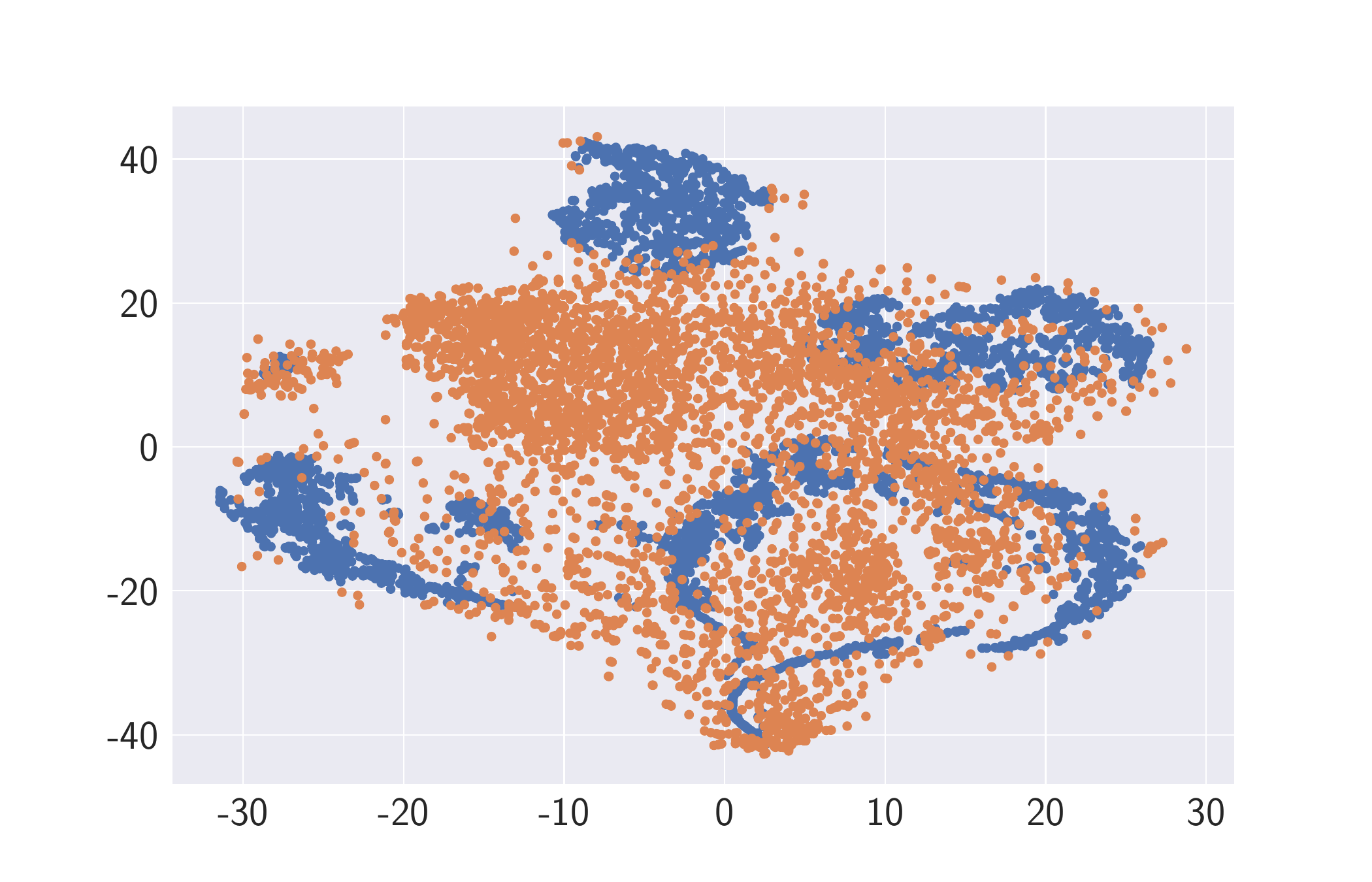}
\caption{ DeepDream }
\label{fig:tsne_deepdream_cifar}
\end{subfigure}
\begin{subfigure}{0.4\columnwidth}
\includegraphics[width=\columnwidth]{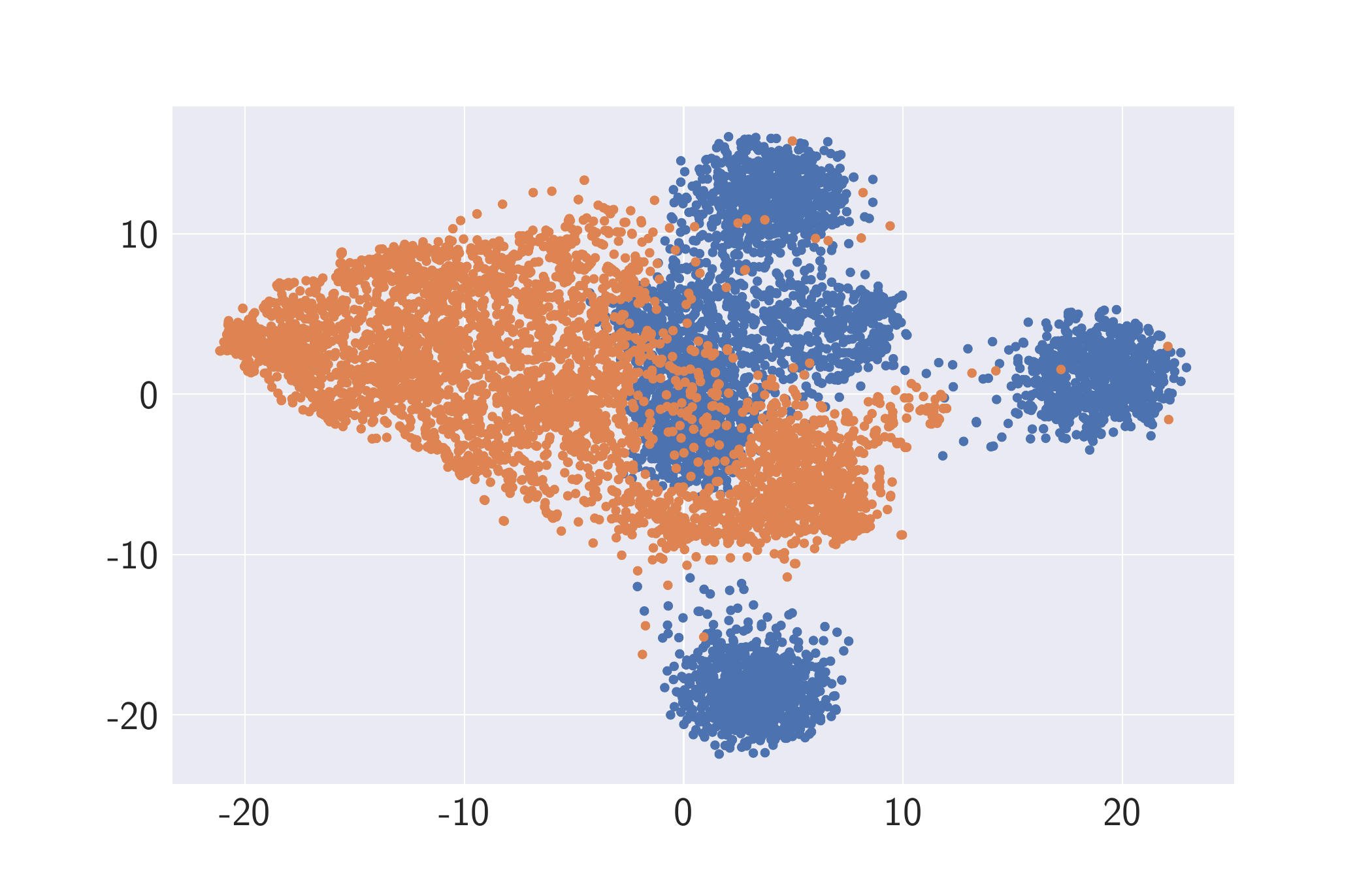}
\caption{ DeepInversion }
\label{fig:tsne_deepinversion_cifar}
\end{subfigure}
\begin{subfigure}{0.4\columnwidth}
\includegraphics[width=\columnwidth]{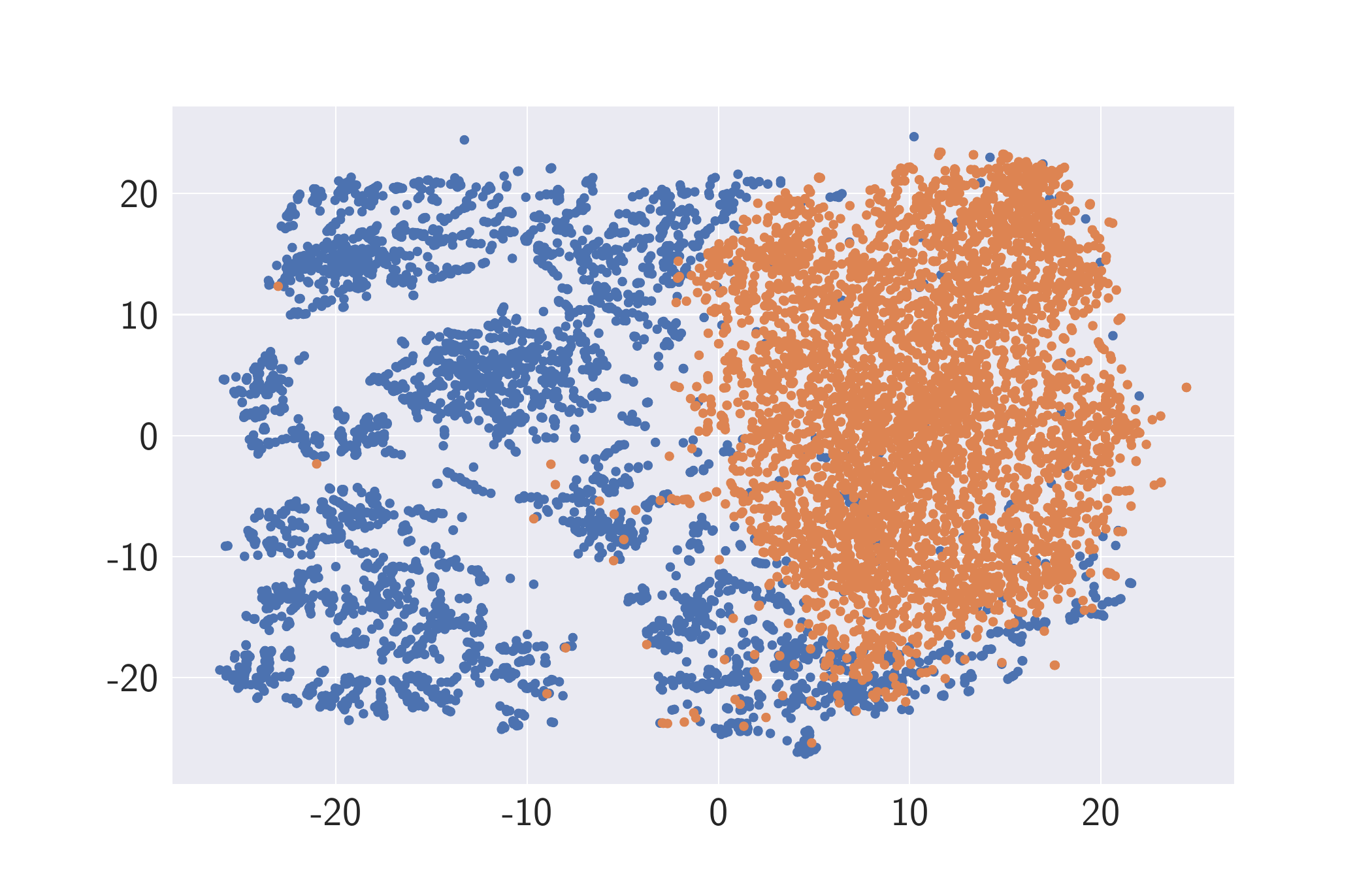}
\caption{ Revealer }
\label{fig:tsne_revealer_cifar}
\end{subfigure}
\begin{subfigure}{0.4\columnwidth}
\includegraphics[width=\columnwidth]{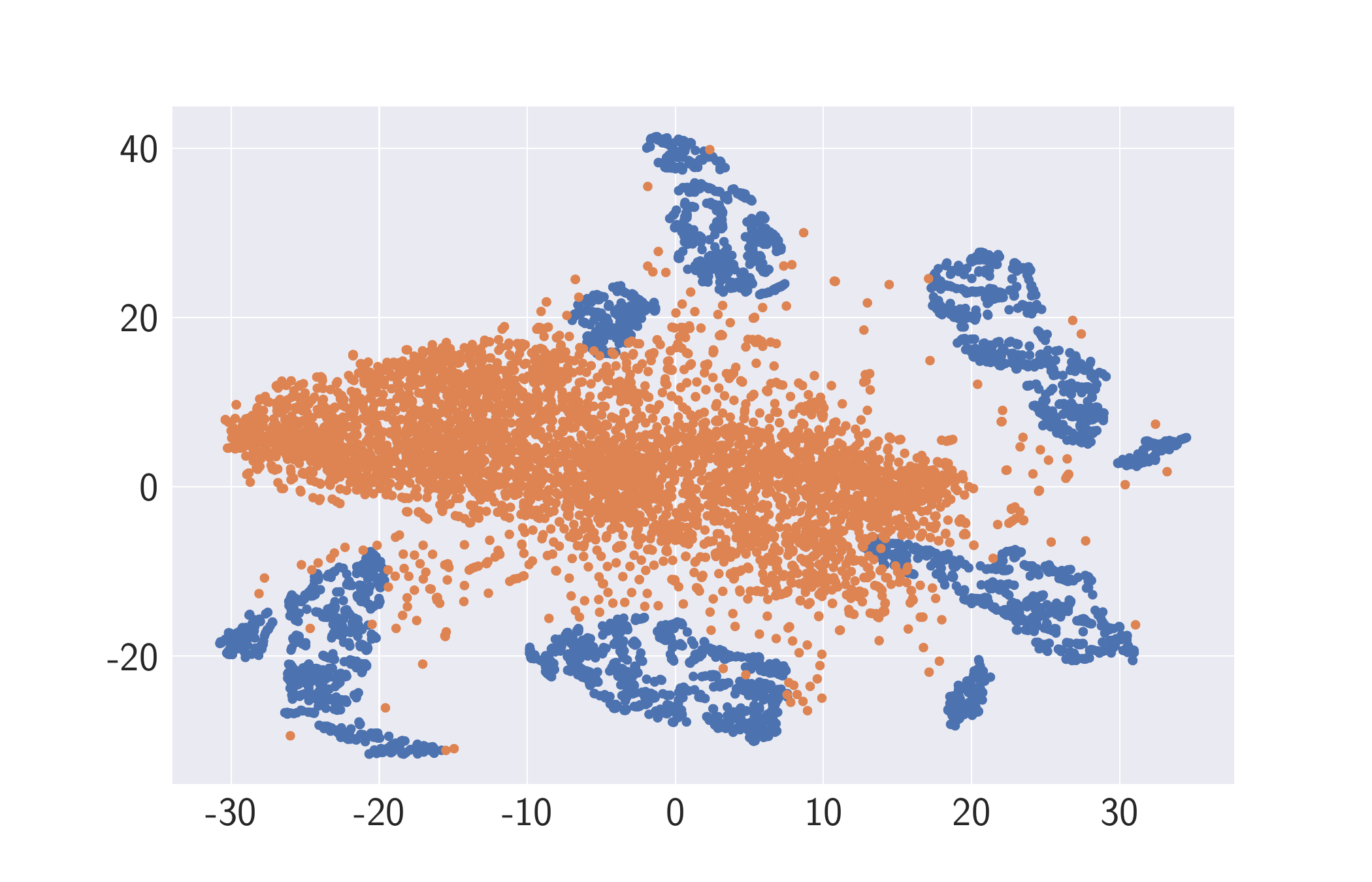}
\caption{ Inv-Alignment }
\label{fig:tsne_inv_cifar}
\end{subfigure}
\begin{subfigure}{0.4\columnwidth}
\includegraphics[width=\columnwidth]{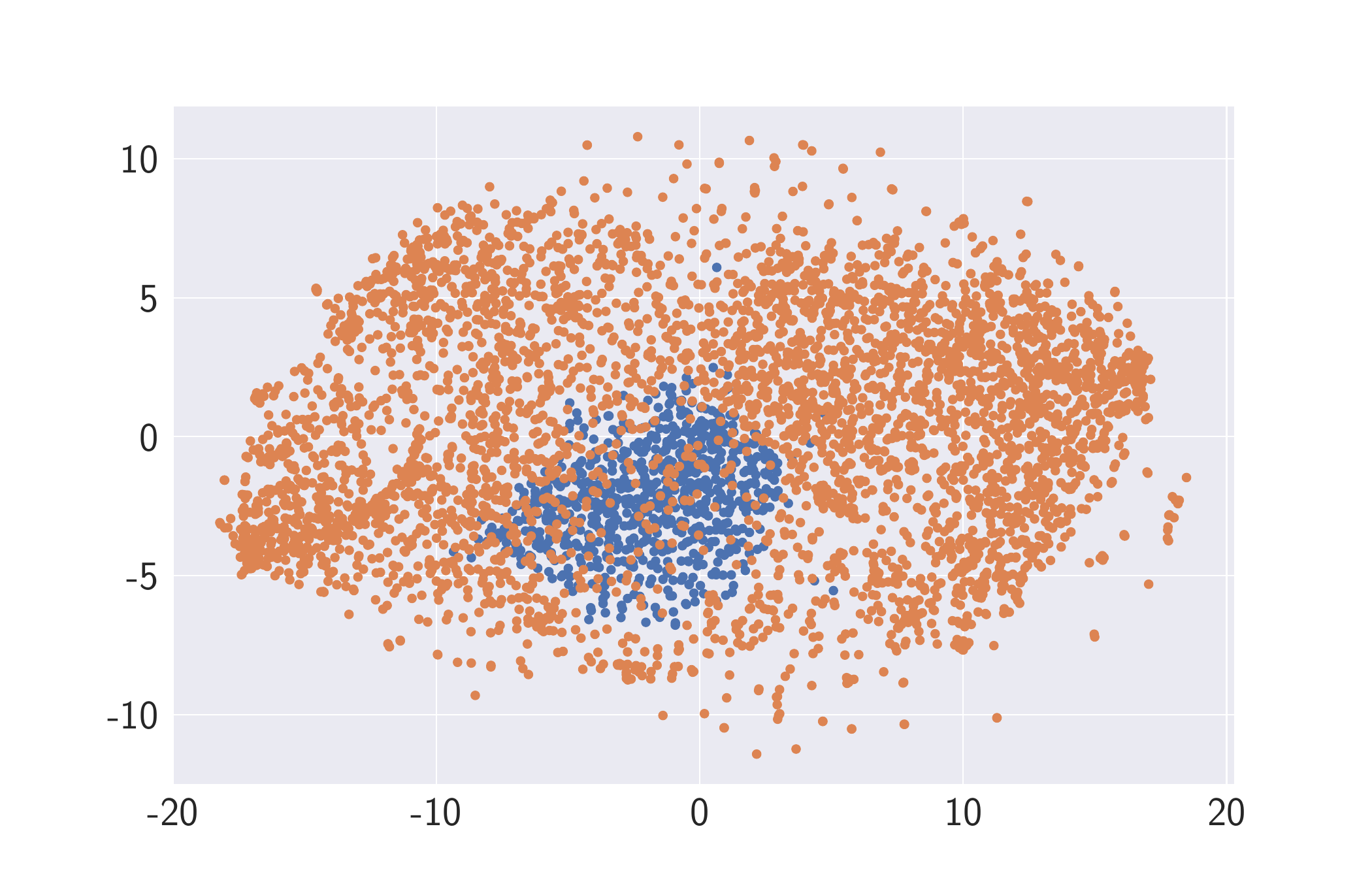}
\caption{ Bias-Rec }
\label{fig:tsne_impbias_cifar}
\end{subfigure}
\begin{subfigure}{0.4\columnwidth}
\includegraphics[width=\columnwidth]{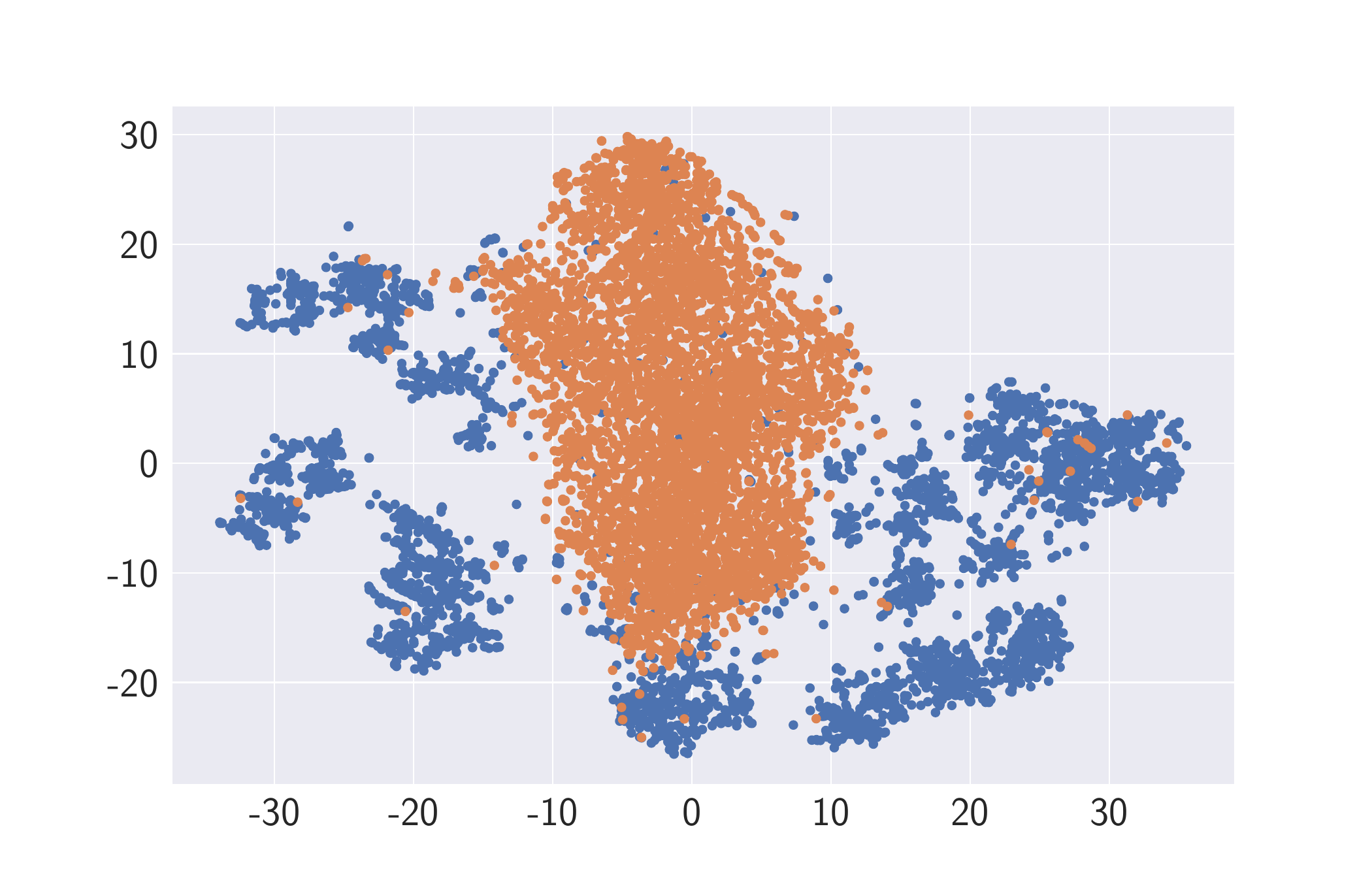}
\caption{ KEDMI }
\label{fig:tsne_kemdmi_cifar}
\end{subfigure}
\begin{subfigure}{0.4\columnwidth}
\includegraphics[width=\columnwidth]{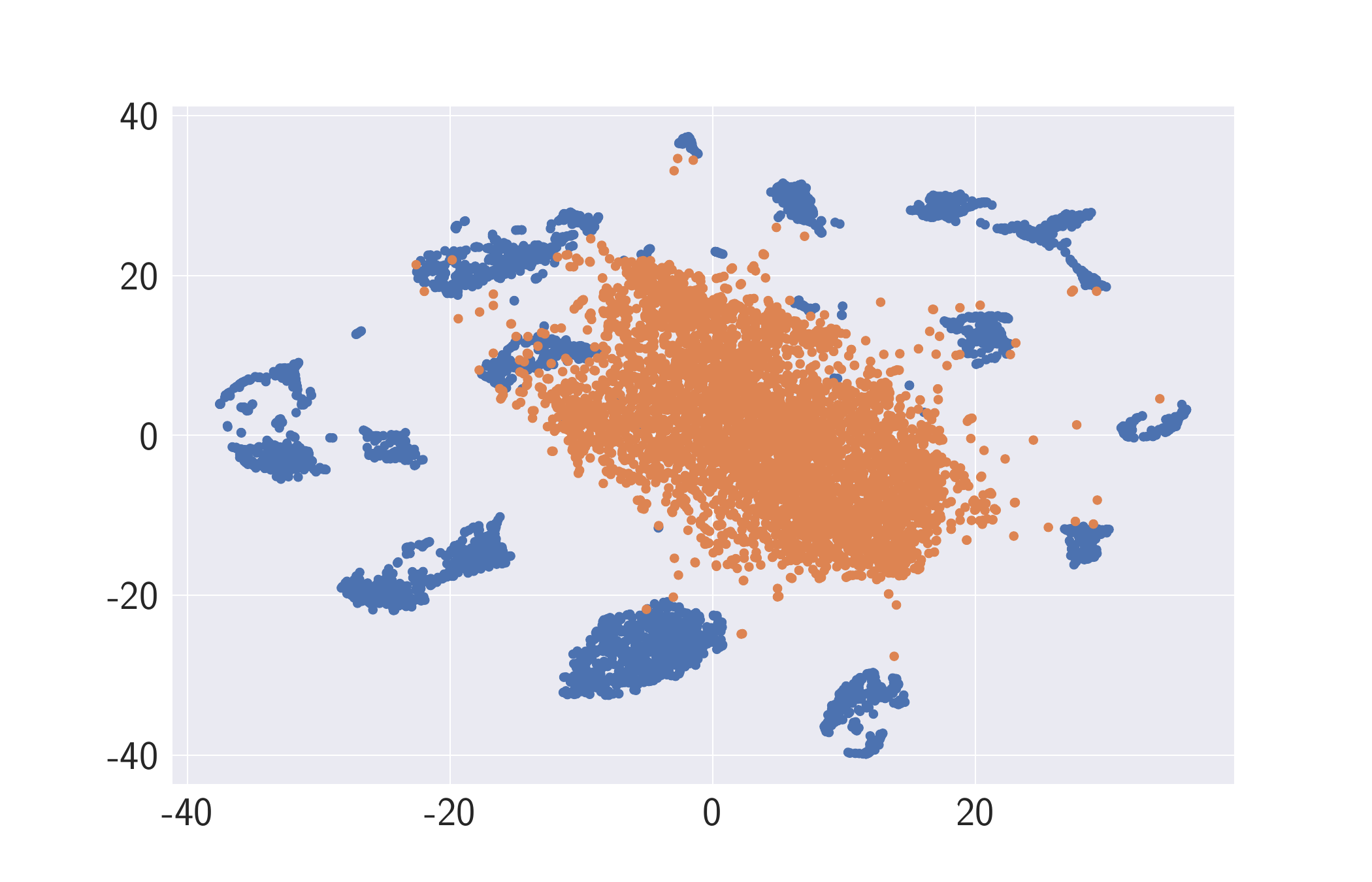}
\caption{ PLGMI }
\label{fig:tsne_plgmi_cifar}
\end{subfigure}
\begin{subfigure}{0.4\columnwidth}
\includegraphics[width=\columnwidth]{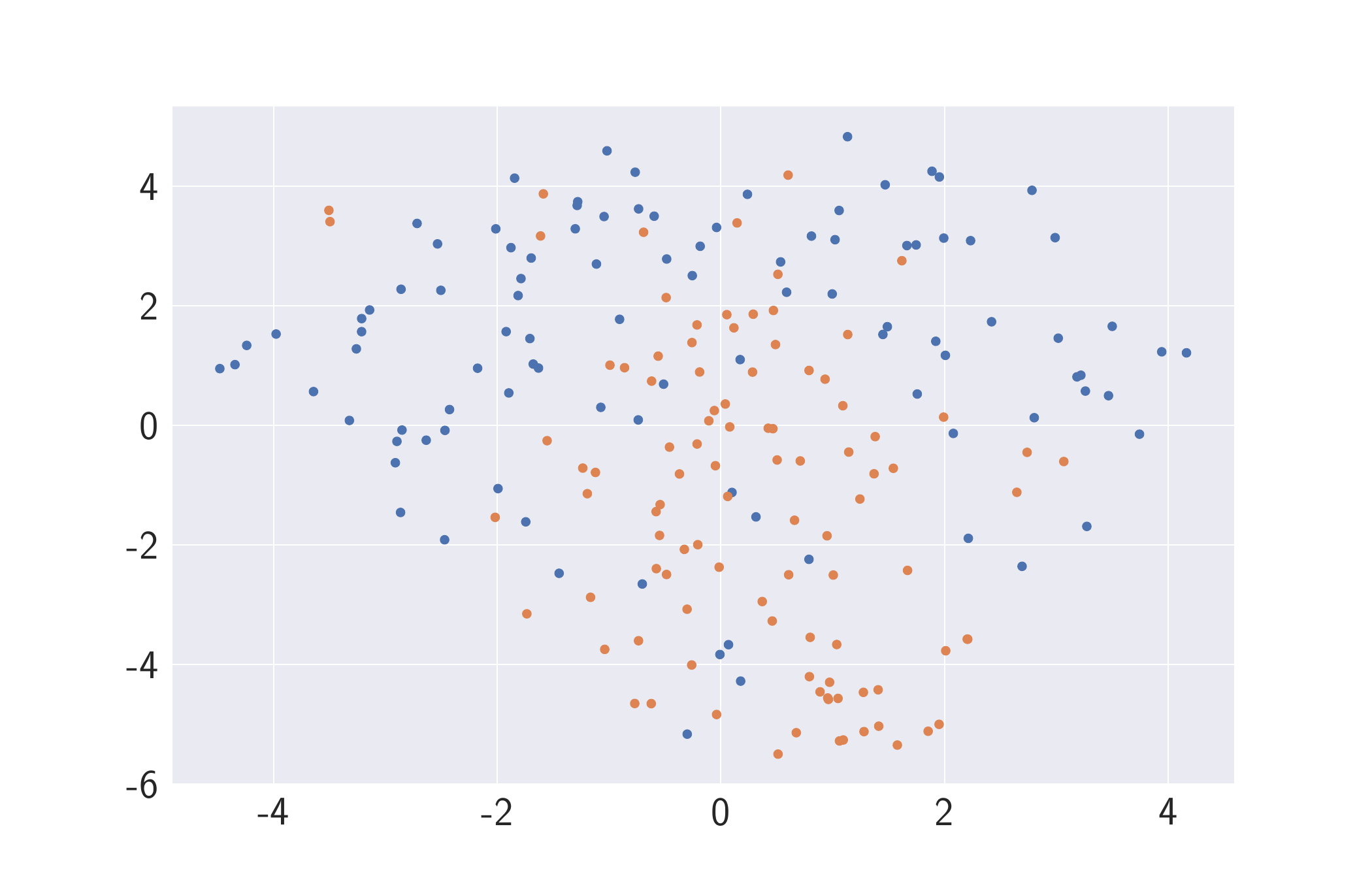}
\caption{ Updates-Leak }
\label{fig:tsne_udleak_cifar}
\end{subfigure}
\begin{subfigure}{0.4\columnwidth}
\includegraphics[width=\columnwidth]{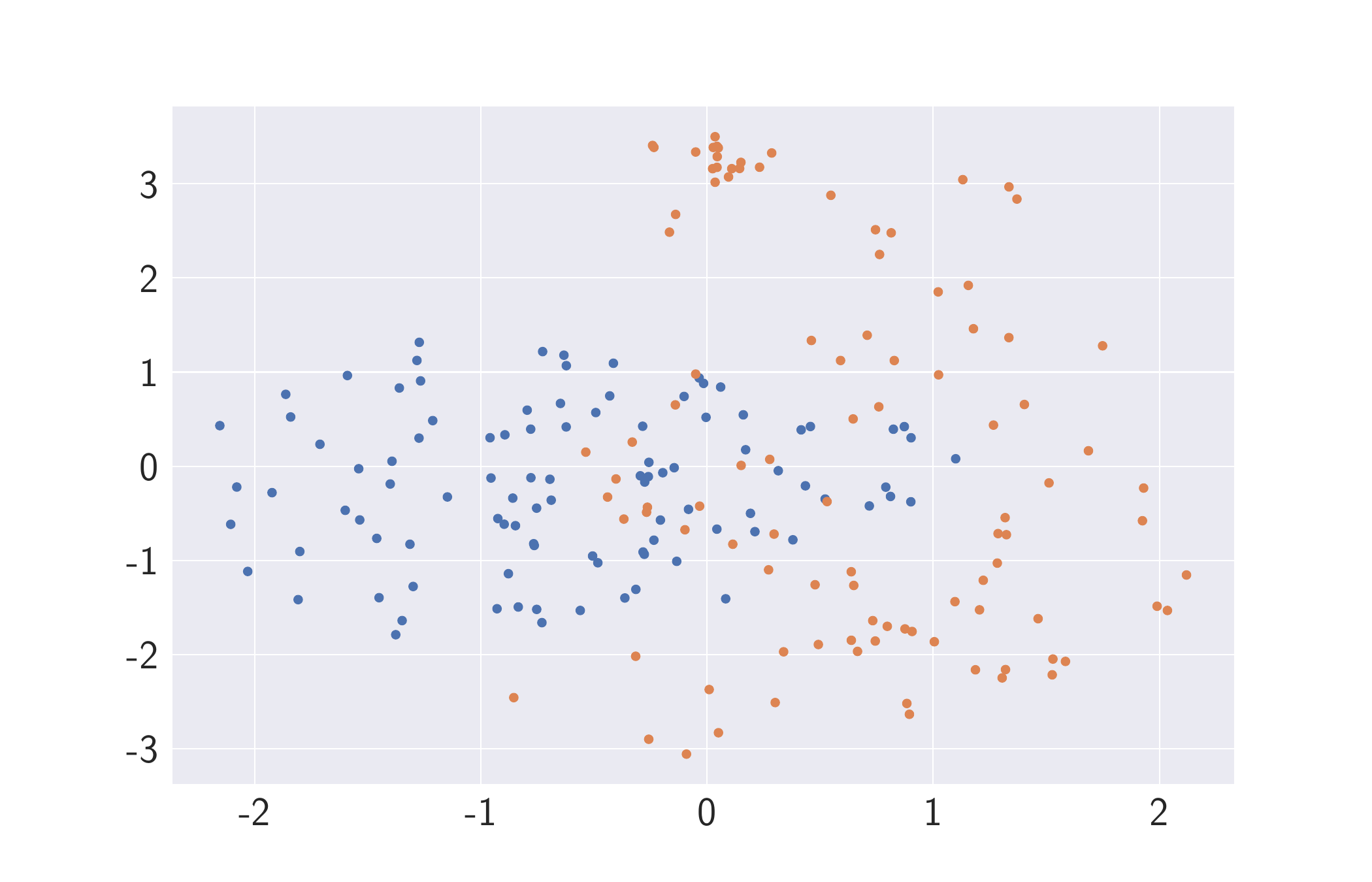}
\caption{ Deep-Leakage }
\label{fig:tsne_gradleak_cifar}
\end{subfigure}
\caption{t-SNE visualization of reconstructions and corresponding target datasets for CIFAR10.}
\label{figure:tsne_reconstruct_cifar10}
\end{figure*}

\begin{figure*}[!t]
\centering
\begin{subfigure}{0.4\columnwidth}
\includegraphics[width=\columnwidth]{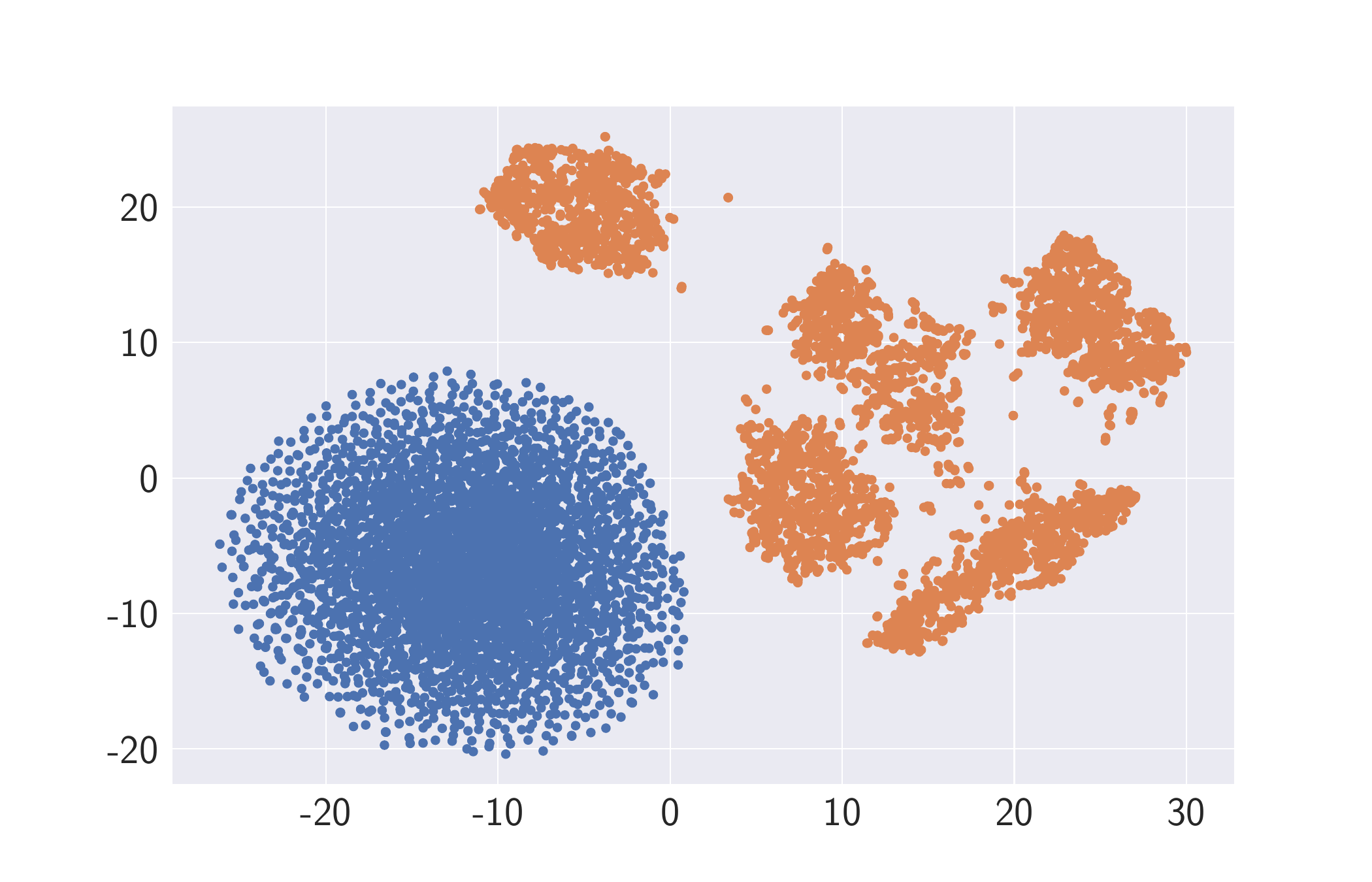}
\caption{ MI-Face}
\label{fig:tsne_mi_mnist}
\end{subfigure}
\begin{subfigure}{0.4\columnwidth}
\includegraphics[width=\columnwidth]{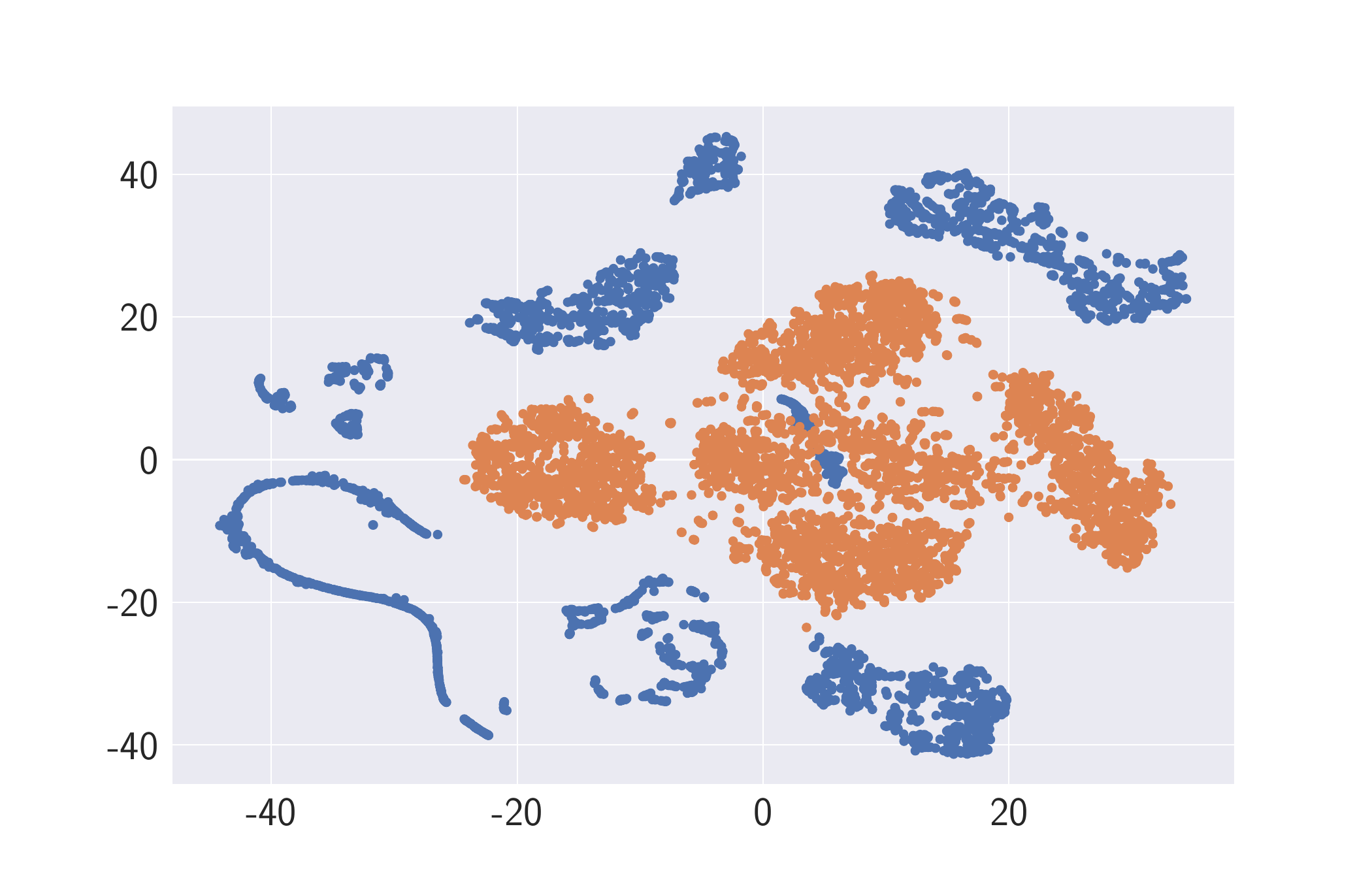}
\caption{ DeepDream }
\label{fig:tsne_deepdream_mnist}
\end{subfigure}
\begin{subfigure}{0.4\columnwidth}
\includegraphics[width=\columnwidth]{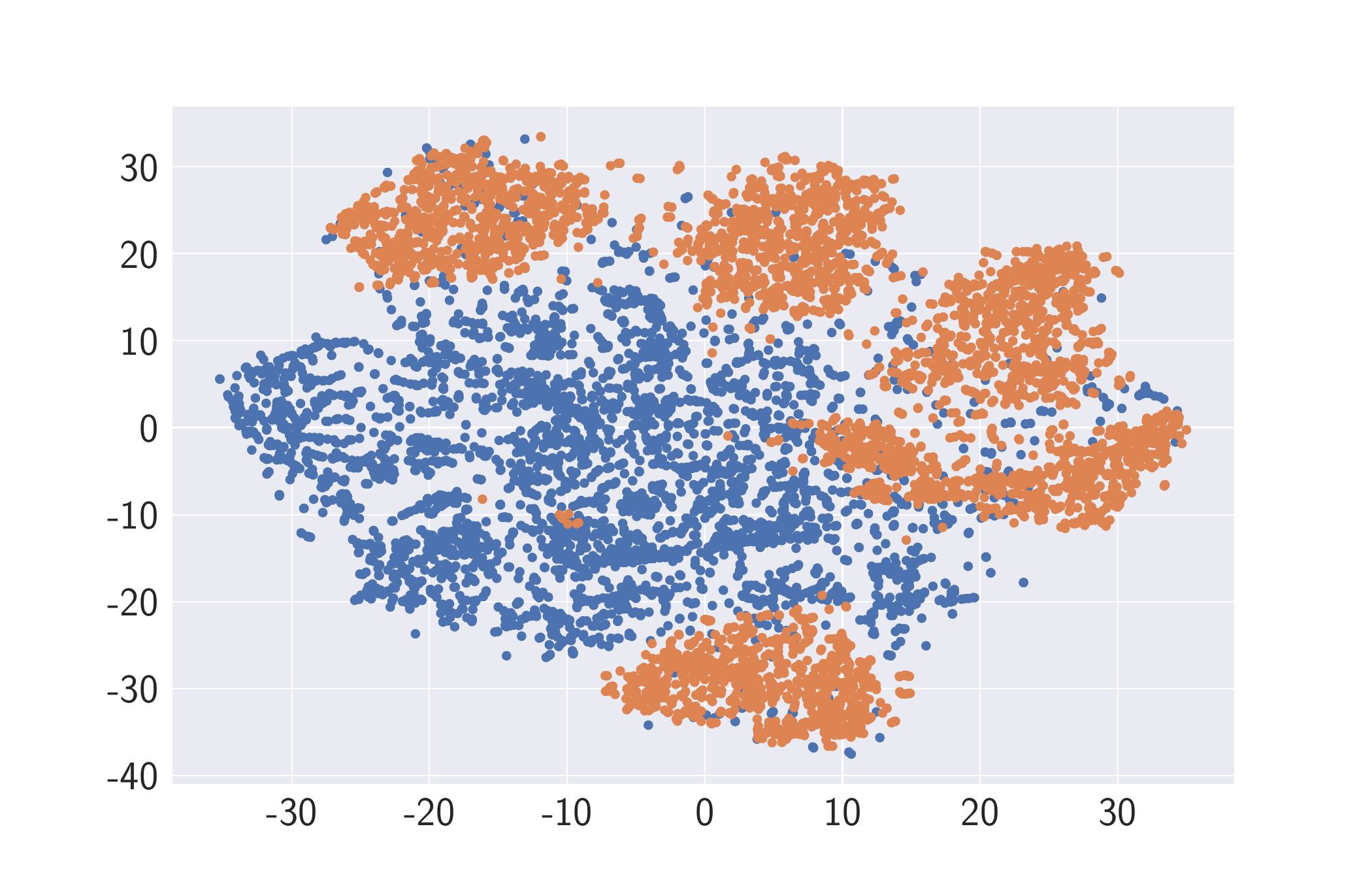}
\caption{ DeepInversion }
\label{fig:tsne_deepinv_mnist}
\end{subfigure}
\begin{subfigure}{0.4\columnwidth}
\includegraphics[width=\columnwidth]{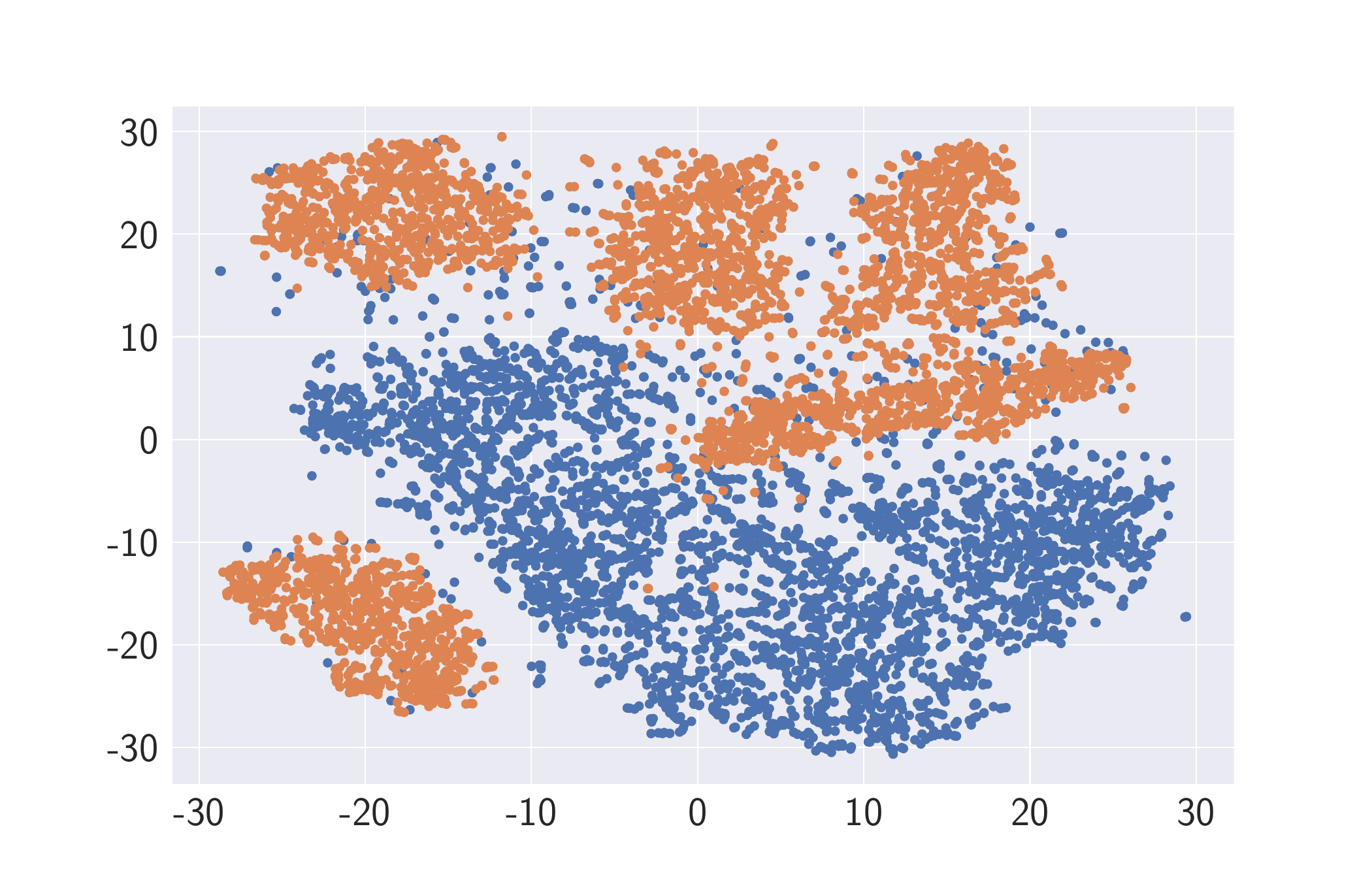}
\caption{ Revealer }
\label{fig:tsne_revealer_mnist}
\end{subfigure}
\begin{subfigure}{0.4\columnwidth}
\includegraphics[width=\columnwidth]{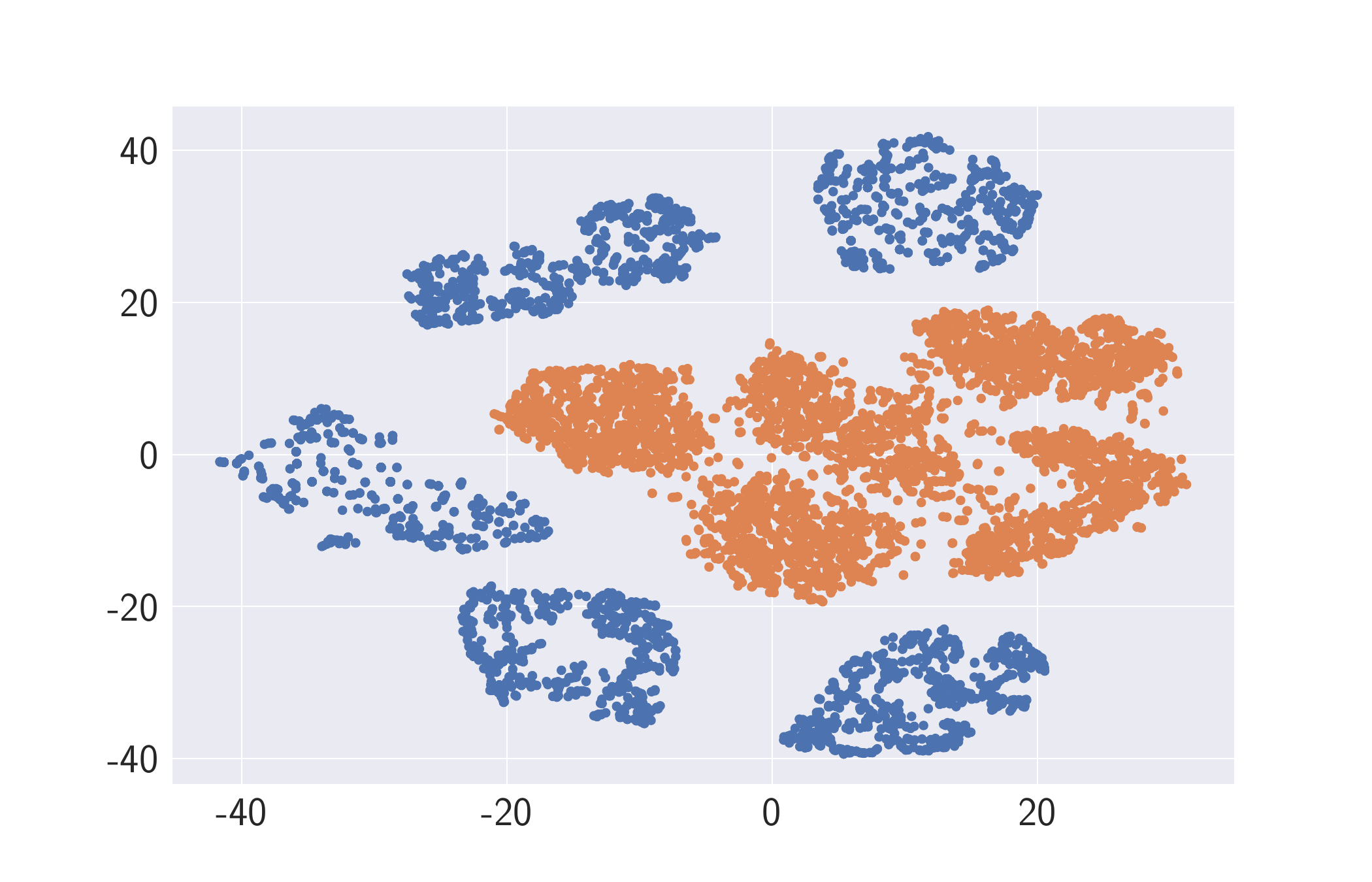}
\caption{ Inv-Alignment }
\label{fig:tsne_inv_mnist}
\end{subfigure}
\begin{subfigure}{0.4\columnwidth}
\includegraphics[width=\columnwidth]{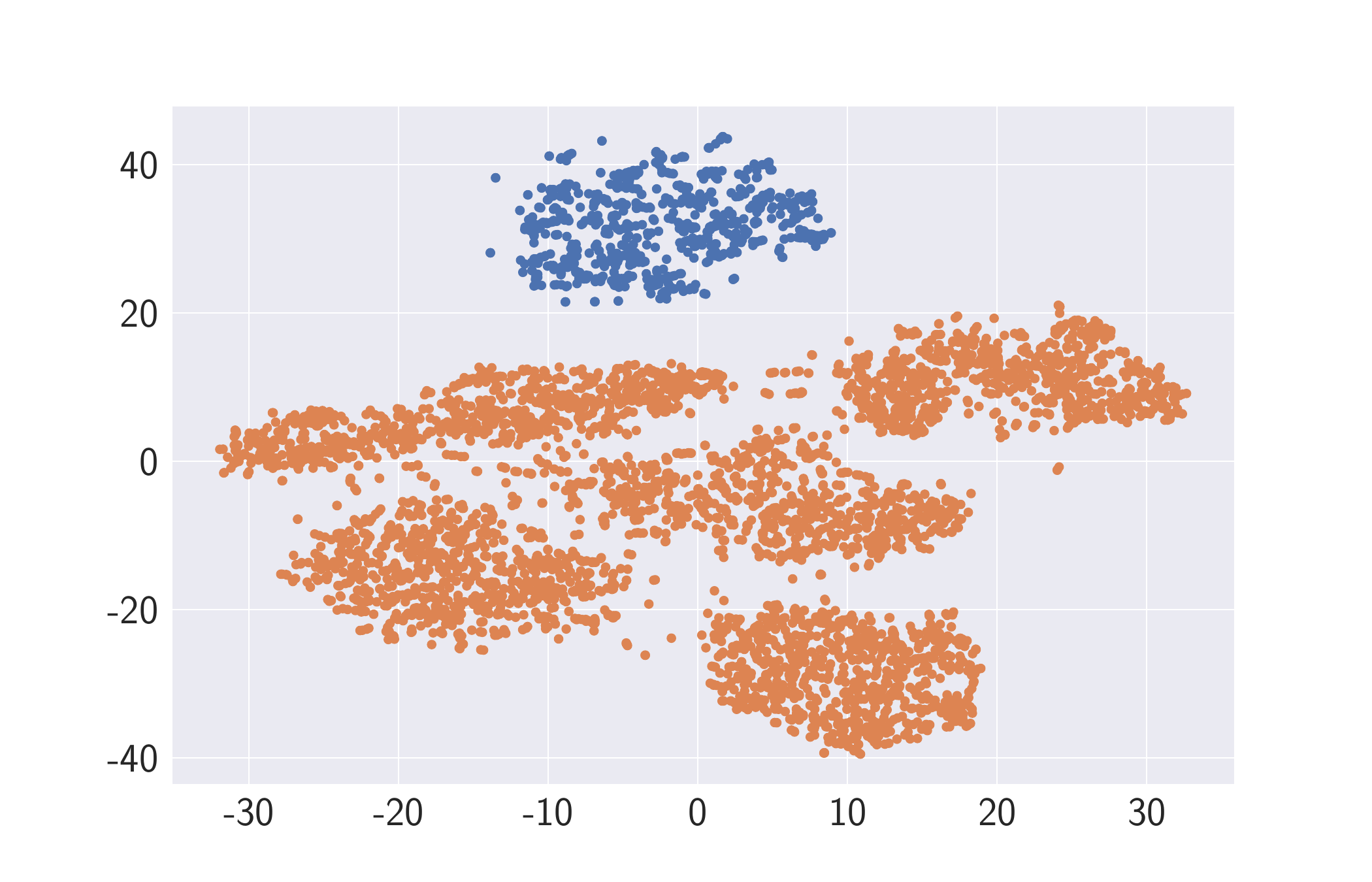}
\caption{ Bias-Rec }
\label{fig:tsne_impbias_mnist}
\end{subfigure}
\begin{subfigure}{0.4\columnwidth}
\includegraphics[width=\columnwidth]{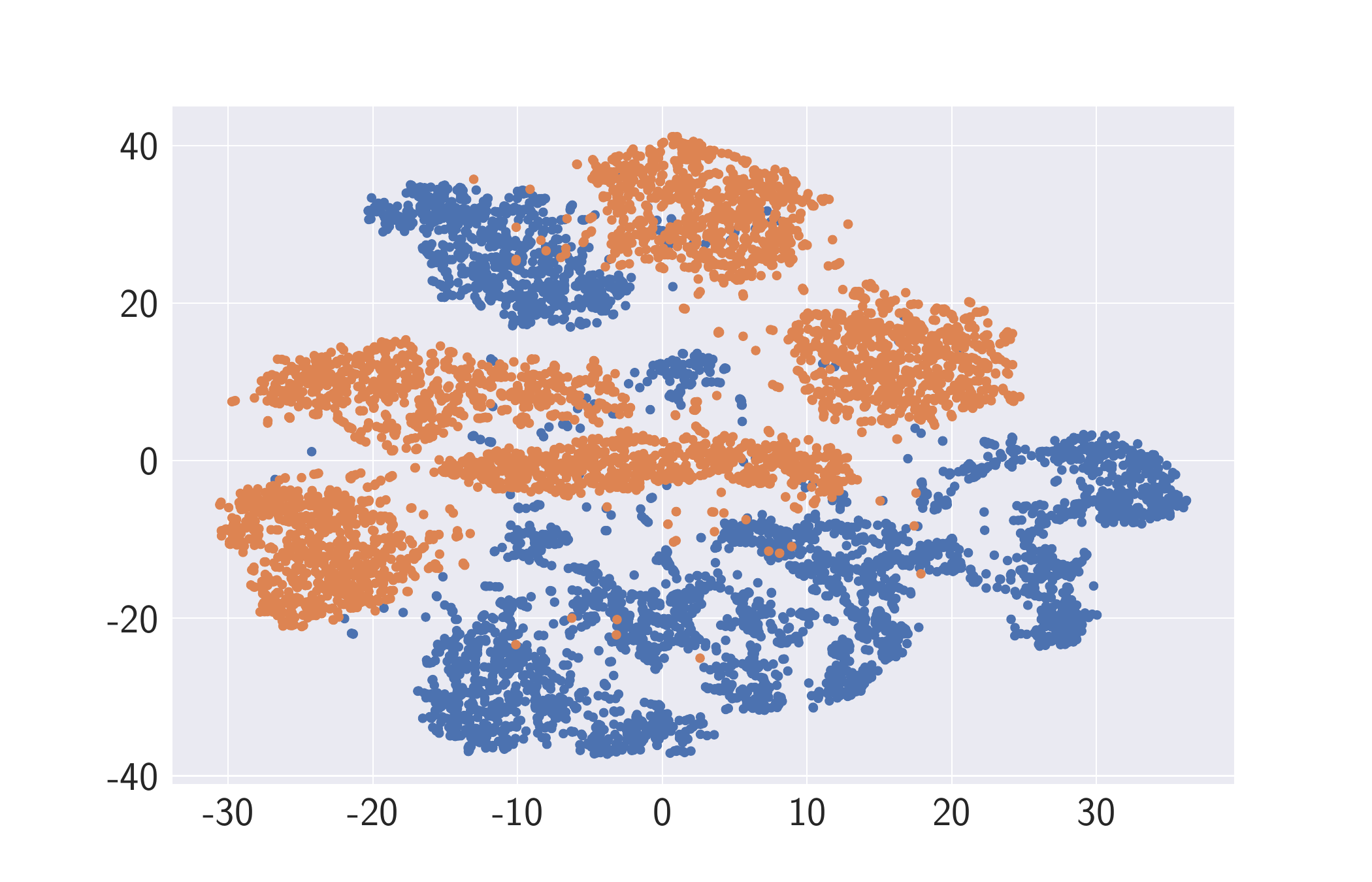}
\caption{ KEDMI }
\label{fig:tsne_kedmi_mnist}
\end{subfigure}
\begin{subfigure}{0.4\columnwidth}
\includegraphics[width=\columnwidth]{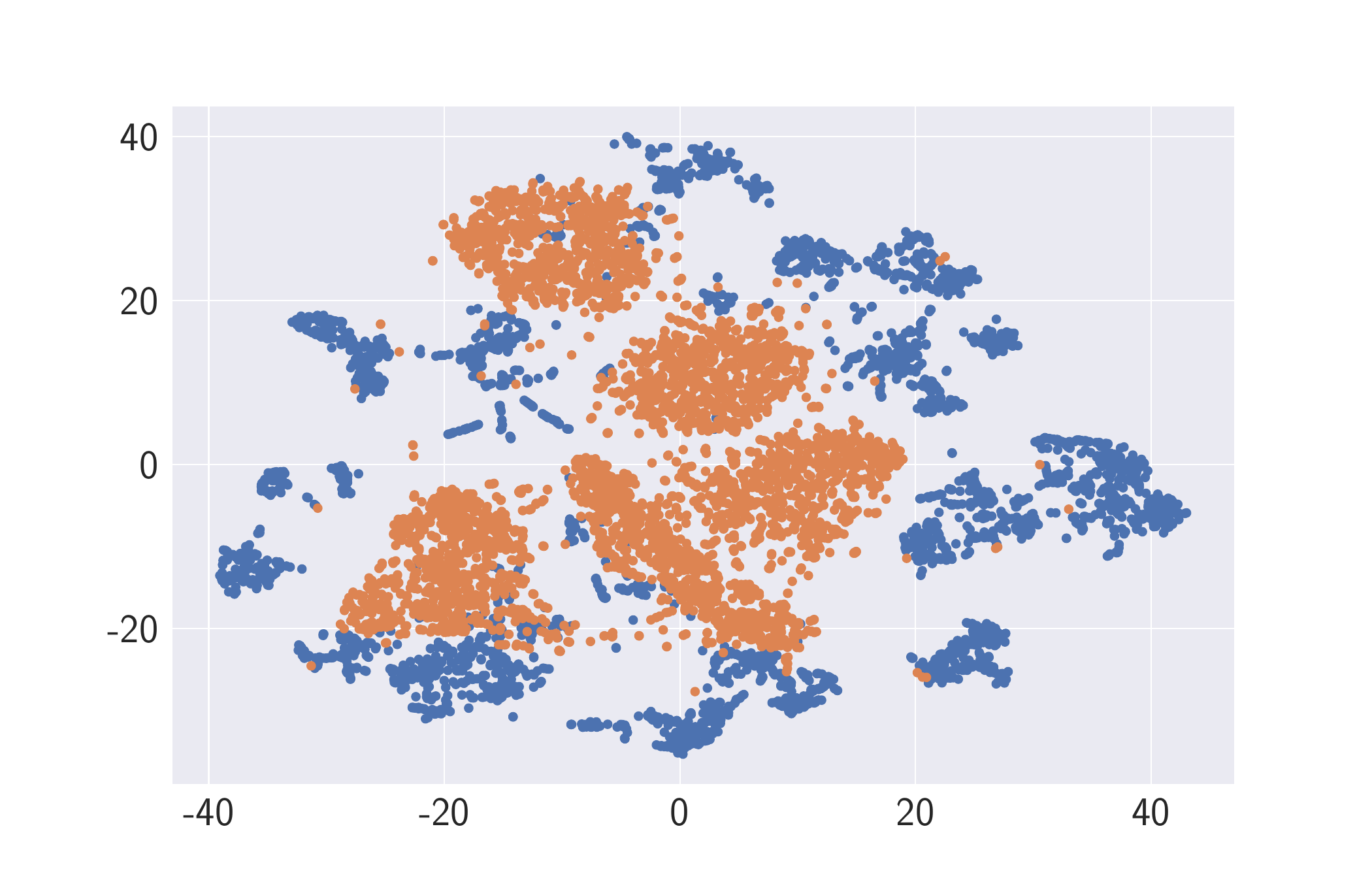}
\caption{ PLGMI }
\label{fig:tsne_pkgmi_mnist}
\end{subfigure}
\begin{subfigure}{0.4\columnwidth}
\includegraphics[width=\columnwidth]{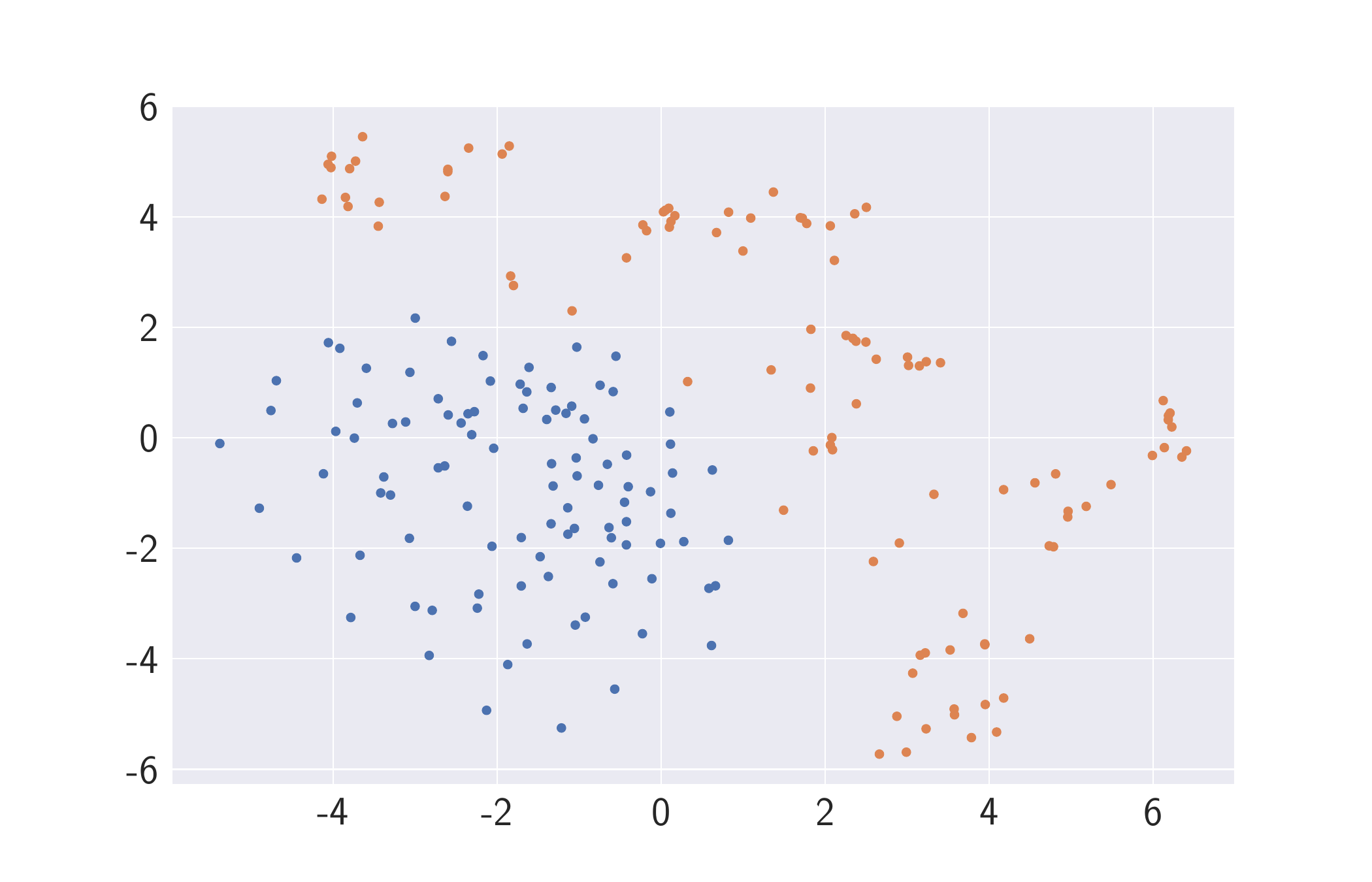}
\caption{ Updates-Leak }
\label{fig:tsne_udleak_mnist}
\end{subfigure}
\begin{subfigure}{0.4\columnwidth}
\includegraphics[width=\columnwidth]{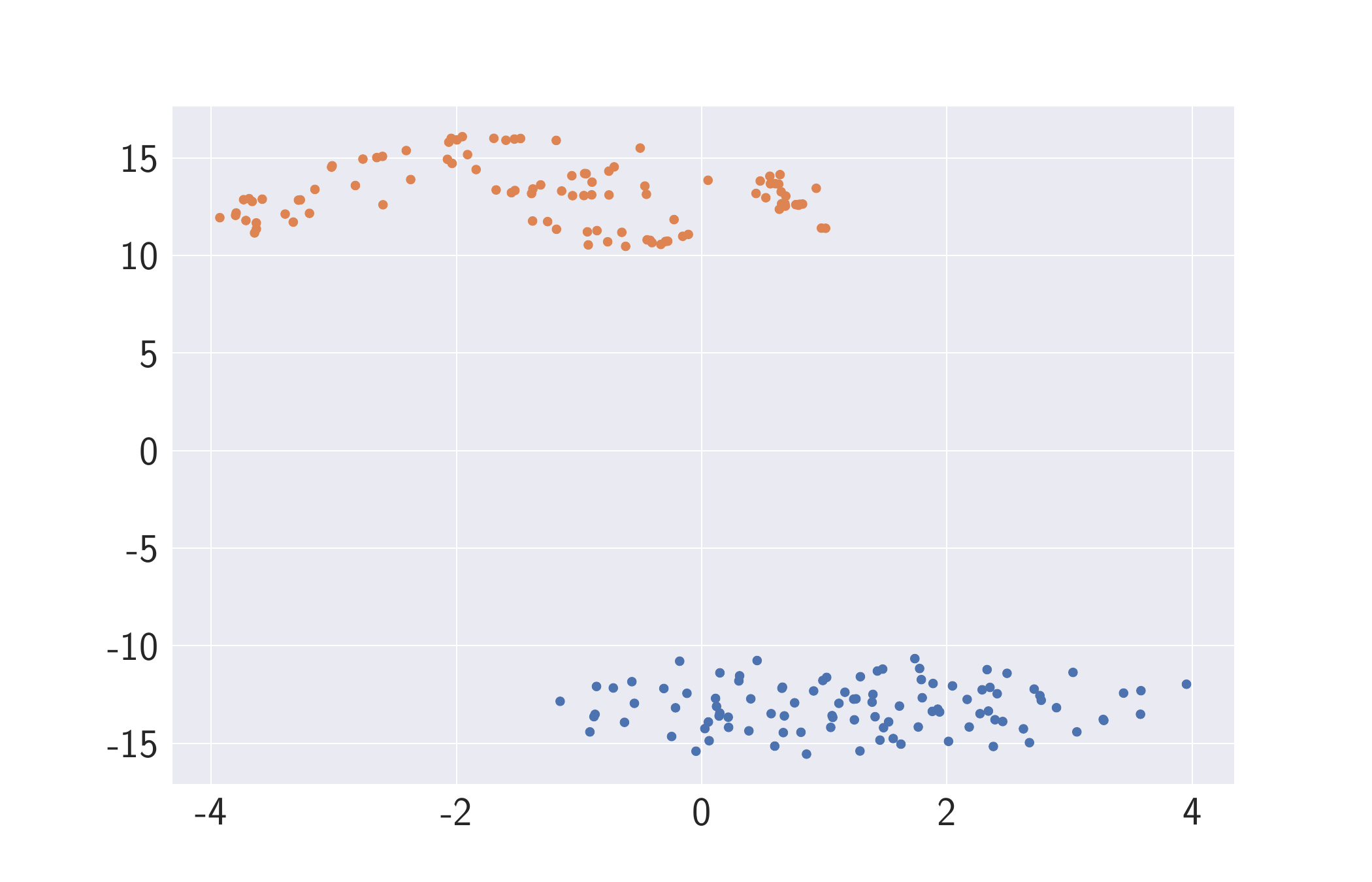}
\caption{ Deep-Leakage }
\label{fig:tsne_gradleak_mnist}
\end{subfigure}
\caption{t-SNE visualization of reconstructions and corresponding target datasets for MNIST.}
\label{figure:tsne_reconstruct_mnist}
\end{figure*}

\section{Counterexamples}
\label{app:counterexample}

We generate counterexamples in~\autoref{subfigure:metric_trainacc} which contribute a lot to the model training, but are vastly dissimilar from the target dataset.

We generate counterexamples that resemble random noise but are still classified with high confidence by the evaluation model in~\autoref{subfigure:metric_testacc}.
These counterexamples, generated using MI-Face~\cite{FJR15}, achieve prediction confidences above 0.9 for their corresponding classes.

\section{t-SNE visualization}
\label{app:tsne}

To provide an overview of the relationship between the reconstructed dataset and the target dataset, we leverage t-distributed stochastic neighbor embedding (t-SNE) to reduce these datasets to two dimensions and exhibit their distributions in~\autoref{figure:tsne_reconstruct_celeba}, \autoref{figure:tsne_reconstruct_cifar10}, and~\autoref{figure:tsne_reconstruct_mnist}.

\section{Setup For Main Experiments}
\label{section:setup}

\mypara{Datasets}
We conduct our evaluation on three benchmark datasets, including CelebA~\cite{LLWT15}, CIFAR10~\cite{CIFAR}, and MNIST~\cite{MNIST}.
CelebA is a large-scale face dataset with $10,177$ identities and $202,599$ images.
CIFAR10 consists of $60,000$ colored images belonging to $10$ classes equally; these classes include common objects in daily life like airplanes, birds, and dogs.
MNIST is a gray-scale dataset containing $10$ classes of digital handwritten numbers. 

\mypara{Attack Details}
We conduct an evaluation of ten representative reconstruction attacks that were introduced in~\autoref{section:existrec}. 
In order to conform to our definition, some of the attacks required modifications. 
Specifically, the MI-Face attack was originally designed to recover one image for each class due to the fixed initialization, but we enable this attack to generate multiple samples for a single class by using random initialization.
Similarly, Inv-Alignment assumes that the adversary had information about the posterior corresponding to the target samples, but we generalize this assumption by randomly creating pseudo-posteriors and using them to reconstruct the target dataset.

We follow the standard setting to split the dataset into two disjoint parts to prevent data leakage. 
For CelebA, training and updating data come from $1,000$ identities, with balanced subsets of sizes $1,000$, $2,000$, $5,000$, $10,000$, $15,000$, and $20,000$ used for training the target model. Smaller subsets are included in larger ones. 
A disjoint subset of size $30,000$ is used as the auxiliary dataset, which can be further used to train the generator models. 
For CIFAR10 and MNIST, we divide each dataset into two non-overlapping parts: one containing samples from the first five classes, designated as the target dataset, and the other comprising samples from the remaining five classes.
The sizes of the target datasets vary from $100$ to $20,000$, while the auxiliary dataset size is consistently set at 20,000. 
For dynamic-type attacks, which involve various versions of the target model, the models are updated with $100$ samples.
Static-type attacks aim to reconstruct the data used for training the target model, while dynamic-type attacks target the $100$ updating samples.

We evaluate all attacks except for Bias-Rec on three different model architectures: VGG16~\cite{SZ15}, MobileNetV2~\cite{SHZZC18}, and ResNet-18~\cite{HZRS16}.
Bias-Rec requires the target model to follow certain architecture constraints, namely that it should not contain any skip-connections or bias terms. 
Therefore, we adhere to the original paper's methodology and construct a model composed of linear layers for this particular attack.

\mypara{Metric Details}
As noted in Definition~\ref{def:samplelevel}, the choice of sample-level metric $d$ in S-Dis can be determined by the auditor. 
In this paper, we use three common metrics to measure sample similarity: Structural Similarity Index Measure (SSIM), Peak Signal-to-Noise Ratio (PSNR), and Mean Squared Error (MSE). 
For SSIM and PSNR, higher values indicate greater similarity, while for MSE, the opposite is true.

For the dataset-level metric, we leverage FID to represent the distribution similarity, and a lower FID score indicates a smaller distance between two distributions.
Note that these metrics can be normalized into $[0,1]$ to incorporate with our definition, for ease of comparison, we present the original value in the following sections.

\section{Coverage Relationship}
\label{app:coverage}

\begin{figure*}[!t]
\centering
\begin{subfigure}{0.49\columnwidth}
\includegraphics[width=\columnwidth]{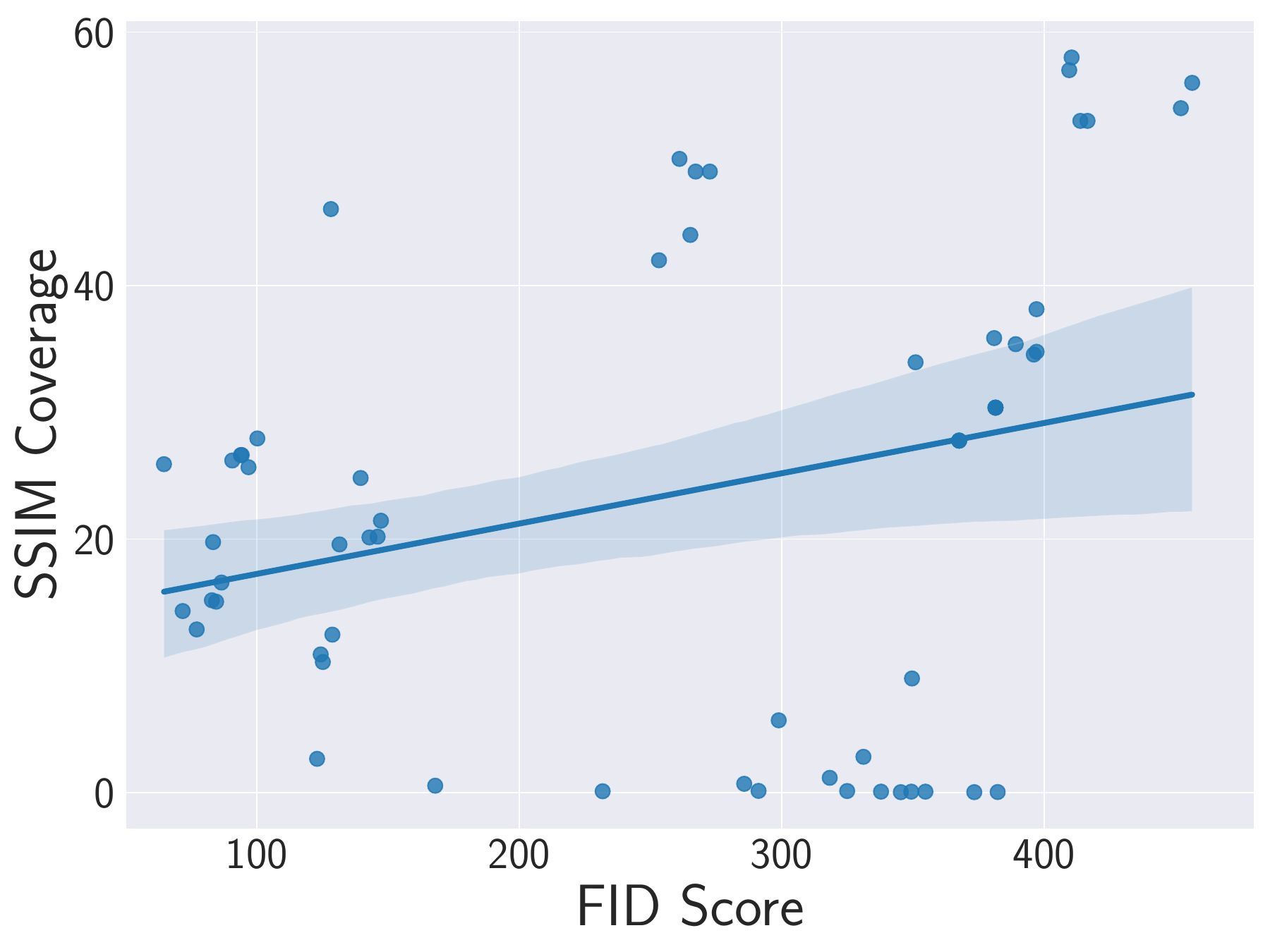}
\caption{FID vs. SSIM}
\label{fig:cov_fidvsssim}
\end{subfigure}
\begin{subfigure}{0.49\columnwidth}
\includegraphics[width=\columnwidth]{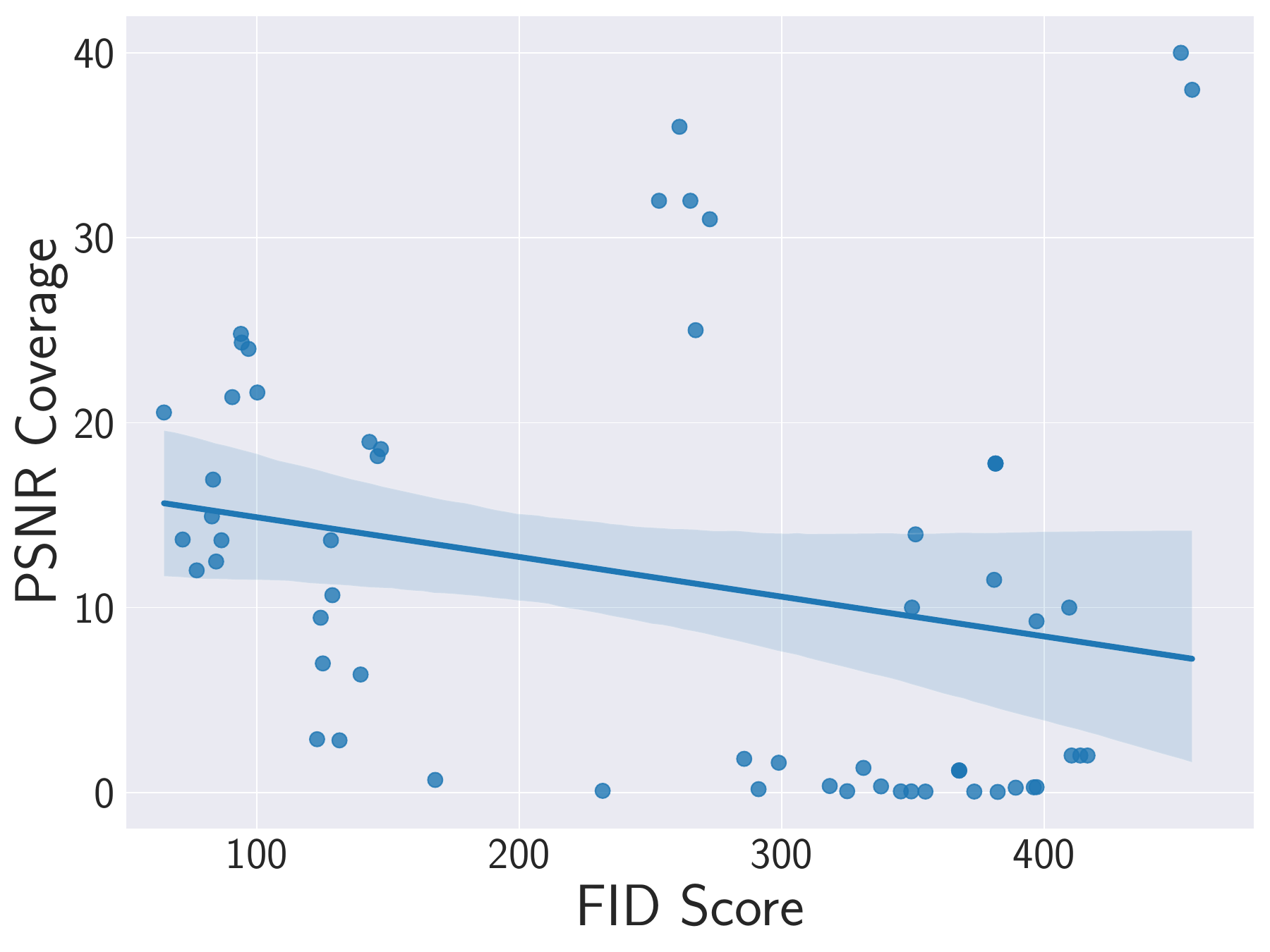}
\caption{FID vs. PSNR}
\label{fig:cov_fidvspsnr_cifar_mnist}
\end{subfigure}
\begin{subfigure}{0.49\columnwidth}
\includegraphics[width=\columnwidth]{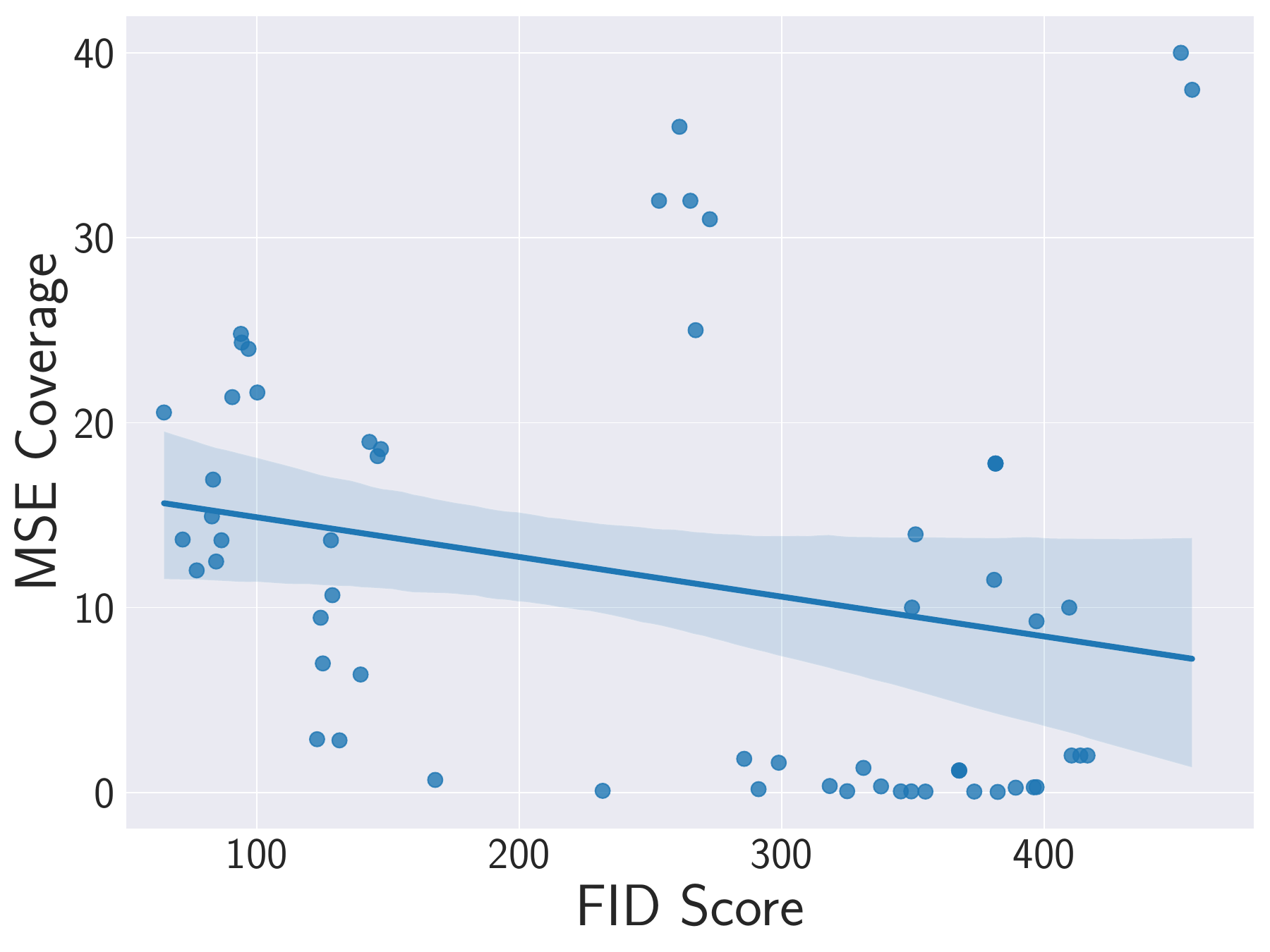}
\caption{FID vs. MSE}
\label{fig:cov_fidvsmse_cifar_mnist}
\end{subfigure}
\begin{subfigure}{0.49\columnwidth}
\includegraphics[width=\columnwidth]{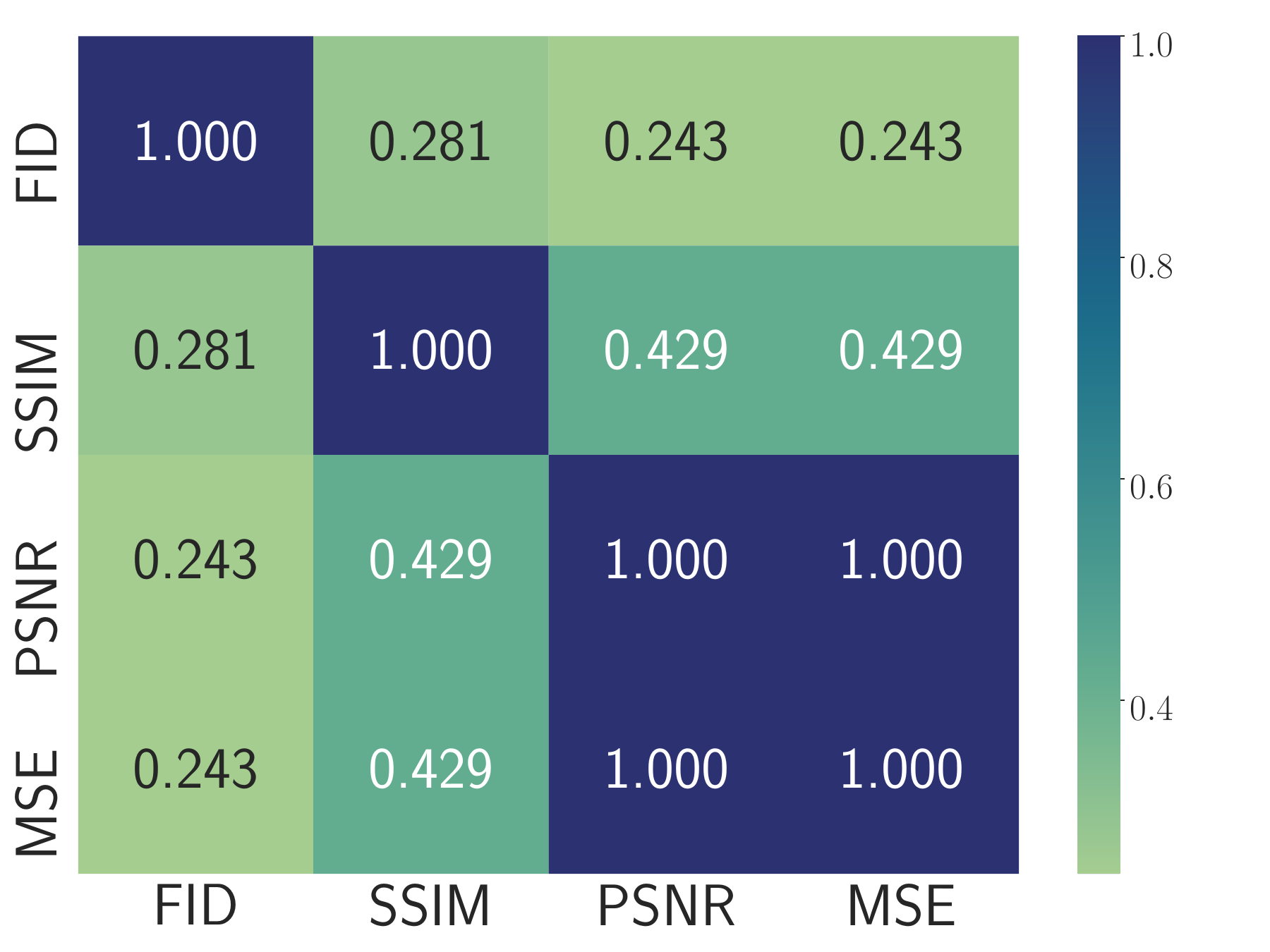}
\caption{Correlation}
\label{fig:coveragecorr_cifar_mnist}
\end{subfigure}
\caption{Relationship between coverages calculated with different sample-level metrics and the dataset-level metric for CIFAR-10 and MNIST, we plot the correlation heatmap using the absolute value of correlation between different metrics.}
\label{figure:coverage_relation_cifar_mnist}
\end{figure*}

\begin{figure*}[!t]
\centering
\begin{subfigure}{0.49\columnwidth}
\includegraphics[width=\columnwidth]{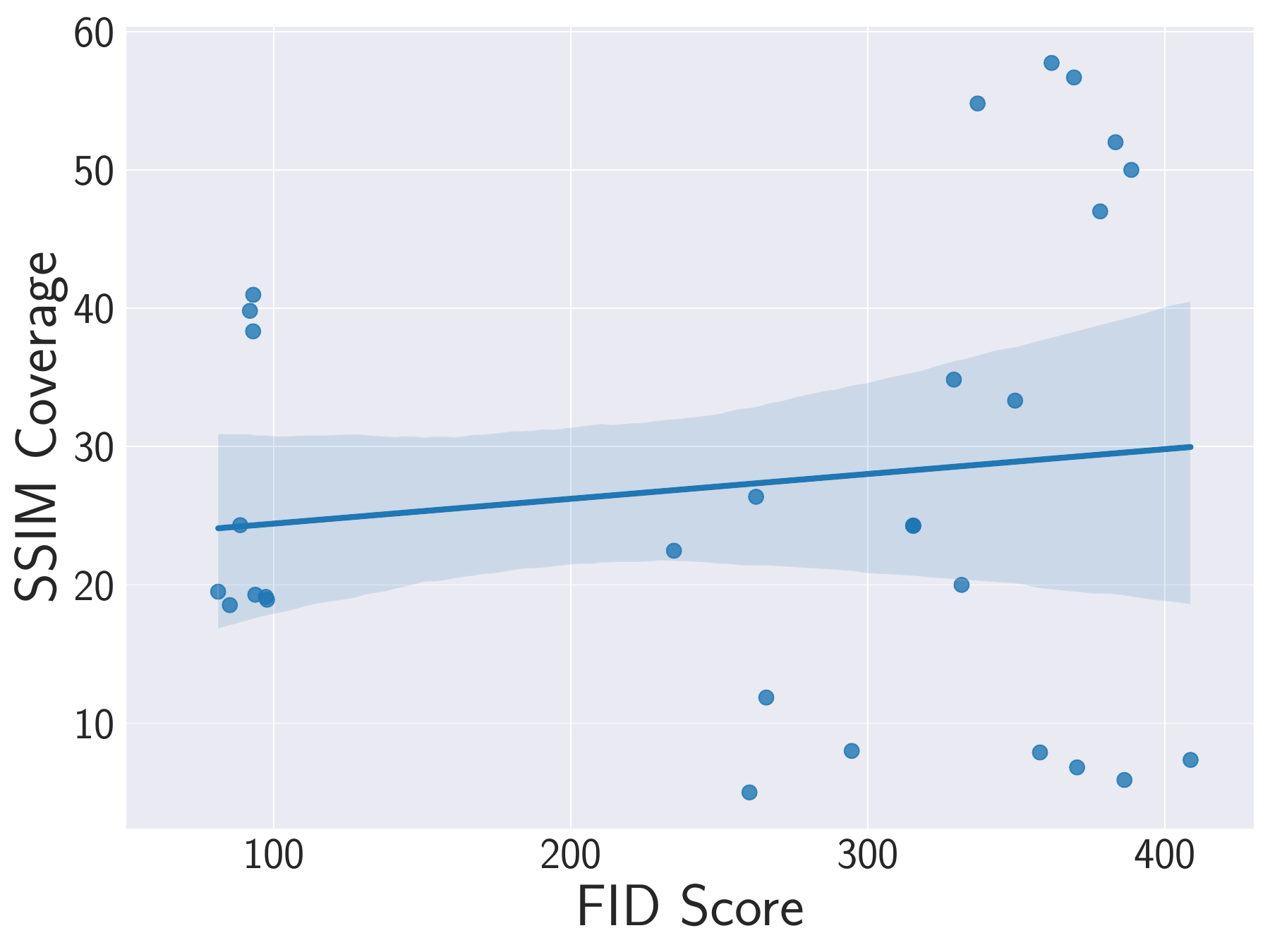}
\caption{FID vs. SSIM}
\label{fig:cov_fidvsssim_celeba}
\end{subfigure}
\begin{subfigure}{0.49\columnwidth}
\includegraphics[width=\columnwidth]{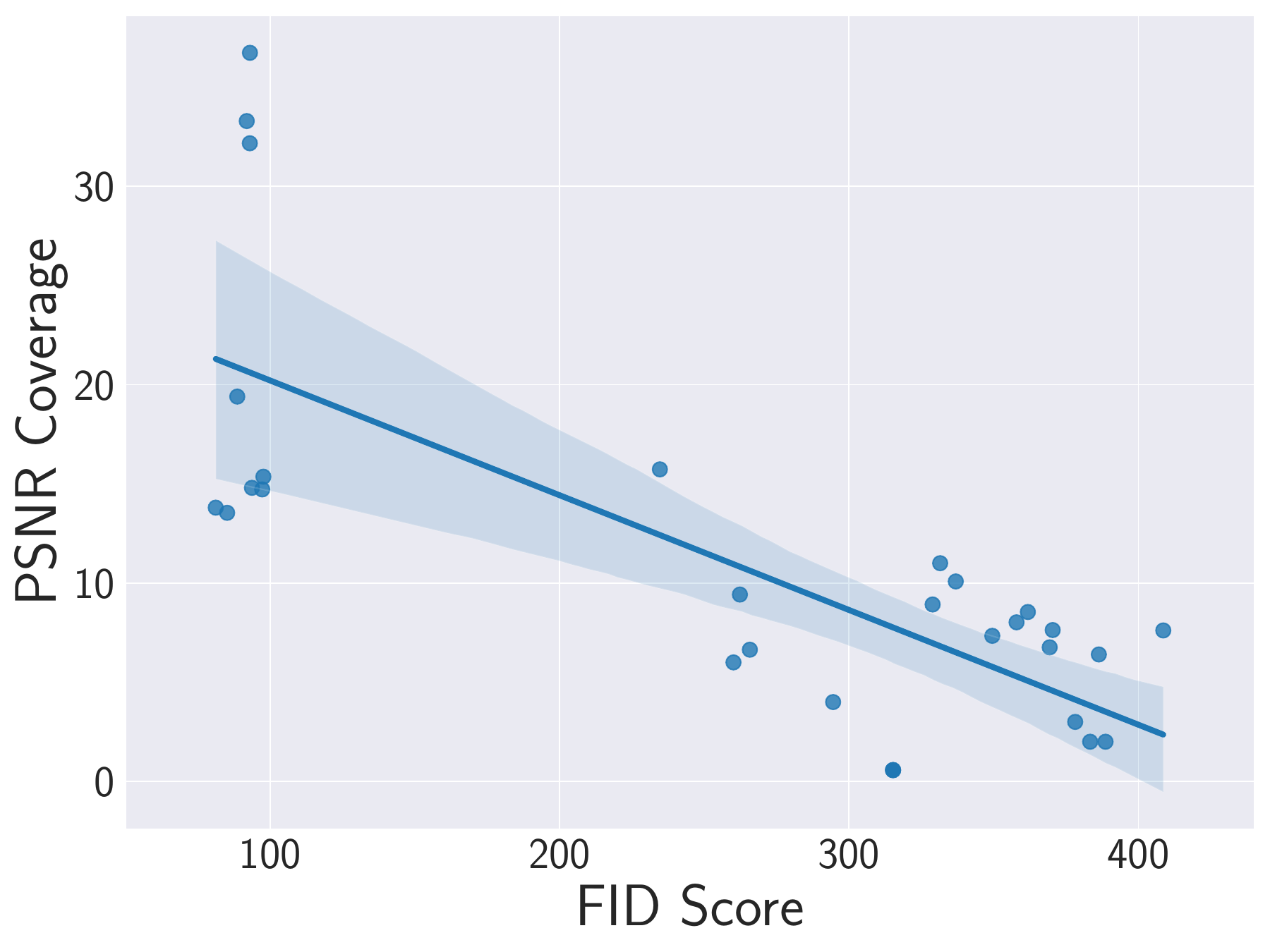}
\caption{FID vs. PSNR}
\label{fig:cov_fidvspsnr_celeba}
\end{subfigure}
\begin{subfigure}{0.49\columnwidth}
\includegraphics[width=\columnwidth]{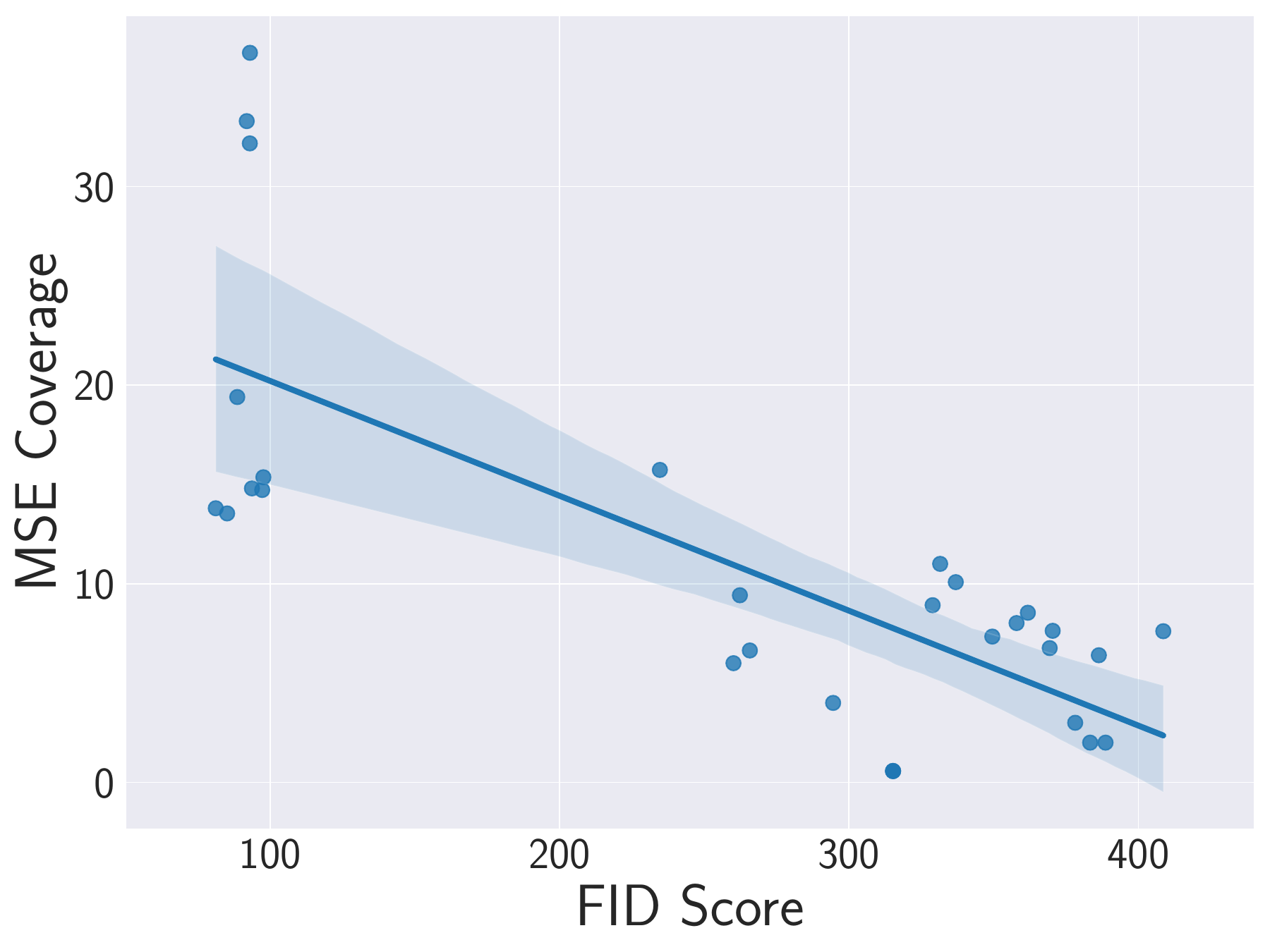}
\caption{FID vs. MSE}
\label{fig:cov_fidvsmse_celeba}
\end{subfigure}
\begin{subfigure}{0.49\columnwidth}
\includegraphics[width=\columnwidth]{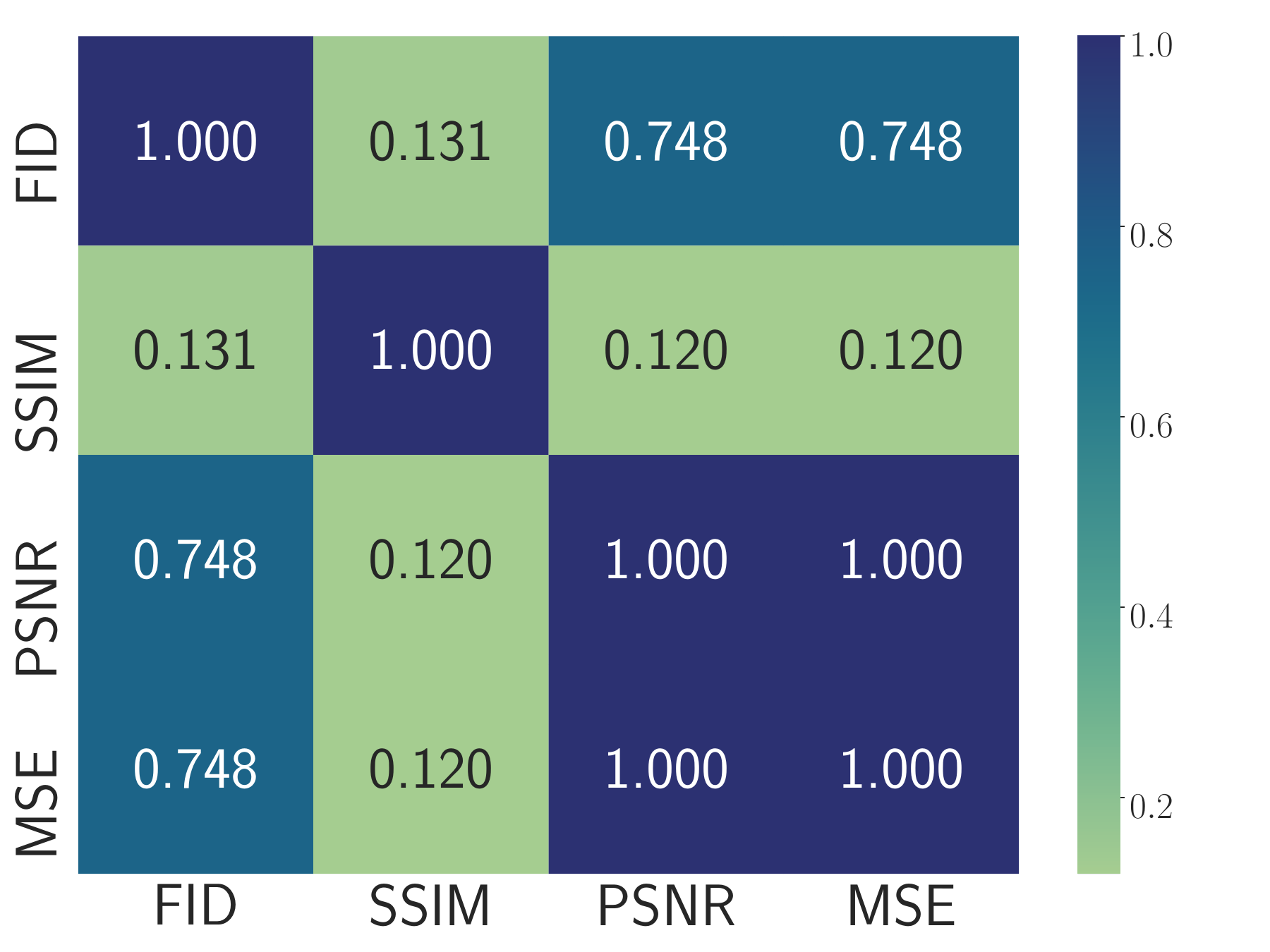}
\caption{Correlation}
\label{fig:coveragecorr_celeba}
\end{subfigure}
\caption{Relationship between coverages calculated with different sample-level metrics and the dataset-level metric for CelebA, we plot the correlation heatmap using the absolute value of correlation between different metrics.}
\label{figure:coverage_relation_celeba}
\end{figure*}

We present the relationship between coverages with different sample-level metrics and the dataset-level metric for CIFAR10 and MNIST together in~\autoref{figure:coverage_relation_cifar_mnist}, and separately for CelebA in~\autoref{figure:coverage_relation_celeba}. 
The findings reveal an intriguing observation: high-quality reconstructions do not necessarily correspond to greater diversity in the reconstructed samples. 

\section{Influence of Reconstruction Quantity}
\label{app:influence_of_quantity}
We explicitly stipulate that the reconstruction algorithm must produce the same number of samples as the target dataset.
In~\autoref{section:recdef}, we provide an intuitive justification for this requirement, and in this section, we present empirical evidence to support our decision.
Specifically, we use DeepInversion with 20,000 target samples as a representative case and reconstruct a total of 40,000 samples. 
We evaluate the performance using different numbers of reconstructions, ranging from 5,000 to 40,000 in steps of 5,000.

\begin{figure*}[!t]
\centering
\begin{subfigure}{0.49\columnwidth}
\includegraphics[width=\columnwidth]{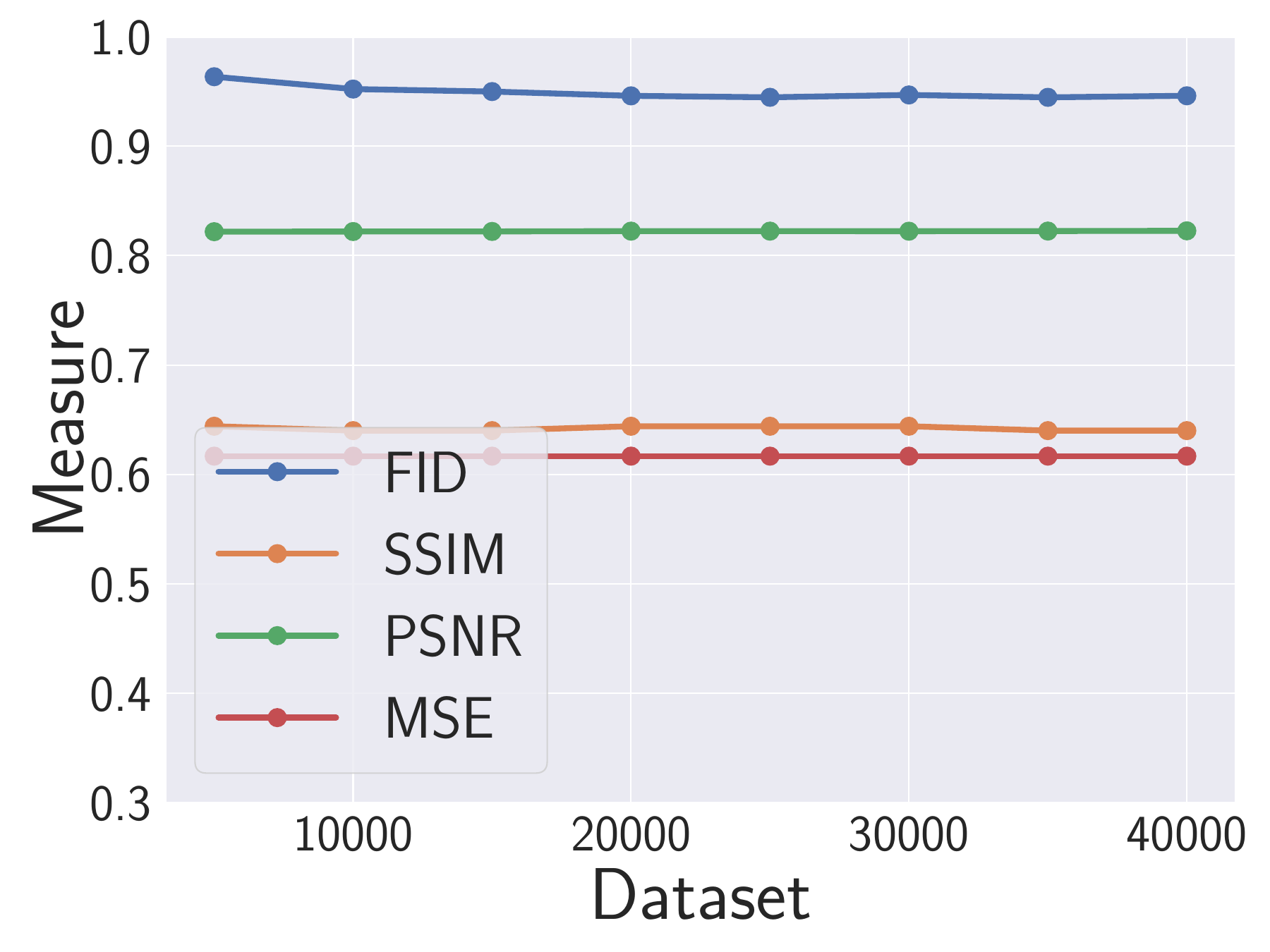}
\caption{Measure (CelebA)}
\end{subfigure}
\begin{subfigure}{0.49\columnwidth}
\includegraphics[width=\columnwidth]{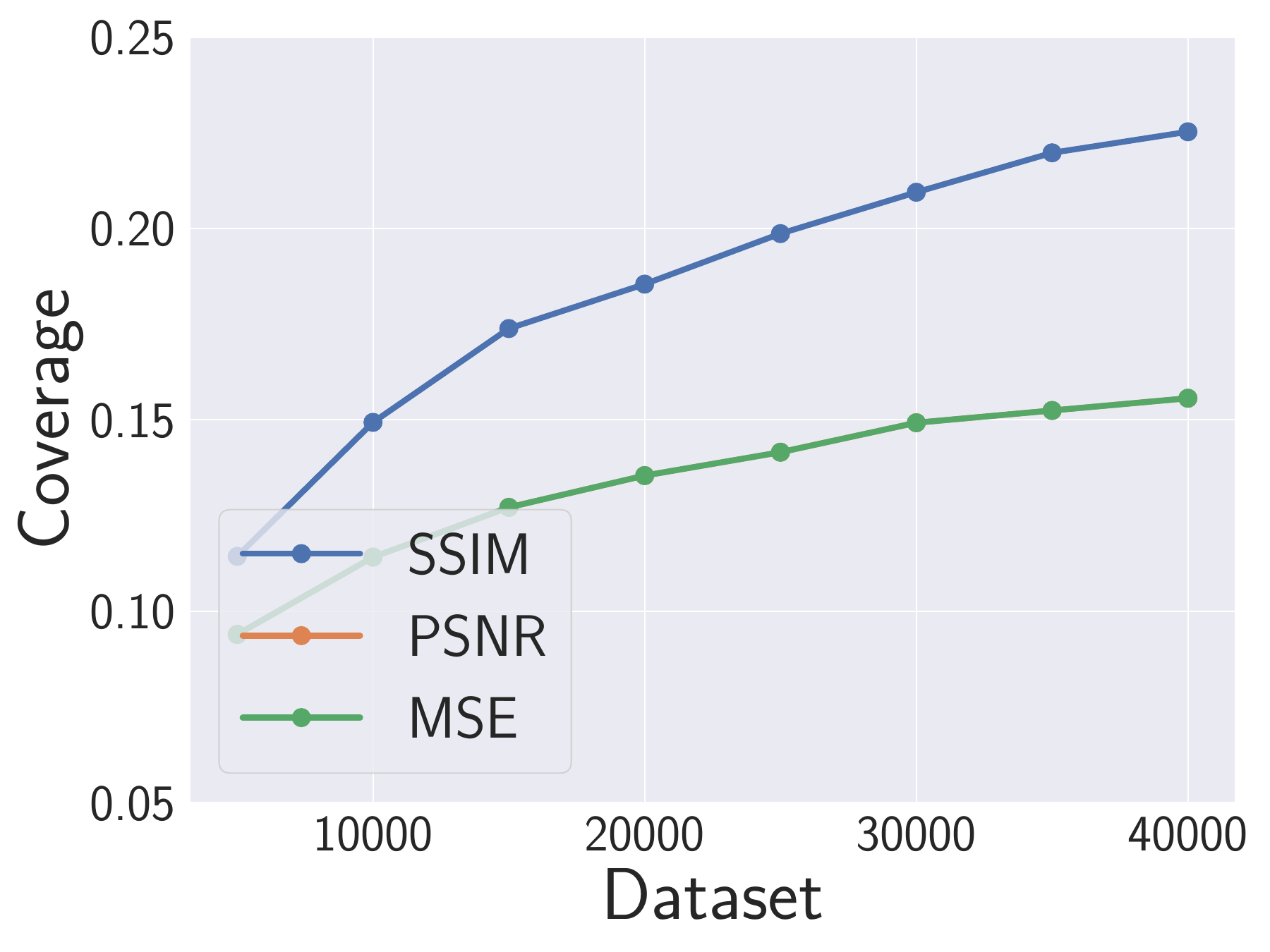}
\caption{Coverage (CelebA)}
\end{subfigure}
\begin{subfigure}{0.49\columnwidth}
\includegraphics[width=\columnwidth]{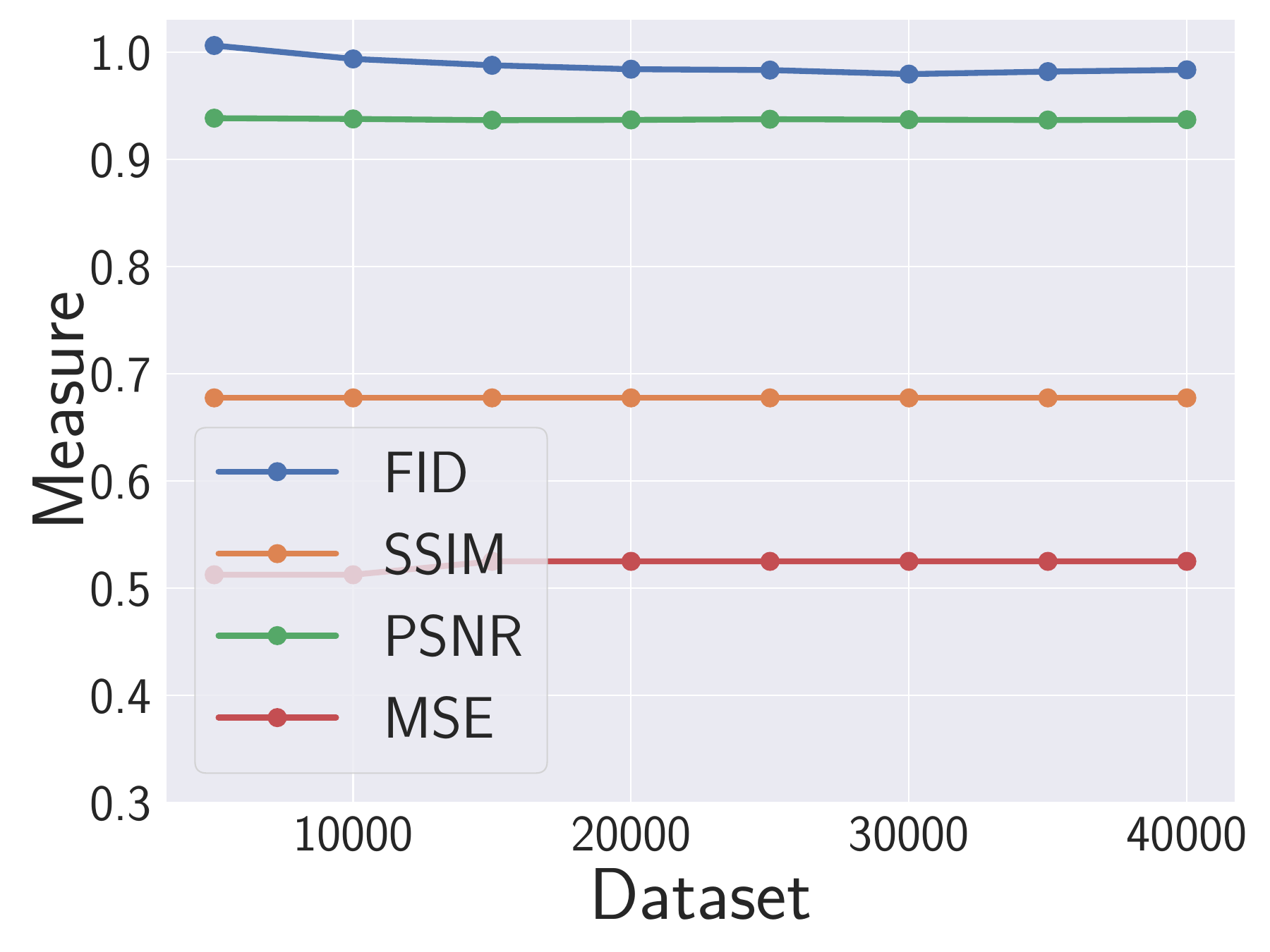}
\caption{Measure (CIFAR10)}
\end{subfigure}
\begin{subfigure}{0.49\columnwidth}
\includegraphics[width=\columnwidth]{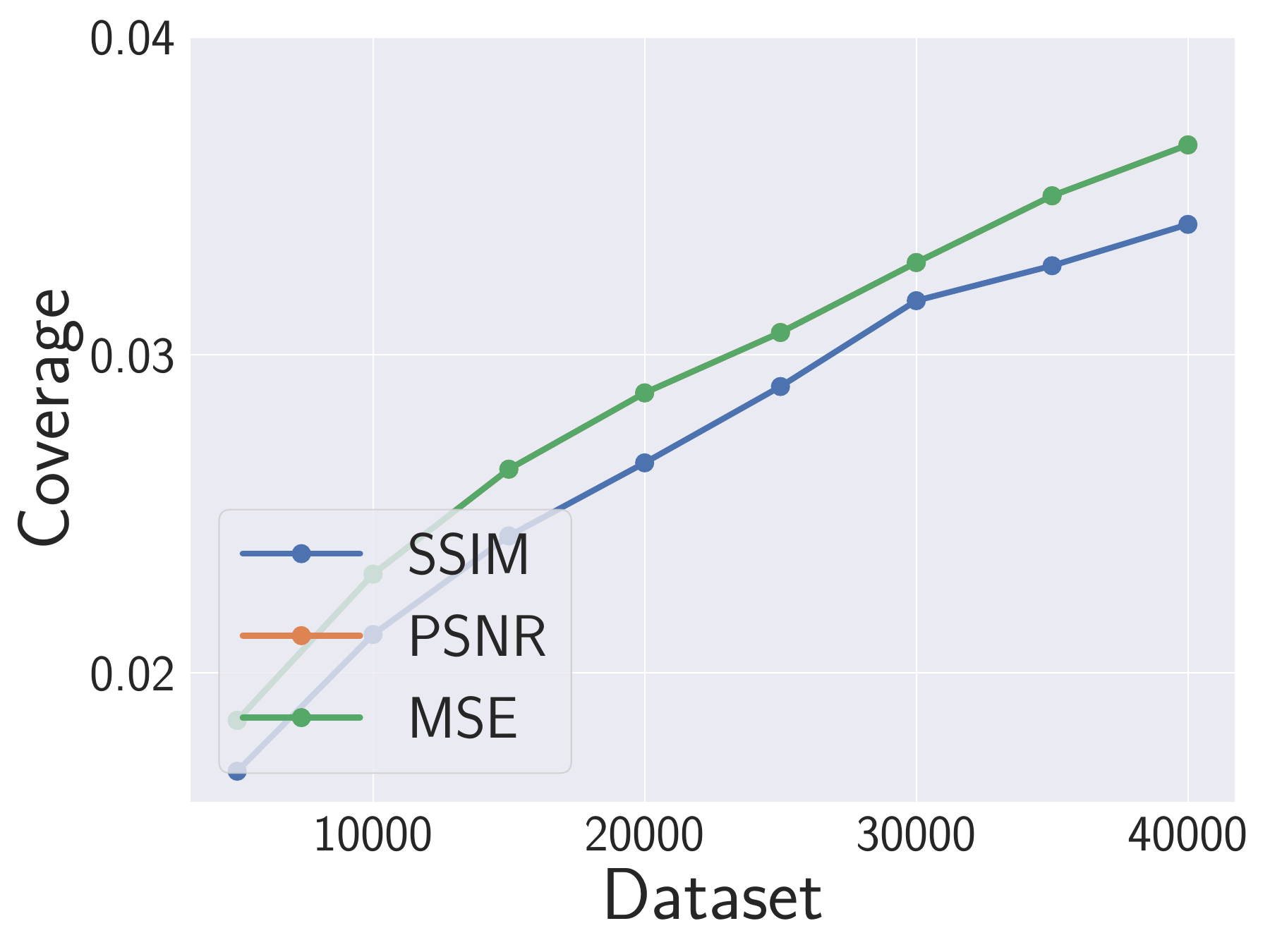}
\caption{Coverage (CIFAR10)}
\end{subfigure}
\begin{subfigure}{0.49\columnwidth}
\includegraphics[width=\columnwidth]{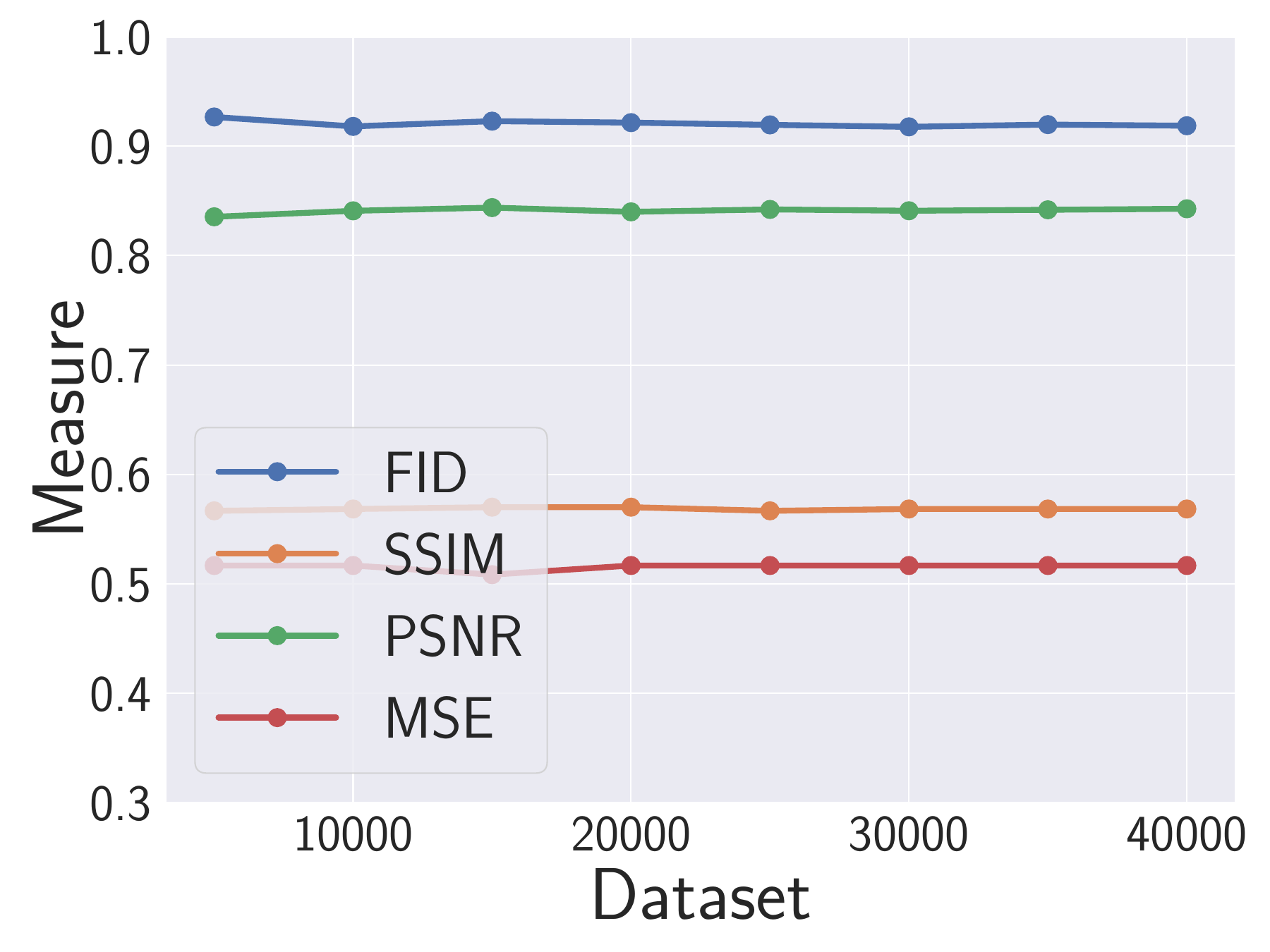}
\caption{Measeure (MNIST)}
\end{subfigure}
\begin{subfigure}{0.49\columnwidth}
\includegraphics[width=\columnwidth]{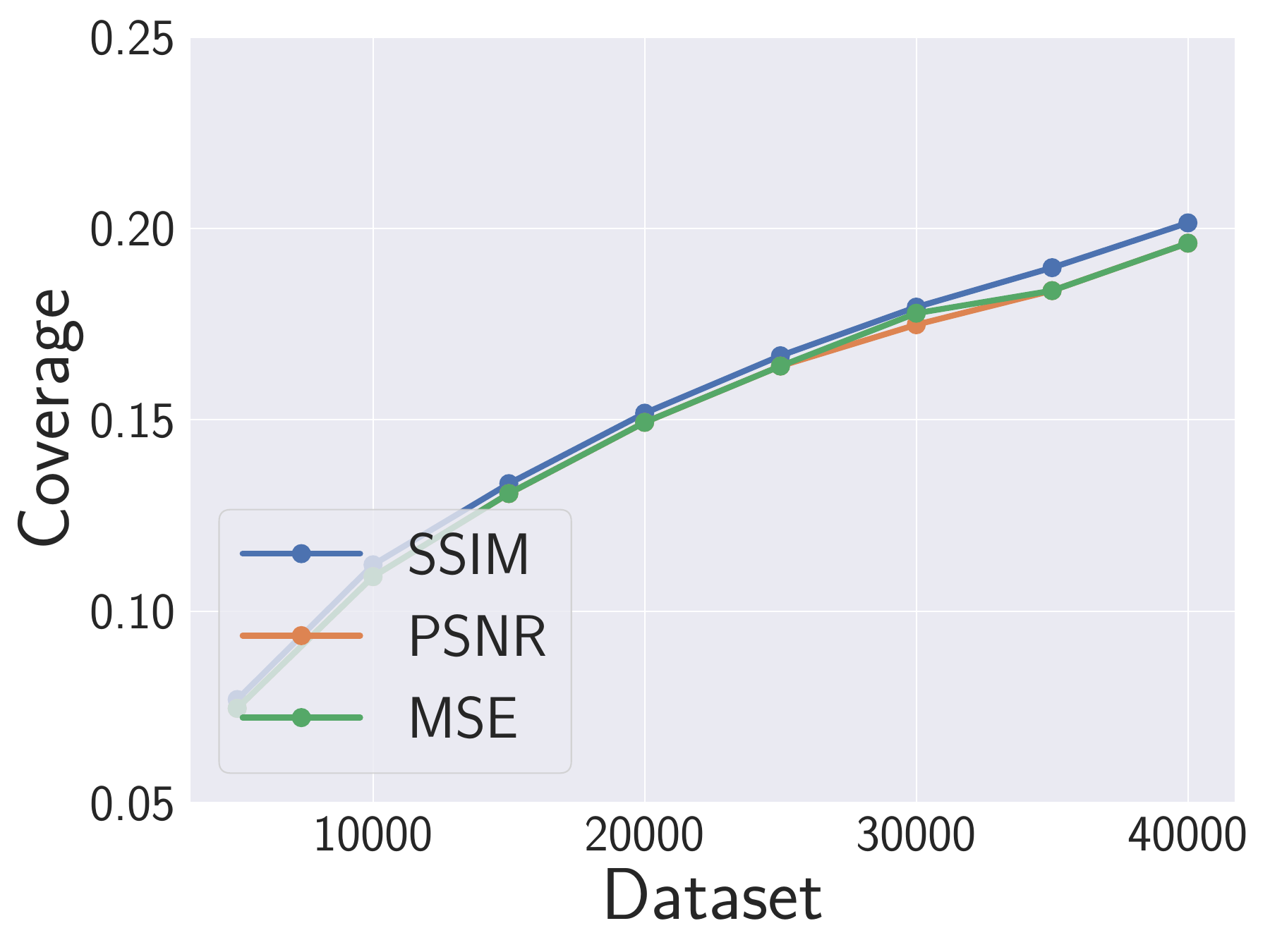}
\caption{Coverage (MNIST)}
\end{subfigure}
\caption{Influence of reconstruction quantity. For ease of presentation, values are normalized to fit the same y-axis.}
\label{figure:quantity_influence_more}
\end{figure*}

As depicted in~\autoref{figure:quantity_influence_more}, the number of reconstructions has a minimal impact on reconstruction quality, both at the dataset level and sample level. 
This suggests that the reconstruction algorithm's ability is not affected by the number of reconstructions.

On the other hand, increasing the number of reconstructions significantly increases coverage.
This supports our earlier argument in~\autoref{section:recdef} that generating an infinite number of samples through random initialization is not reasonable, as it only increases coverage without improving reconstruction quality. 
Furthermore, without the guidance of the target dataset, the adversary has no way to determine which reconstructions are of high quality, and the increased diversity does not contribute to better performance at either the dataset level or sample level.

\section{Influence of Attack Knowledge}
\label{app:ablationstudy}
\begin{figure*}[!t]
\centering
\begin{subfigure}{0.4\columnwidth}
\includegraphics[width=\columnwidth]{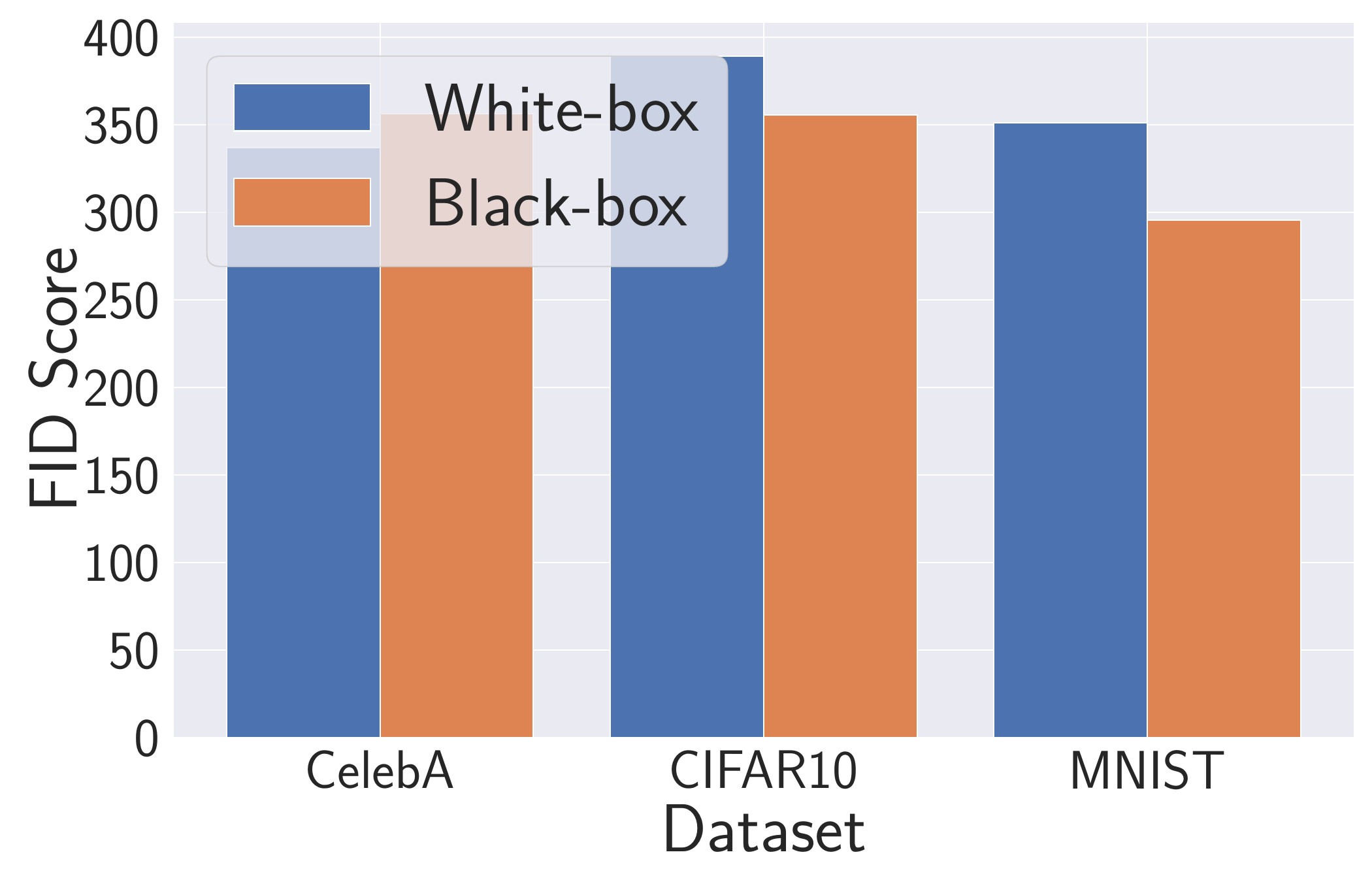}
\caption{MI-Face}
\label{figure:influence_of_model_access_mi}
\end{subfigure}
\begin{subfigure}{0.4\columnwidth}
\includegraphics[width=\columnwidth]{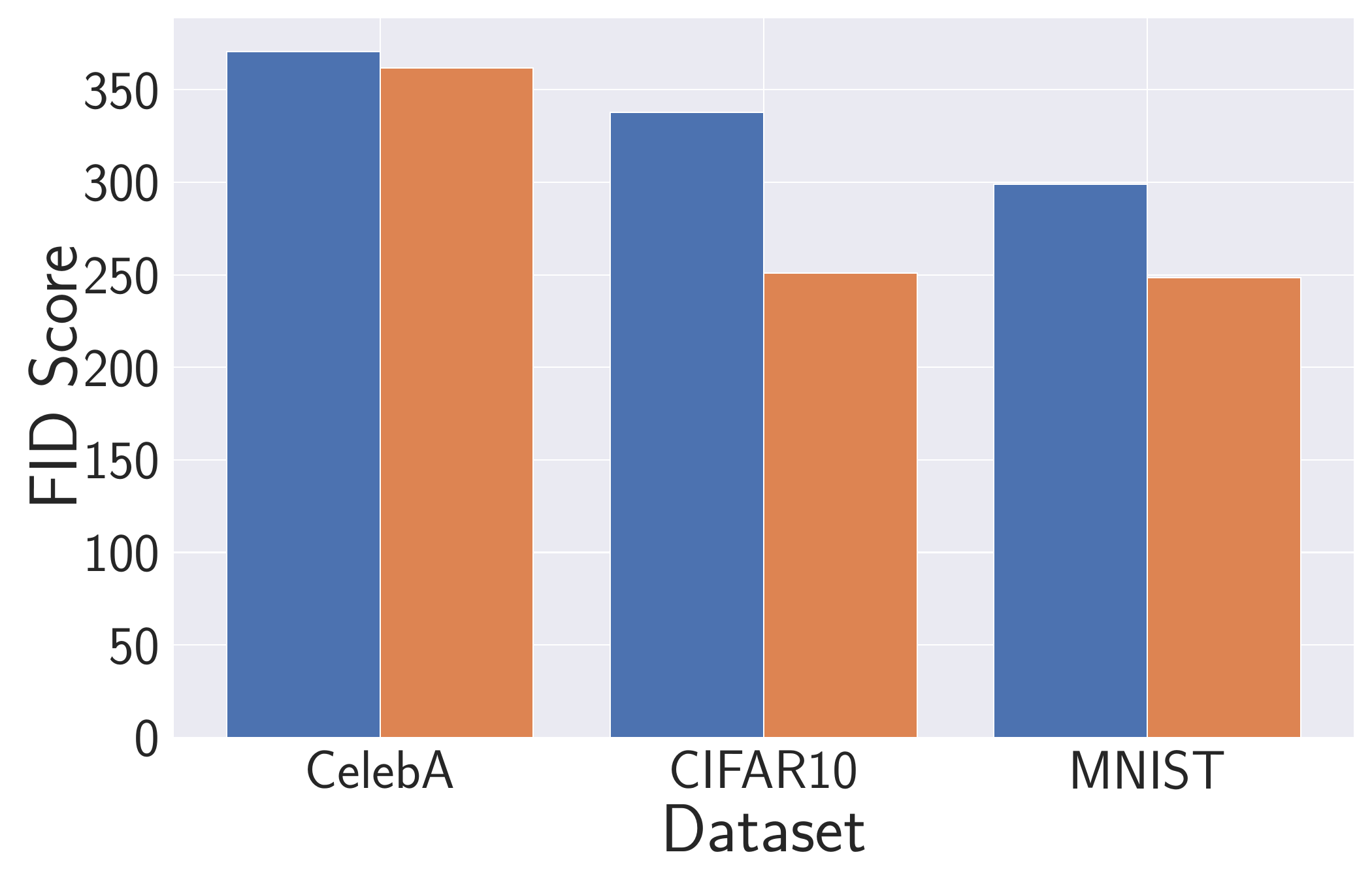}
\caption{DeepDream}
\label{figure:influence_of_model_access_deepdream}
\end{subfigure}
\begin{subfigure}{0.4\columnwidth}
\includegraphics[width=\columnwidth]{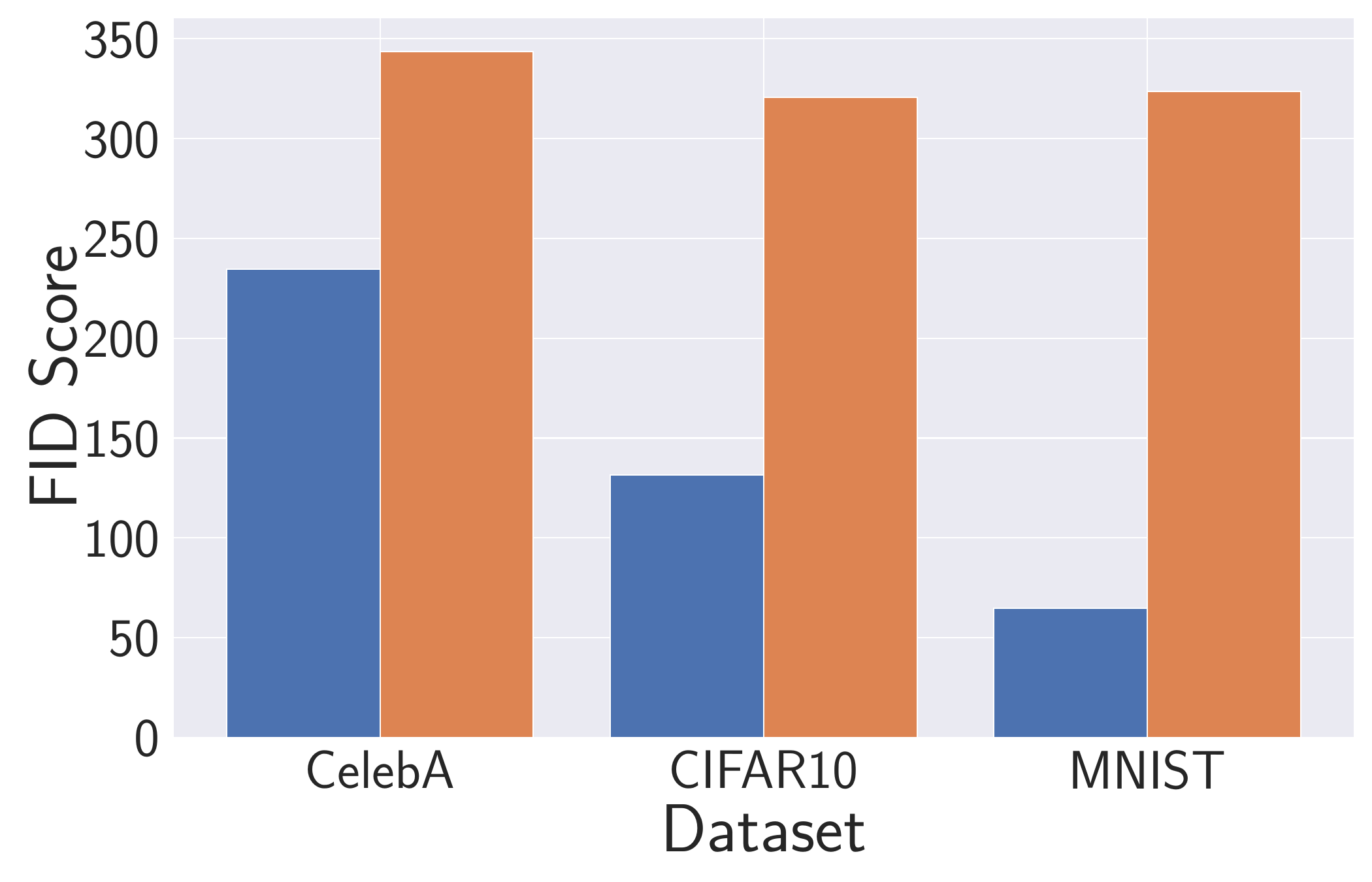}
\caption{DeepInversion}
\label{figure:influence_of_model_access_deepinv}
\end{subfigure}
\begin{subfigure}{0.4\columnwidth}
\includegraphics[width=\columnwidth]{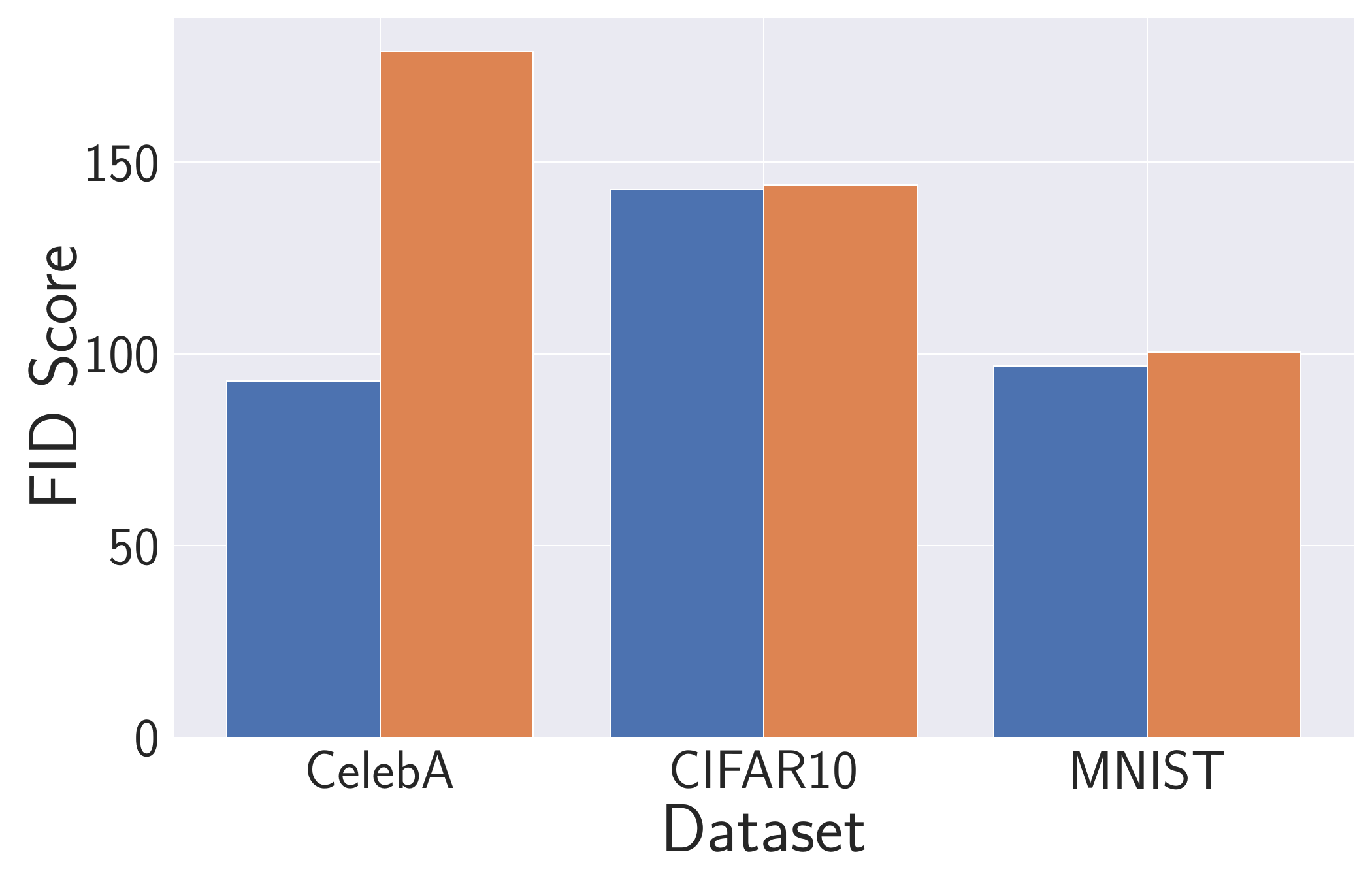}
\caption{Revealer}
\label{figure:influence_of_model_access_revealer}
\end{subfigure}
\begin{subfigure}{0.4\columnwidth}
\includegraphics[width=\columnwidth]{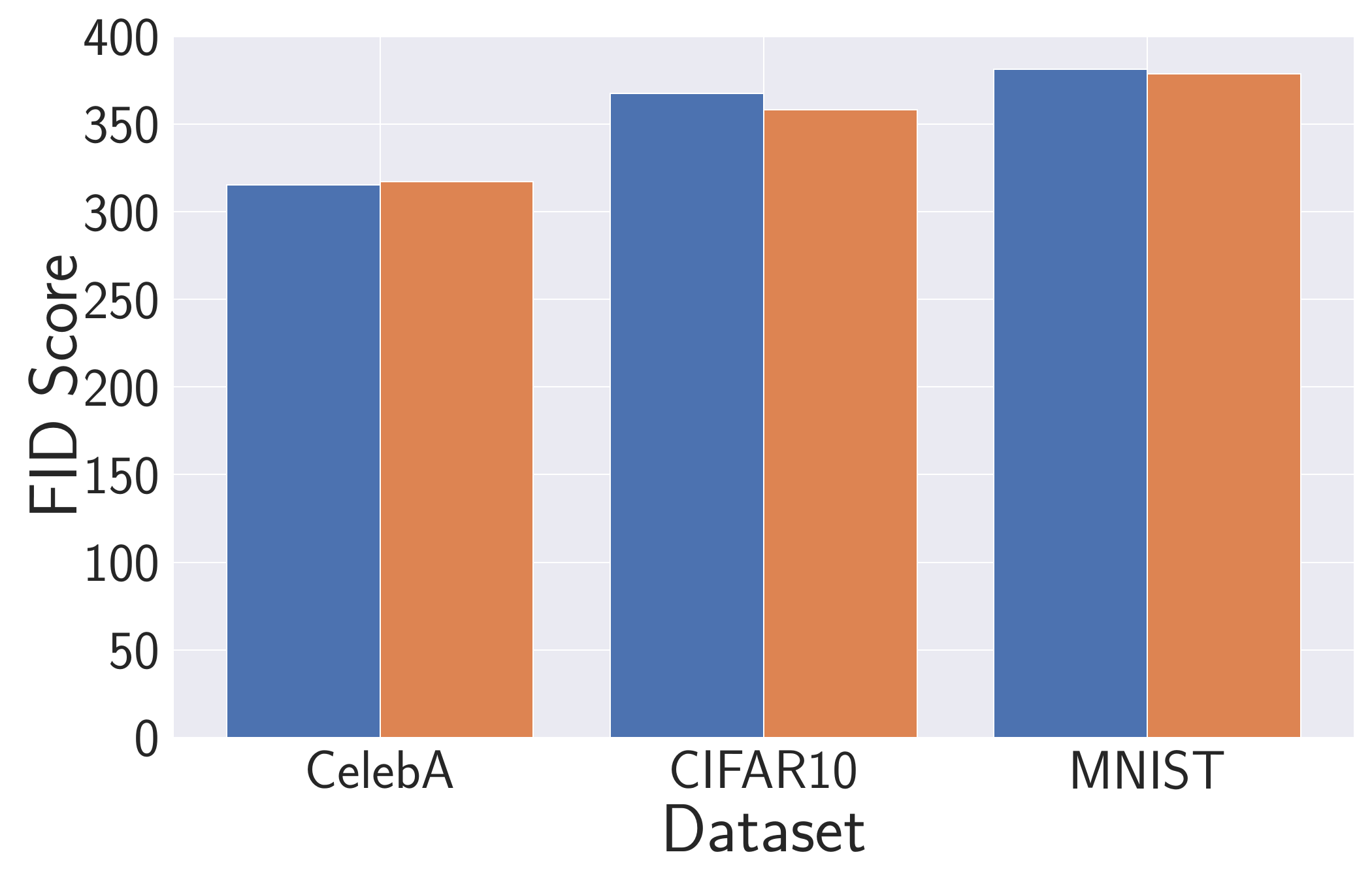}
\caption{Bias-Rec}
\label{figure:influence_of_model_access_impbias}
\end{subfigure}
\begin{subfigure}{0.4\columnwidth}
\includegraphics[width=\columnwidth]{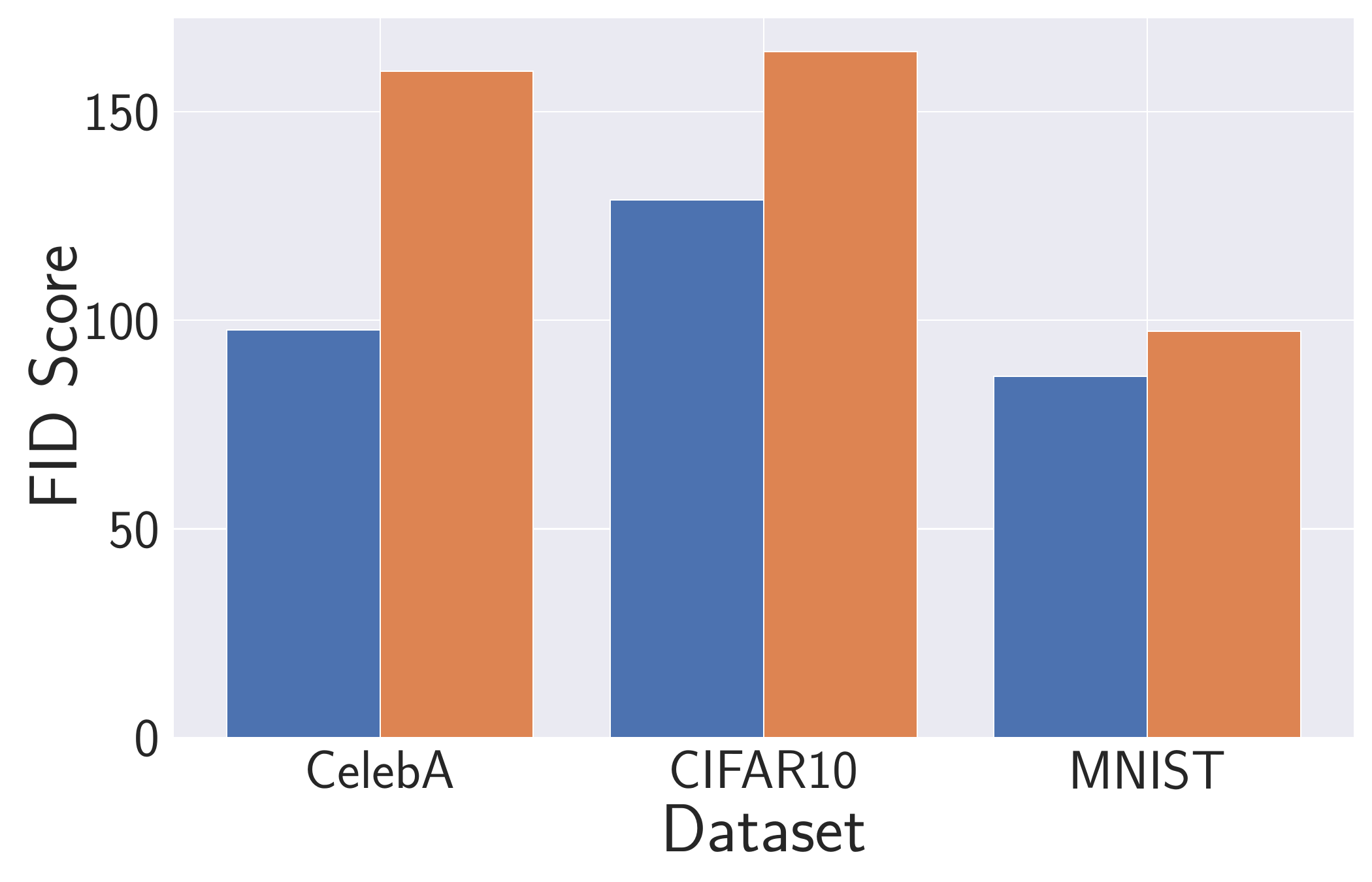}
\caption{KEDMI}
\label{figure:influence_of_model_access_kedmi}
\end{subfigure}
\begin{subfigure}{0.4\columnwidth}
\includegraphics[width=\columnwidth]{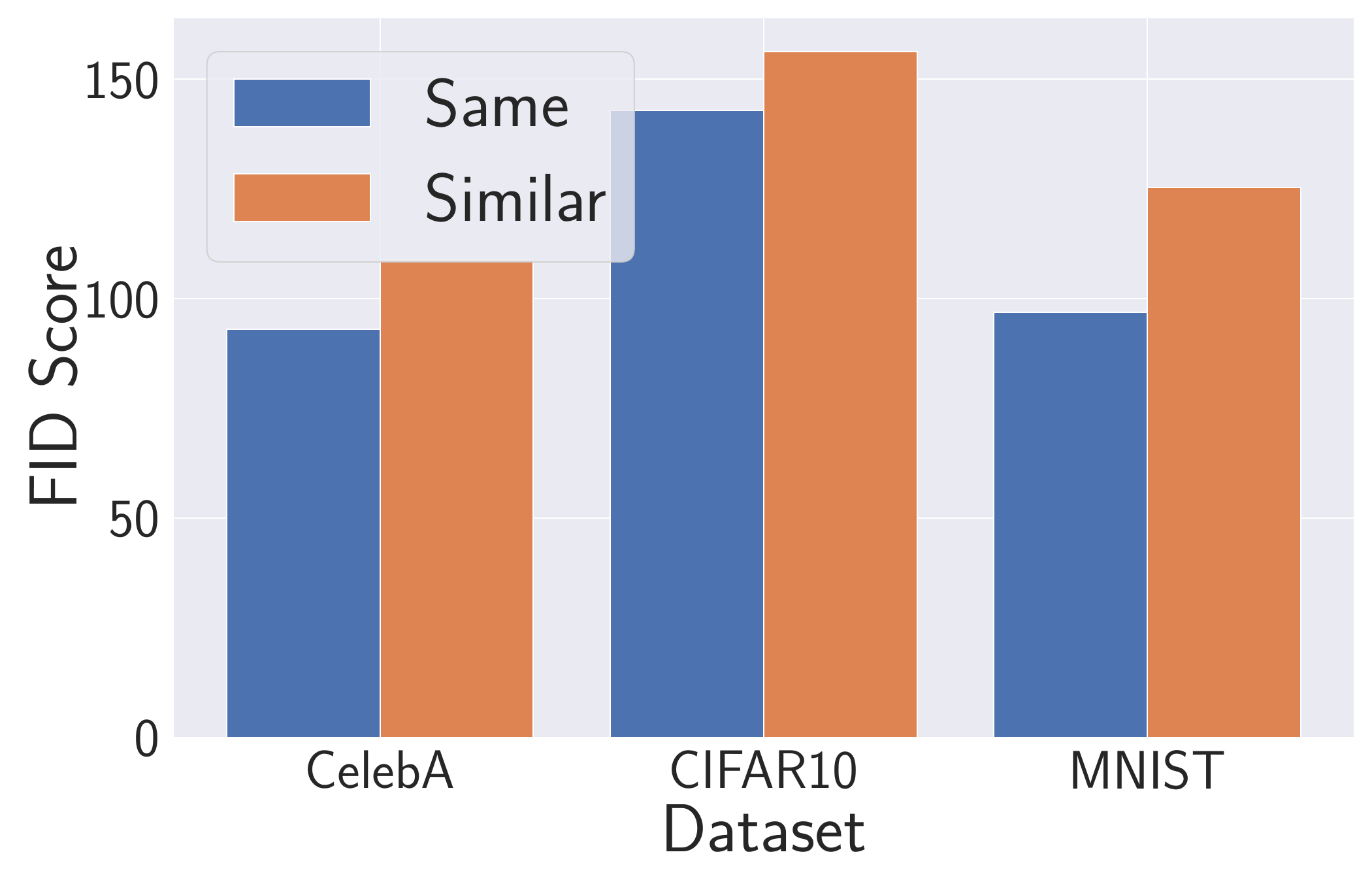}
\caption{Revealer}
\label{figure:influence_of_data_access_revealer}
\end{subfigure}
\begin{subfigure}{0.4\columnwidth}
\includegraphics[width=\columnwidth]{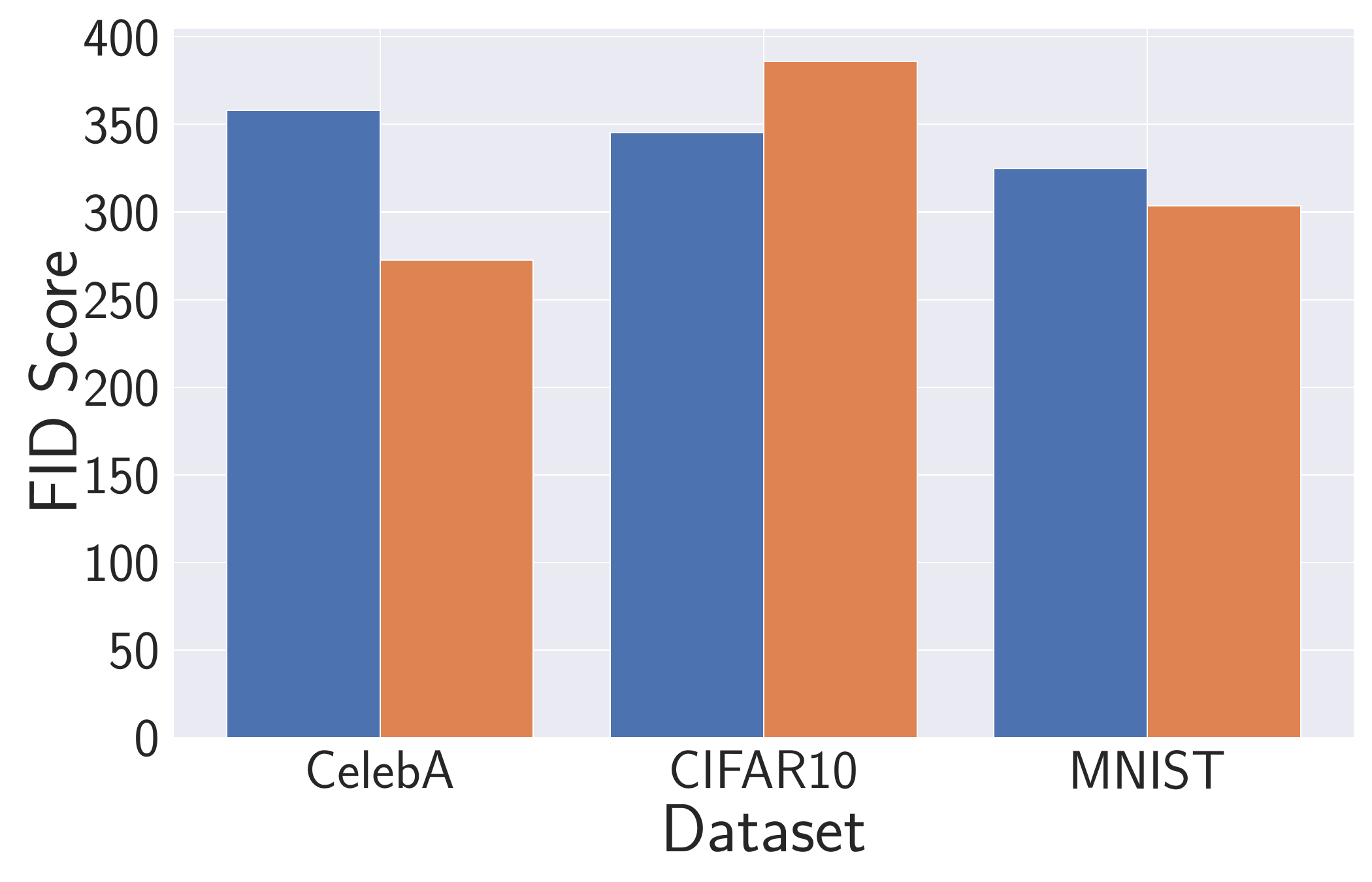}
\caption{Inv-Alignment}
\label{figure:influence_of_data_access_invalign}
\end{subfigure}
\begin{subfigure}{0.4\columnwidth}
\includegraphics[width=\columnwidth]{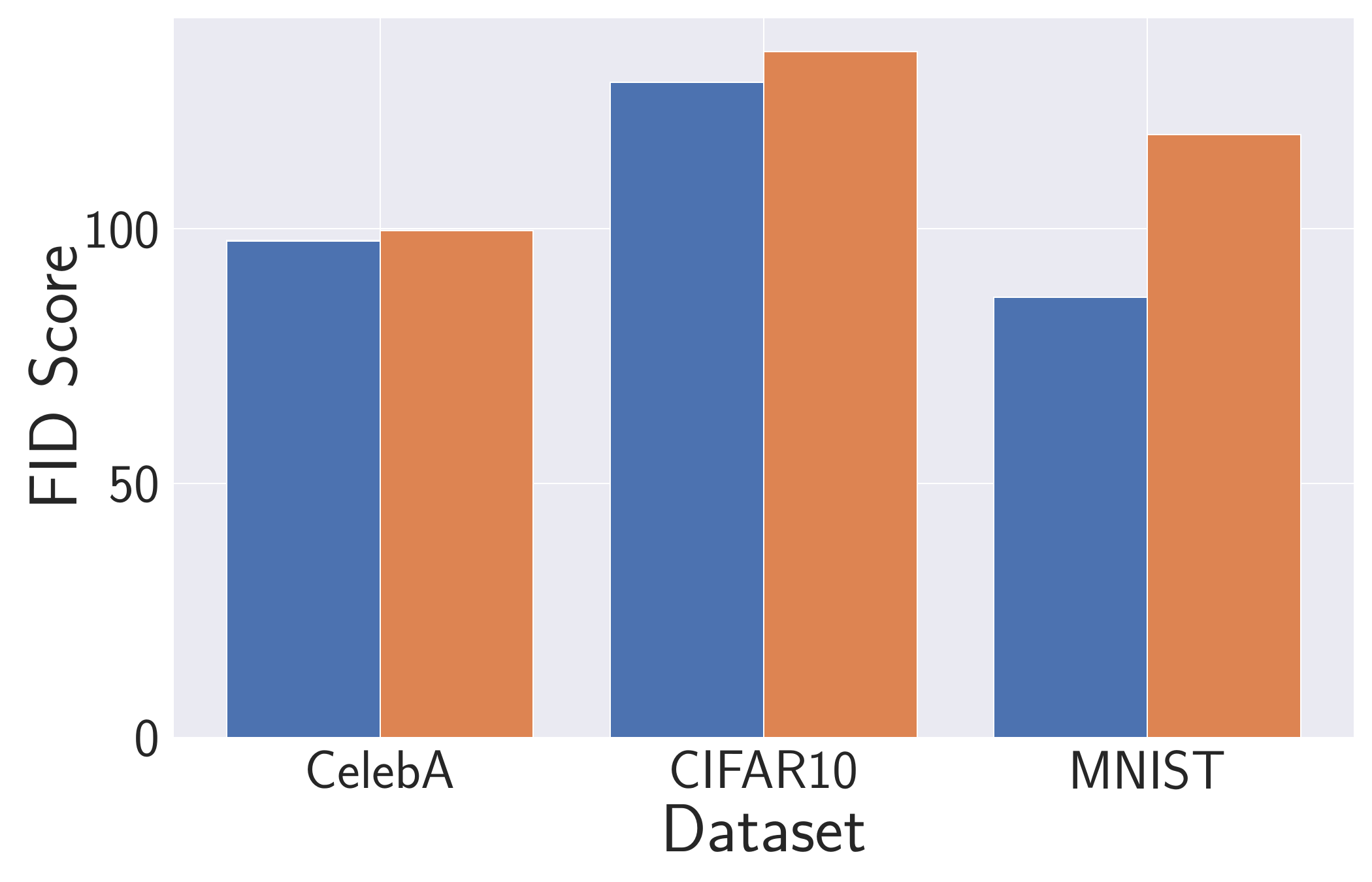}
\caption{KEDMI}
\label{figure:influence_of_data_access_kedmi}
\end{subfigure}
\caption{More results on the influence of auxiliary information.}
\label{figure:influence_of_knowledge_more}
\end{figure*}

We plot the influence of attack knowledge in~\autoref{figure:influence_of_knowledge_more}.

\section{Experimental Setup For The Text Modality}
\label{app:nlp_setup}
Our experimental evaluation is performed on three established text datasets: SST2~\cite{SPWCMNP13}, AGNews~\cite{ZZL15}, and IMDB~\cite{MDPHNP11}.
For fine-tuning the target models, we utilize subsets of these datasets ranging from 1,000 to 20,000 samples.
This focus on fine-tuning, rather than training language models from scratch, is adopted due to the substantial data requirements of the latter and aligns with common practices in recent literature~\cite{WWBZS23,DDYPB23}.

We investigate two distinct attack methodologies:
\begin{enumerate}
    \item \textbf{Vec2Text Attack~\cite{MKSR23}:} In this configuration, the \texttt{sentence-transformers/gtr-t5-base} model is fine-tuned. 
    The corresponding decoder is a \texttt{T5-base} model, which is further trained using an auxiliary dataset drawn from the same data distribution as the fine-tuning set.
    \item \textbf{Complete Attack~\cite{CTWJHLRBSEOR21,WLBZ24}:} For this attack, the \texttt{GPT2} model is fine-tuned for 20 epochs. 
    To reconstruct the original input, the fine-tuned model is queried using the first three words of the said input as a prompt.
\end{enumerate}

Detailed results of these attack evaluations are presented in~\autoref{table:eval_results_vec2text} and~\autoref{table:eval_results_complete}.
To assess the quality of the reconstructed text, we employ several metrics.
Semantic similarity is quantified by the cosine similarity between sentence embeddings; these embeddings are generated using the \texttt{sentence-transformers/all-MiniLM-L6-v2} model.
Additionally, we report BLEU scores to measure n-gram precision and ROUGE-L scores to evaluate the longest common subsequence, thereby providing insights into the lexical overlap and fluency of the reconstructed outputs.

\begin{table}[!t]
\centering
\caption{Evaluation of the Complete~\cite{CTWJHLRBSEOR21,WLBZ24} attack performance across varying target data sizes on different datasets, measured by BLEU, Sim. (Similarity), and R-L (ROUGE-L). 
Higher scores on these metrics correspond to better attack performance.}
\setlength{\tabcolsep}{3.pt}
\scalebox{0.8}{
\begin{tabular}{@{}llcccccc@{}}
\toprule
\multirow{2}{*}{Dataset} & \multirow{2}{*}{Metric} & \multicolumn{6}{c}{Target Data Size} \\
\cmidrule(l){3-8}
& & $1,000$ & $2,000$ & $5,000$ & $10,000$ & $15,000$ & $20,000$ \\
\midrule
\multirow{3}{*}{SST2} & BLEU & $0.606$ & $0.527$ & $0.450$ & $0.115$ & $0.103$ & $0.097$ \\
& Sim. & $0.875$ & $0.774$ & $0.721$ & $0.496$ & $0.449$ & $0.417$ \\
& R-L & $0.777$ & $0.651$ & $0.575$ & $0.229$ & $0.205$ & $0.203$ \\
\midrule
\multirow{3}{*}{AGNews} & BLEU & $0.412$ & $0.397$ & $0.370$ & $0.328$ & $0.291$ & $0.258$ \\
& Sim. & $0.865$ & $0.859$ & $0.848$ & $0.825$ & $0.806$ & $0.784$ \\
& R-L & $0.576$ & $0.563$ & $0.536$ & $0.493$ & $0.452$ & $0.415$ \\
\midrule
\multirow{3}{*}{IMDB} & BLEU & $0.344$ & $0.282$ & $0.123$ & $0.051$ & $0.035$ & $0.029$ \\
& Sim. & $0.747$ & $0.697$ & $0.552$ & $0.463$ & $0.442$ & $0.431$ \\
& R-L & $0.557$ & $0.481$ & $0.272$ & $0.182$ & $0.164$ & $0.157$ \\
\bottomrule
\end{tabular}
}
\label{table:eval_results_complete}
\end{table}

\section{Defenses Against Reconstruction Attacks}
\label{app:defense}
We test Mutual Information Regularization (MID)~\cite{WZJ21} and Differential Privacy (DP)~\cite{ACGMMTZ16} as countermeasures against data reconstruction attacks. 
The experimental results, summarized in \autoref{table:eval_results_celeba_defense}, demonstrate that both MID and DP can mitigate model vulnerability in most cases. 
This is evidenced by higher Fréchet Inception Distance (FID) scores, indicating successful defense, across varied training dataset sizes.

However, the protective capabilities of MID and DP are not uniformly significant across all evaluated conditions. 
This variability in performance underscores the pressing need for the research community to develop more consistently robust and broadly effective defense mechanisms. 
Such advancements are crucial for adequately countering the sophisticated threats posed by advanced reconstruction attacks.

\section{Transferability To Larger Architectures}
\label{app:transfer2transformer}

We extend our analysis of the relationship between memorization and reconstruction performance to larger, contemporary model architectures—specifically, transformer-based models that are widely used in modern machine learning practice. 
Our experiments with Swin~\cite{LLCHWZLG21} and MAE~\cite{HCXLDG22} architectures, as presented in~\autoref{table:eval_swin_gpt4o} and~\autoref{table:eval_mae_gpt4o}, reveal consistent trends: models trained on smaller datasets tend to be more vulnerable, in line with our previous findings.

We also observe that the worst-case performance does not occur at the smallest dataset size. 
This can be attributed to the nature of transformer-based models, which typically require substantially larger datasets to train effectively. 
As the dataset grows, the model begins to capture sample-specific features, leading to increased unnecessary memorization. 
This nuanced behavior aligns with our analysis in~\autoref{section:deep_inves_metric}.

\begin{table}[!t]
\centering
\setlength{\tabcolsep}{3.0pt}
\caption{Evaluation on MAE with GPT-4o.}
\label{table:eval_mae_gpt4o}
\scalebox{0.8}{
\begin{tabular}{@{}cccccccc@{}}
\toprule
\multirow{2}{*}{Attack}& \multirow{2}{*}{Metrics} &\multicolumn{6}{c}{Target Data Size}\\ 
\cmidrule(l){3-8}& & $1,000$ & $2,000$ & $5,000$ & $10,000$ & $15,000$ & $20,000$\\
\midrule
\multirow{3}{*}{PLGMI}& \# of Major & $79$ &$233$ &$312$ &$275$ &$76$ &$25$ \\
& \# of All & $7$ &$12$ &$24$ &$24$ &$0$ &$1$ \\
& Pred Rate & $0.097$ &$0.222$ &$0.299$ &$0.267$ &$0.090$ &$0.025$ \\
\bottomrule
\end{tabular}
}
\end{table}

\section{Related Work}
\label{section:related}

\subsection{Membership Inference Attack}
\label{section:related_mia}
Membership inference attack reveals the membership status of a target sample, i.e., whether the target sample is in the training dataset or not, which leads to a direct privacy breach.

Shokri et al.~\cite{SSSS17} proposed the seminal work on membership inference attack against machine learning models, wherein several shadow models were trained to imitate the behavior of the target model.
This attack requires access to data from the same distribution as the training dataset.
Later, Salem et al.~\cite{SZHBFB19} relax the assumption of the same distribution and demonstrate the effectiveness of using only one shadow model, largely reducing the computational cost required.
Subsequent research~\cite{CTCP21,LZ21} explores a more challenging setting where the adversary only has hard-label access to the target model.
Specifically, Li and Zhang~\cite{LZ21} utilize adversary examples to approximate the distance between the target sample to its decision boundary in order to make decisions based on this distance. 
Recently, more work~\cite{CCNSTT22,LZBZ22,TSJLJHC22} aims at enhancing the performance of membership inference attacks.
For example, Carlini et al.~\cite{CCNSTT22} take advantage of the discrepancy of models trained with and without the target sample.
Liu et al.~\cite{LZBZ22} demonstrate the effectiveness of loss trajectory.

\subsection{Other Privacy Attacks}
Property inference attack differs from data reconstruction and membership inference attack as it aims to infer macro-level information about the target dataset, such as the gender proportion. 
Ateniese et al.~\cite{AMSVVF15} presented the first property inference attacks against Hidden Markov Models (HMMs) and Support Vector Machine (SVM), which was later extended to Fully Connected Neural Networks (FCNNs) by Ganju et al.~\cite{GWYGB18}.
Both attacks rely on a meta-classifier to infer the property of the training dataset using white-box access to the target model. 
Training the meta-classifier requires multiple shadow models, making it computationally expensive.

Suri and Evans~\cite{SE21} first formalized the property inference attack and provided a method for conducting the attack with black-box access to the target model. 
Subsequent research~\cite{MGC22,CAOJTU23}, has focused on improving the performance of property inference attacks by adding a small amount of ``poisoned'' data to the training dataset.
For example, Chaudhari et al.~\cite{CAOJTU23} select a limited number of samples in one class, and flip their labels to increase the discrepancy of posterior distributions for different properties.

Additionally, model stealing attacks aim to construct a local surrogate model from the target model.
Tram{\`e}r et al.~\cite{TZJRR16} propose the first attack against neural networks by querying the target model to construct the training dataset.
However, their attack demands high-quality data to be effective, which motivates recent research to relax this assumption through the development of data-free attack paradigms~\cite{KPQ21,TMWP21,SAB22}.

\subsection{Memorization and Privacy Leakage}
Song et al.~\cite{SRS17} demonstrate that malicious trainers can easily encode training samples into the model parameters, which can later be extracted.
This highlights the potential threats for model leakage and emphasizes the need for researchers to consider the security implications of memorization in their models. 
Song and Shmatikov~\cite{SS20} further develop this point by uncovering the intrinsic behavior of models, specifically their tendency to retain information that is not pertinent to the designed classification task.
Despite attempts to eliminate this extraneous information, the unintentional disclosure of sensitive data remains a persistent issue.

Several studies have leveraged memorization to enhance the attack performance. 
Tram{\`e}r et al.~\cite{TSJLJHC22} show that with access to a tiny fraction of the training dataset, the adversary can boost the performance of membership inference by poisoning data samples as the memorization of these poisoned samples increases. 
Wen et al.~\cite{WBZ25} investigate the relationship between data importance (used as a proxy for memorization) and privacy attacks, observing that data samples with higher importance exhibit increased vulnerability to certain attacks.

\begin{table*}[!t]
\centering
\caption{Evaluation results of existing reconstruction attacks on larger models. 
The target model is Swin Transformer trained on CelebA with 6 different sizes.}
\scalebox{0.75}{
\begin{tabular}{@{}cccccccccc@{}}
\toprule
\multirow{2}{*}{Attack}& \multicolumn{2}{c}{\multirow{2}{*}{Metrics}} &\multicolumn{6}{c}{Target Data Size}\\ \cmidrule(l){4-9}& & & $1,000$ & $2,000$ & $5,000$ & $10,000$ & $15,000$ & $20,000$\\ \midrule
\multirow{4}{*}{MI-Face} & Dataset-level & FID $\downarrow$  & $359.896$ &$357.629$ &$348.142$ &$364.073$ &$366.656$ &$360.530$ \\ \cmidrule(l){3-9}
& \multirow{3}{*}{Sample-level} & SSIM $\uparrow$ & $0.013 (100.00\%)$ & $0.017 (70.65\%)$ & $0.021 (59.00\%)$ & $0.023 (54.89\%)$ & $0.023 (54.49\%)$ & $0.022 (55.33\%)$ \\
& & PSNR $\uparrow$ & $8.292 (100.00\%)$ & $8.921 (53.70\%)$ & $8.977 (26.72\%)$ & $9.444 (14.25\%)$ & $9.526 (10.59\%)$ & $9.800 (9.31\%)$ \\
& & MSE $\downarrow$  & $0.152 (100.00\%)$ & $0.130 (53.70\%)$ & $0.128 (26.72\%)$ & $0.115 (14.25\%)$ & $0.112 (10.59\%)$ & $0.106 (9.31\%)$ \\ \cmidrule(l){1-9}
\multirow{4}{*}{DeepDream} & Dataset-level & FID $\downarrow$  & $344.805$ &$365.463$ &$357.851$ &$368.678$ &$344.333$ &$320.711$ \\ \cmidrule(l){3-9}
& \multirow{3}{*}{Sample-level} & SSIM $\uparrow$ & $0.055 (100.00\%)$ & $0.069 (68.80\%)$ & $0.099 (53.52\%)$ & $0.193 (29.45\%)$ & $0.191 (24.78\%)$ & $0.209 (17.25\%)$ \\
\bottomrule
\end{tabular}
}
\label{table:eval_results_celeba_swin}
\end{table*}

\begin{table*}[!t]
\centering
\caption{Evaluation results of existing reconstruction attacks on larger models. 
The target model is Masked AutoEncoder (MAE) trained on CelebA with 6 different sizes.}
\scalebox{0.75}{
\begin{tabular}{@{}cccccccccc@{}}
\toprule
\multirow{2}{*}{Attack}& \multicolumn{2}{c}{\multirow{2}{*}{Metrics}} &\multicolumn{6}{c}{Target Data Size}\\ \cmidrule(l){4-9}& & & $1,000$ & $2,000$ & $5,000$ & $10,000$ & $15,000$ & $20,000$\\ \midrule
\multirow{4}{*}{MI-Face} & Dataset-level & FID $\downarrow$  & $325.226$ & $304.771$ & $311.452$ & $322.519$ & $337.842$ & $351.695$  \\ \cmidrule(l){3-9}
& \multirow{3}{*}{Sample-level} & SSIM $\uparrow$ & $0.014 (100.00\%)$ & $0.018 (70.05\%)$ & $0.021 (57.19\%)$ & $0.026 (49.44\%)$ & $0.025 (49.80\%)$ & $0.022 (49.90\%)$ \\
& & PSNR $\uparrow$ & $8.191 (100.00\%)$ & $8.750 (55.85\%)$ & $9.188 (29.80\%)$ & $9.737 (16.51\%)$ & $9.823 (13.58\%)$ & $9.541 (9.24\%)$  \\
& & MSE $\downarrow$ & $0.160 (100.00\%)$ & $0.138 (55.85\%)$ & $0.123 (29.80\%)$ & $0.107 (16.51\%)$ & $0.106 (13.58\%)$ & $0.104 (9.24\%)$ \\ \cmidrule(l){1-9}
\multirow{4}{*}{DeepDream} & Dataset-level & FID $\downarrow$  & $288.219$ & $258.241$ & $321.577$ & $303.685$ & $289.375$ & $313.243$  \\ \cmidrule(l){3-9}
& \multirow{3}{*}{Sample-level} & SSIM $\uparrow$ & $0.063 (100.00\%)$ & $0.131 (58.35\%)$ & $0.056 (49.86\%)$ & $0.102 (35.16\%)$ & $0.228 (18.79\%)$ & $0.199 (19.16\%)$ \\
\bottomrule
\end{tabular}
}
\label{table:eval_results_celeba_mae}
\end{table*}

\begin{table*}[!t]
\centering
\caption{Evaluation results of existing reconstruction attacks. 
The target model is VGG16 trained on CIFAR100 with 6 different sizes.
Attacks with a gray background belong to the dynamic training type.
For FID and MSE, a lower score indicates better reconstruction quality; while for SSIM and PSNR, a higher score indicates better performance. 
}
\scalebox{0.75}{
\begin{tabular}{@{}cccccccccc@{}}
\toprule
\multirow{2}{*}{Attack}& \multicolumn{2}{c}{\multirow{2}{*}{Metrics}} &\multicolumn{6}{c}{Target Data Size}\\ \cmidrule(l){4-9}& & & $100$ & $500$ & $1,000$ & $5,000$ & $10,000$ & $20,000$\\ \midrule
& \multicolumn{2}{c}{Memorization} & $\textbf{0.970}$ & $\textbf{0.800}$ & $\textbf{0.760}$ & $\textbf{0.460}$ & $\textbf{0.390}$ & $\textbf{0.310}$ \\ \midrule
\multirow{4}{*}{MI-Face} & Dataset-level & FID $\downarrow$  & $391.454$ &$348.294$ &$334.498$ &$302.592$ &$315.456$ &$319.278$ \\ \cmidrule(l){3-9}
& \multirow{3}{*}{Sample-level} & SSIM $\uparrow$ & $0.028 (76.00\%)$ & $0.042 (59.20\%)$ & $0.050 (53.40\%)$ & $0.060 (47.00\%)$ & $0.064 (46.54\%)$ & $0.065 (44.27\%)$ \\
& & PSNR $\uparrow$ & $9.591 (57.00\%)$ & $9.896 (19.20\%)$ & $10.425 (12.80\%)$ & $11.414 (3.44\%)$ & $11.887 (1.94\%)$ & $11.947 (1.34\%)$ \\
& & MSE $\downarrow$  & $0.113 (57.00\%)$ & $0.104 (19.20\%)$ & $0.092 (12.80\%)$ & $0.073 (3.44\%)$ & $0.065 (1.94\%)$ & $0.064 (1.34\%)$ \\ \cmidrule(l){1-9}
\multirow{4}{*}{DeepDream} & Dataset-level & FID $\downarrow$  & $295.639$ &$288.325$ &$265.255$ &$256.581$ &$250.863$ &$251.972$ \\
\bottomrule
\end{tabular}
\label{table:eval_results_cifar100_vgg16}
}
\end{table*}

\begin{table*}[!t]
\centering
\caption{Evaluation results of existing reconstruction attacks with defenses MID and DP. 
The target model is VGG16 trained on CelebA with 2 different sizes.
Attacks with a gray background belong to the dynamic training type.
For FID and MSE, a lower score indicates better reconstruction quality; while for SSIM and PSNR, a higher score indicates better performance. 
}
\scalebox{0.75}{
\begin{tabular}{@{}cccccccccc@{}}
\toprule
\multirow{2}{*}{Attack}& \multicolumn{2}{c}{\multirow{2}{*}{Metrics}} &\multicolumn{2}{c}{No Defense} & \multicolumn{2}{c}{MID} & \multicolumn{2}{c}{DP} \\ \cmidrule(l){4-9}& & & $1,000$ & $20,000$ & $1,000$ & $20,000$ & $1,000$ & $20,000$\\ \midrule
\multirow{4}{*}{MI-Face} & Dataset-level & FID $\downarrow$  & $362.361$ &$336.906$ &$370.543$ &$363.668$ &$378.974$ &$370.767$ \\ \cmidrule(l){3-9}
& \multirow{3}{*}{Sample-level} & SSIM $\uparrow$ & $0.014 (100.00\%)$ & $0.023 (54.80\%)$ & $0.015 (100.00\%)$ & $0.027 (56.02\%)$ & $0.010 (100.00\%)$ & $0.021 (48.32\%)$ \\
& & PSNR $\uparrow$ & $7.978 (100.00\%)$ & $9.037 (10.08\%)$ & $8.066 (100.00\%)$ & $9.018 (9.82\%)$ & $7.868 (100.00\%)$ & $9.686 (6.69\%)$ \\
& & MSE $\downarrow$  & $0.163 (100.00\%)$ & $0.126 (10.08\%)$ & $0.169 (100.00\%)$ & $0.126 (9.82\%)$ & $0.171 (100.00\%)$ & $0.128 (6.69\%)$ \\ \cmidrule(l){1-9}
\multirow{4}{*}{DeepDream} & Dataset-level & FID $\downarrow$  & $337.139$ &$370.396$ &$373.527$ &$388.262$ &$365.830$ &$352.024$ \\ \cmidrule(l){3-9}
& \multirow{3}{*}{Sample-level} & SSIM $\uparrow$ & $0.079 (100.00\%)$ & $0.346 (6.81\%)$ & $0.020 (100.00\%)$ & $0.150 (27.58\%)$ & $0.020 (100.00\%)$ & $0.149 (23.12\%)$ \\
\bottomrule
\end{tabular}
}
\label{table:eval_results_celeba_defense}
\end{table*}

\begin{table*}[!t]
\centering
\caption{Evaluation results of existing reconstruction attacks. 
The target model is MobileNetV2 trained on CelebA with 6 different sizes.
Attacks with a gray background belong to the dynamic training type.
For FID and MSE, a lower score indicates better reconstruction quality; while for SSIM and PSNR, a higher score indicates better performance.
}
\scalebox{0.75}{
\begin{tabular}{@{}cccccccccc@{}}
\toprule
\multirow{2}{*}{Attack}& \multicolumn{2}{c}{\multirow{2}{*}{Metrics}} &\multicolumn{6}{c}{Target Data Size}\\ \cmidrule(l){4-9}& & & $1,000$ & $2,000$ & $5,000$ & $10,000$ & $15,000$ & $20,000$\\ \midrule
& \multicolumn{2}{c}{Memorization} & $1.000$ & $0.987$ & $0.864$ & $0.626$ & $0.474$ & $0.377$  \\ \midrule
\multirow{4}{*}{MI-Face} & Dataset-level & FID $\downarrow$ & $366.886$ &$371.293$ &$359.870$ &$365.360$ &$352.610$ &$361.813$ \\ \cmidrule(l){3-9}
& \multirow{3}{*}{Sample-level} & SSIM $\uparrow$ & $0.010 (100.00\%)$ & $0.015 (72.45\%)$ & $0.019 (63.14\%)$ & $0.020 (60.78\%)$ & $0.020 (58.72\%)$ & $0.020 (57.73\%)$ \\
& & PSNR $\uparrow$ & $8.047 (100.00\%)$ & $8.283 (53.15\%)$ & $8.683 (26.08\%)$ & $9.178 (15.79\%)$ & $9.242 (11.93\%)$ & $9.562 (8.54\%)$ \\
& & MSE $\downarrow$ & $0.161 (100.00\%)$ & $0.151 (53.15\%)$ & $0.137 (26.08\%)$ & $0.122 (15.79\%)$ & $0.120 (11.93\%)$ & $0.112 (8.54\%)$ \\ \cmidrule(l){1-9}
\multirow{4}{*}{DeepDream} & Dataset-level & FID $\downarrow$ & $343.531$ &$332.206$ &$352.673$ &$351.912$ &$356.788$ &$349.529$ \\
\bottomrule
\end{tabular}
}
\label{table:eval_results_celeba_mobilenetv2}
\end{table*}

\begin{table*}[!t]
\centering
\caption{Evaluation results of existing reconstruction attacks. 
The target model is ResNet-18 trained on CelebA with 6 different sizes.
Attacks with a gray background belong to the dynamic training type.
For FID and MSE, a lower score indicates better reconstruction quality; while for SSIM and PSNR, a higher score indicates better performance.}
\scalebox{0.75}{
\begin{tabular}{@{}cccccccccc@{}}
\toprule
\multirow{2}{*}{Attack}& \multicolumn{2}{c}{\multirow{2}{*}{Metrics}} &\multicolumn{6}{c}{Target Data Size}\\ \cmidrule(l){4-9}& & & $1,000$ & $2,000$ & $5,000$ & $10,000$ & $15,000$ & $20,000$\\ \midrule
& \multicolumn{2}{c}{Memorization} & $0.999$ & $0.984$ & $0.871$ & $0.560$ & $0.374$ & $0.317$ \\ \midrule
\multirow{4}{*}{MI-Face} & Dataset-level & FID $\downarrow$ & $364.926$ &$326.776$ &$313.911$ &$363.683$ &$371.507$ &$369.358$ \\ \cmidrule(l){3-9}
& \multirow{3}{*}{Sample-level} & SSIM $\uparrow$ & $0.020 (100.00\%)$ & $0.016 (69.55\%)$ & $0.023 (56.92\%)$ & $0.021 (58.08\%)$ & $0.021 (57.39\%)$ & $0.022 (56.68\%)$ \\
& & PSNR $\uparrow$ & $9.243 (100.00\%)$ & $9.120 (50.95\%)$ & $9.796 (22.46\%)$ & $9.998 (12.14\%)$ & $10.108 (8.38\%)$ & $10.087 (6.76\%)$ \\
& & MSE $\downarrow$ & $0.125 (100.00\%)$ & $0.125 (50.95\%)$ & $0.107 (22.46\%)$ & $0.101 (12.14\%)$ & $0.098 (8.38\%)$ & $0.099 (6.76\%)$ \\ \cmidrule(l){1-9}
\multirow{4}{*}{DeepDream} & Dataset-level & FID $\downarrow$ & $284.058$ &$283.468$ &$291.180$ &$255.410$ &$269.499$ &$265.756$ \\
\bottomrule
\end{tabular}
}
\label{table:eval_results_celeba_resnet18}
\end{table*}

\begin{table*}[!t]
\centering
\caption{Evaluation results of existing reconstruction attacks. 
The target model is VGG16 trained on CIFAR10 with 6 different sizes.
Attacks with a gray background belong to the dynamic training type.
For FID and MSE, a lower score indicates better reconstruction quality; while for SSIM and PSNR, a higher score indicates better performance.}
\scalebox{0.75}{
\begin{tabular}{@{}cccccccccc@{}}
\toprule
\multirow{2}{*}{Attack}& \multicolumn{2}{c}{\multirow{2}{*}{Metrics}} &\multicolumn{6}{c}{Target Data Size}\\ \cmidrule(l){4-9}& & & 100 & 500 & $1,000$ & $5,000$ & $10,000$ & $20,000$\\ \midrule
& \multicolumn{2}{c}{Memorization} & $0.512$ & $0.339$ & $0.289$ & $0.140$ & $0.090$ & $0.080$ \\ \midrule
\multirow{4}{*}{MI-Face} & Dataset-level & FID $\downarrow$ & $402.753$ &$392.789$ &$384.108$ &$384.251$ &$392.885$ &$389.169$ \\ \cmidrule(l){3-9}
& \multirow{3}{*}{Sample-level} & SSIM $\uparrow$ & $0.038 (58.00\%)$ & $0.048 (54.20\%)$ & $0.054 (48.70\%)$ & $0.058 (40.64\%)$ & $0.060 (38.09\%)$ & $0.064 (35.38\%)$ \\
& & PSNR $\uparrow$ & $10.924 (11.00\%)$ & $11.240 (4.00\%)$ & $11.334 (1.90\%)$ & $11.487 (0.64\%)$ & $11.364 (0.41\%)$ & $11.474 (0.26\%)$ \\
& & MSE $\downarrow$ & $0.083 (11.00\%)$ & $0.076 (4.00\%)$ & $0.074 (1.90\%)$ & $0.071 (0.64\%)$ & $0.073 (0.41\%)$ & $0.071 (0.26\%)$ \\ \cmidrule(l){1-9}
\multirow{4}{*}{DeepDream} & Dataset-level & FID $\downarrow$ & $466.706$ &$411.230$ &$421.903$ &$313.134$ &$322.337$ &$337.856$ \\
\bottomrule
\end{tabular}
}

\label{table:eval_results_cifar_vgg16}
\end{table*}

\begin{table*}[!t]
\centering
\caption{Evaluation results of existing reconstruction attacks. 
The target model is MobileNetV2 trained on CIFAR10 with 6 different sizes.
Attacks with a gray background belong to the dynamic training type.
For FID and MSE, a lower score indicates better reconstruction quality; while for SSIM and PSNR, a higher score indicates better performance.}
\scalebox{0.75}{
\begin{tabular}{@{}cccccccccc@{}}
\toprule
\multirow{2}{*}{Attack}& \multicolumn{2}{c}{\multirow{2}{*}{Metrics}} &\multicolumn{6}{c}{Target Data Size}\\ \cmidrule(l){4-9}& & & 100 & 500 & $1,000$ & $5,000$ & $10,000$ & $20,000$\\ \midrule
& \multicolumn{2}{c}{Memorization} & $0.613$ & $0.369$ & $0.319$ & $0.140$ & $0.090$ & $0.020$ \\ \midrule
\multirow{4}{*}{MI-Face} & Dataset-level & FID $\downarrow$ & $422.825$ &$399.889$ &$402.926$ &$390.659$ &$392.791$ &$397.124$ \\ \cmidrule(l){3-9}
& \multirow{3}{*}{Sample-level} & SSIM $\uparrow$ & $0.037 (60.00\%)$ & $0.049 (51.80\%)$ & $0.049 (49.20\%)$ & $0.057 (40.60\%)$ & $0.060 (37.94\%)$ & $0.062 (34.78\%)$ \\
& & PSNR $\uparrow$ & $10.368 (15.00\%)$ & $11.117 (3.40\%)$ & $11.015 (1.50\%)$ & $11.358 (0.72\%)$ & $11.333 (0.42\%)$ & $11.287 (0.29\%)$ \\
& & MSE $\downarrow$ & $0.093 (15.00\%)$ & $0.078 (3.40\%)$ & $0.079 (1.50\%)$ & $0.073 (0.72\%)$ & $0.074 (0.42\%)$ & $0.075 (0.29\%)$ \\ \cmidrule(l){1-9}
\multirow{4}{*}{DeepDream} & Dataset-level & FID $\downarrow$ & $408.101$ &$371.711$ &$355.894$ &$287.888$ &$287.213$ &$291.196$ \\
\bottomrule
\end{tabular}
}

\label{table:eval_results_cifar_mobilenetv2}
\end{table*}

\begin{table*}[!t]
\centering
\caption{Evaluation results of existing reconstruction attacks. 
The target model is ResNet-18 trained on CIFAR10 with 6 different sizes.
Attacks with a gray background belong to the dynamic training type.
For FID and MSE, a lower score indicates better reconstruction quality; while for SSIM and PSNR, a higher score indicates better performance.}
\scalebox{0.75}{
\begin{tabular}{@{}cccccccccc@{}}
\toprule
\multirow{2}{*}{Attack}& \multicolumn{2}{c}{\multirow{2}{*}{Metrics}} &\multicolumn{6}{c}{Target Data Size}\\ \cmidrule(l){4-9}& & & 100 & 500 & $1,000$ & $5,000$ & $10,000$ & $20,000$\\ \midrule
& \multicolumn{2}{c}{Memorization} & $0.504$ & $0.305$ & $0.349$ & $0.140$ & $0.100$ & $0.050$ \\ \midrule
\multirow{4}{*}{MI-Face} & Dataset-level & FID $\downarrow$ & $410.967$ &$390.645$ &$388.878$ &$395.207$ &$392.996$ &$396.056$ \\ \cmidrule(l){3-9}
& \multirow{3}{*}{Sample-level} & SSIM $\uparrow$ & $0.038 (58.00\%)$ & $0.051 (50.40\%)$ & $0.054 (49.00\%)$ & $0.057 (40.72\%)$ & $0.060 (37.79\%)$ & $0.062 (34.56\%)$ \\
& & PSNR $\uparrow$ & $10.628 (16.00\%)$ & $11.295 (3.00\%)$ & $11.378 (1.70\%)$ & $11.355 (0.66\%)$ & $11.339 (0.43\%)$ & $11.360 (0.28\%)$  \\
& & MSE $\downarrow$ & $0.087 (16.00\%)$ & $0.075 (3.00\%)$ & $0.073 (1.70\%)$ & $0.074 (0.66\%)$ & $0.074 (0.43\%)$ & $0.073 (0.28\%)$ \\ \cmidrule(l){1-9}
\multirow{4}{*}{DeepDream} & Dataset-level & FID $\downarrow$ & $487.742$ &$456.497$ &$407.365$ &$313.767$ &$316.199$ &$285.730$ \\
\bottomrule
\end{tabular}
}
\label{table:eval_results_cifar_resnet18}
\end{table*}

\begin{table*}[!t]
\centering
\caption{Evaluation results of existing reconstruction attacks. 
The target model is VGG16 trained on MNIST with 6 different sizes.
Attacks with a gray background belong to the dynamic training type.
For FID and MSE, a lower score indicates better reconstruction quality; while for SSIM and PSNR, a higher score indicates better performance.}
\scalebox{0.75}{
\begin{tabular}{@{}cccccccccc@{}}
\toprule
\multirow{2}{*}{Attack}& \multicolumn{2}{c}{\multirow{2}{*}{Metrics}} &\multicolumn{6}{c}{Target Data Size}\\ \cmidrule(l){4-9}& & & 100 & 500 & $1,000$ & $5,000$ & $10,000$ & $20,000$\\ \midrule
& \multicolumn{2}{c}{Memorization} & $0.078$ & $0.028$ & $0.009$ & $0.000$ & $0.000$ & $0.000$ \\ \midrule
\multirow{4}{*}{MI-Face} & Dataset-level & FID $\downarrow$ & $415.896$ &$397.556$ &$392.129$ &$354.141$ &$375.405$ &$351.019$ \\ \cmidrule(l){3-9}
& \multirow{3}{*}{Sample-level} & SSIM $\uparrow$ & $0.031 (68.00\%)$ & $0.043 (48.80\%)$ & $0.046 (49.60\%)$ & $0.054 (41.66\%)$ & $0.054 (39.33\%)$ & $0.069 (33.95\%)$ \\
& & PSNR $\uparrow$ & $5.956 (34.00\%)$ & $6.418 (21.80\%)$ & $6.476 (21.40\%)$ & $6.554 (15.12\%)$ & $6.319 (12.78\%)$ & $6.734 (13.96\%)$ \\
& & MSE $\downarrow$ & $0.255 (34.00\%)$ & $0.229 (21.80\%)$ & $0.226 (21.40\%)$ & $0.222 (15.12\%)$ & $0.234 (12.78\%)$ & $0.214 (13.96\%)$ \\ \cmidrule(l){1-9}
\multirow{4}{*}{DeepDream} & Dataset-level & FID $\downarrow$ & $333.082$ &$348.128$ &$369.028$ &$309.106$ &$308.409$ &$298.882$ \\
\bottomrule
\end{tabular}
}
\label{table:eval_results_mnist_vgg16}
\end{table*}

\begin{table*}[!t]
\centering
\caption{Evaluation results of existing reconstruction attacks. 
The target model is MobileNetV2 trained on MNIST with 6 different sizes.
Attacks with a gray background belong to the dynamic training type.
For FID and MSE, a lower score indicates better reconstruction quality; while for SSIM and PSNR, a higher score indicates better performance.}
\scalebox{0.75}{
\begin{tabular}{@{}cccccccccc@{}}
\toprule
\multirow{2}{*}{Attack}& \multicolumn{2}{c}{\multirow{2}{*}{Metrics}} &\multicolumn{6}{c}{Target Data Size}\\ \cmidrule(l){4-9}& & & 100 & 500 & $1,000$ & $5,000$ & $10,000$ & $20,000$\\ \midrule
& \multicolumn{2}{c}{Memorization} & $0.059$ & $0.038$ & $0.040$ & $0.009$ & $0.000$ & $0.000$ \\ \midrule
\multirow{4}{*}{MI-Face} & Dataset-level & FID $\downarrow$ & $443.258$ &$418.851$ &$414.673$ &$405.400$ &$394.948$ &$397.093$ \\ \cmidrule(l){3-9}
& \multirow{3}{*}{Sample-level} & SSIM $\uparrow$ & $0.031 (54.00\%)$ & $0.038 (56.40\%)$ & $0.043 (53.60\%)$ & $0.050 (45.44\%)$ & $0.052 (39.79\%)$ & $0.055 (38.14\%)$ \\
& & PSNR $\uparrow$ & $5.908 (32.00\%)$ & $6.044 (21.60\%)$ & $6.425 (18.50\%)$ & $6.342 (13.42\%)$ & $6.389 (11.55\%)$ & $6.299 (9.26\%)$ \\
& & MSE $\downarrow$ & $0.257 (32.00\%)$ & $0.249 (21.60\%)$ & $0.229 (18.50\%)$ & $0.233 (13.42\%)$ & $0.230 (11.55\%)$ & $0.235 (9.26\%)$ \\ \cmidrule(l){1-9}
\multirow{4}{*}{DeepDream} & Dataset-level & FID $\downarrow$ & $341.891$ &$325.367$ &$359.402$ &$254.154$ &$293.305$ &$331.156$ \\
\bottomrule
\end{tabular}
}
\label{table:eval_results_mnist_mobilenetv2}
\end{table*}

\begin{table*}[!t]
\centering
\caption{Evaluation results of existing reconstruction attacks. 
The target model is ResNet-18 trained on MNIST with 6 different sizes.
Attacks with a gray background belong to the dynamic training type.
For FID and MSE, a lower score indicates better reconstruction quality; while for SSIM and PSNR, a higher score indicates better performance.}
\scalebox{0.75}{
\begin{tabular}{@{}cccccccccc@{}}
\toprule
\multirow{2}{*}{Attack}& \multicolumn{2}{c}{\multirow{2}{*}{Metrics}} &\multicolumn{6}{c}{Target Data Size}\\ \cmidrule(l){4-9}& & & 100 & 500 & $1,000$ & $5,000$ & $10,000$ & $20,000$\\ \midrule
& \multicolumn{2}{c}{Memorization} & $0.045$ & $0.031$ & $0.025$ & $0.016$ & $0.009$ & $0.008$ \\ \midrule
\multirow{4}{*}{MI-Face} & Dataset-level & FID $\downarrow$ & $422.880$ &$428.743$ &$411.834$ &$356.822$ &$371.393$ &$380.956$ \\ \cmidrule(l){3-9}
& \multirow{3}{*}{Sample-level} & SSIM $\uparrow$ & $0.036 (59.00\%)$ & $0.043 (52.00\%)$ & $0.047 (50.50\%)$ & $0.061 (38.18\%)$ & $0.063 (36.32\%)$ & $0.063 (35.86\%)$ \\
& & PSNR $\uparrow$ & $6.250 (37.00\%)$ & $6.305 (21.60\%)$ & $6.425 (19.60\%)$ & $6.475 (15.48\%)$ & $6.491 (13.22\%)$ & $6.544 (11.50\%)$ \\
& & MSE $\downarrow$ & $0.238 (37.00\%)$ & $0.235 (21.60\%)$ & $0.228 (19.60\%)$ & $0.226 (15.48\%)$ & $0.225 (13.22\%)$ & $0.222 (11.50\%)$ \\ \cmidrule(l){1-9}
\multirow{4}{*}{DeepDream} & Dataset-level & FID $\downarrow$ & $362.900$ &$332.212$ &$342.366$ &$314.468$ &$301.383$ &$318.317$ \\
\bottomrule
\end{tabular}
}
\label{table:eval_results_mnist_resnet18}
\end{table*}

\end{document}